\documentclass[prb,aps,twocolumn,amssymb,amsmath]{revtex4-2}

\usepackage{graphicx} 

\begin{document}
\title{Emergent $\mathrm{SU(3)}$ symmetry  in a four leg spin tube} 
\author{Edmond Orignac}
\affiliation{ENSL, CNRS, Laboratoire de Physique, F-69342 Lyon, France}
\date{\today}
\begin{abstract}
  We consider an antiferromagnetic four leg spin-1/2 tube using abelian and non-abelian bosonization. 
  We show that in the limit of weak interchain coupling, the most relevant interaction gives rise to an emergent $\mathrm{SU(3)}$ symmetry, broken only by marginal interactions that can be canceled by diagonal interchain couplings. We discuss the low energy spectrum in the semiclassical limit and using a mapping to a trimerized $\mathrm{SU(3)}$ spin chain. We establish that the correlation functions of ferroquadrupolar operators can be used to reveal the emergent symmetry. 
\end{abstract} 
\maketitle

\section{Introduction}
\label{sec:intro}

Emergent symmetry\cite{gomes_2016} is a symmetry that is obtained when
restricting to a particular subspace of the Hilbert space, such as the
low energy subspace, and differs from the symmetry of the full
Hamiltonian. A well known example is furnished by quantum critical
points of one-dimensional many-body systems, where scale and conformal
invariance\cite{difrancesco_book_conformal} determine the low-energy
spectrum and the low-energy long-wavelength response functions. In
such case, only a finite number of operators are relevant at the
critical point, and when the system is tuned to the critical point,
all the irrelevant symmetry breaking fields are canceled by the
renormalization group flow. As an example of unexpected emergent symmetry,  a critical point described by the $\mathrm{SU(3)_1}$ Wess-Zumino-Novikov-Witten\cite{difrancesco_book_conformal,witten_wz} model has been proposed\cite{chen_2015} in a spin-2 chain model in which only $\mathrm{SU(2)}$ symmetry should be present.  However, the critical character of the correlations in that model remains debated\cite{li_2022}. Examples of fully gapped or partially
gapped interacting system with emergent symmetry are more
scarce. One example is a two-leg spin-1/2 ladder with biquadratic interaction that shows $\mathrm{SO(4)}$ symmetry\cite{nersesyan_biquad}. A mechanism for emergent symmetry in
gapped systems is Dynamical Symmetry Enlargement (DSE) symmetry which is
is produced by a marginally relevant renormalization \cite{lin_so8}
flow in which the relevant parameters converge asymptotically to those
of the $\mathrm{SO(2N)}$ Gross-Neveu model.\cite{gross_neveu} In the
the two-leg Hubbard ladder, DSE has been predicted from
$\mathrm{SU(2)}$ in the lattice model to $\mathrm{SO(8)}$ in the low
energy theory at
half-filling\cite{lin_so8,konik_exact_commensurate_ladder,essler04_condmat_exact_review}
and from $\mathrm{SU(2)}$ in the lattice model to $\mathrm{SO(6)}$ at
low energy away from half-filling\cite{schulz_son,essler_2007}. A DSE
from $\mathrm{SU(4)}\sim \mathrm{SO(6)}$ to $\mathrm{SO(8)}$ has also
been found in a generalization of the Hubbard model with
$\mathrm{SU(4)}$ symmetry\cite{assaraf_2004}, and a DSE from
$\mathrm{SU(2)}$ to $\mathrm{SO(6)}$ in a model of zigzag carbon
nanotubes.\cite{bunder_dynamical_2007} In all these models, DSE allows
to take advantage of the integrability of the Gross-Neveu
model\cite{zamolodchikov79_smatrices,karowski81_gross_neveu} to obtain
form-factor expansions\cite{karowski_ff} of the correlation and
response functions in the two-leg Hubbard
ladder.\cite{konik_exact_commensurate_ladder,essler_2007} In the
present paper, we wish to propose an example of emergent symmetry in a
four leg spin tube system.  Spin-1/2 tube
systems\cite{cabra_ladders,kawano_3legs,tandon_3legs,arlego_4leg_2011}
are made of antiferromagnetic spin-1/2 chains with a transverse
coupling satisfying periodic boundary conditions, whereas in planar
spin ladders the transverse coupling obeys open boundary
conditions\cite{dagotto_ladder_review,dagotto_supra_ladder_review}.
Experimentally, three leg spin tubes were proposed in
$\mathrm{Na_2V_3O_7}$ \cite{gavilano03_spintube}
$\mathrm{CsCrF _4}$\cite{manaka2009} and
$\mathrm{[(CuCl_2 tachH)_3 Cl]Cl_2}$ and four-leg spin tubes in
$\mathrm{Sul-CuCl4}$
\cite{garlea2008,zheludev_universality,garlea_2009,schrettle_2013,glazkov_2020}.
Planar antiferromagnetic spin-1/2 ladder systems exhibit an even-odd
alternation of ground state magnetic properties: ladders with an odd
number of legs present a ground state with quasi long range order and
a gap branch of linearly dispersing excitations, while ladders with an
even number of legs present a ground state with short range order and
gapped
excitations\cite{dagotto_ladder_review,dagotto_supra_ladder_review,lecheminant_revue_1d}. Such
result is analogous to the alternation between short range order for
integer spin and quasi-long range order for half-odd integer spin in
antiferromagnetic spin chains\cite{haldane_gap,affleck_houches} and
can be understood in terms of a topological contribution to the action
\cite{sierra_nchains,cabra_field-theoretical_2004}. In spin tubes with
an odd number of legs, the periodic character of the transverse
interaction can modify the nature of the ground
state\cite{cabra_ladders,kawano_3legs,totsuka_3legs,sakai_2010}. In
contrast with the gapless three leg ladder, the three leg spin tube
presents short range order and a spin
gap\cite{cabra_ladders,kawano_3legs} as a result of frustration in the
rung direction.  With an even number of legs, the transverse
interaction is not frustrating, and the spin gap phase of the tube is
analogous to the one the planar ladder. In the case of the four-leg
spin tube, series expansion studies have confirmed the presence of a
spin gap, in the limit of strong rung coupling, but found a richer
excitation spectrum than in the two-leg ladder\cite{arlego_4leg_2011}.
In the present manuscript, we consider the 4 leg spin-1/2 tube at weak
coupling using
bosonization\cite{lecheminant_revue_1d,giamarchi_book_1d} and
conformal field theory methods. We find that although the ground state
has the same gap and short range order as in a planar four leg ladder,
the excitation spectrum presents an emergent $\mathrm{SU(3)}$
symmetry. Going beyond the spectrum, we also show that some
ferroquadrupolar\cite{hikihara_2008} (or nematic) order parameters can
reveal the emergent symmetry via their correlation functions. The
microscopic model is introduced in Sec.~\ref{sec:model}, in
Sec.~\ref{sec:bosonization} the non-abelian bosonization is used to
reveal the emergent symmetry, in Sec.~\ref{sec:abelian} a more
detailed abelian bosonization treatment allows to describe the
operators whose correlations reveal the $\mathrm{SU(3)}$ symmetry of
the low energy theory. We present our conclusions in
Sec.~\ref{sec:ccl}.


\section{Model and Hamiltonian}
\label{sec:model}
We consider a four leg spin tube made of four antiferromagnetic spin-1/2 chains with intrachain exchange interaction $J_\parallel$  and interchain exchange $J_\perp$. Its Hamiltonian reads 
\begin{eqnarray}
  \label{eq:4leg-tube}
  H=\sum_{j=1,N \atop p=1,4} J_\parallel \mathbf{S}_{j,p}\cdot\mathbf{S}_{j+1,p} + J_\perp  S_{j,p}\cdot S_{j,p+1},   
\end{eqnarray}
with the identification $\mathbf{S}_{j,5}=\mathbf{S}_{j,1}$.   
We see that the interchain exchange interaction can be rewritten
\begin{equation}
  \label{eq:interchain}
   J_\perp \sum_{j=1,N} (\mathbf{S}_{j,1}+\mathbf{S}_{j,3})\cdot    (\mathbf{S}_{j,2}+\mathbf{S}_{j,4}), 
 \end{equation}
 or
 \begin{eqnarray}
   \label{eq:multiplets} 
  &&  \sum_{j=1,N} \frac {J_\perp} 2 \left[    (\mathbf{S}_{j,1}+\mathbf{S}_{j,2}  +\mathbf{S}_{j,3}+\mathbf{S}_{j,4})^2 -(\mathbf{S}_{j,1}+\mathbf{S}_{j,3})^2 \right.\nonumber \\
   && \left. -(\mathbf{S}_{j,2}+\mathbf{S}_{j,4})^2\right]. 
 \end{eqnarray}
 When the squares are decoupled, $J_\parallel=0$, the spins on the odd
 and on the even chains add up forming either a 
 spin $0$ or a spin $1$ state. When at least one of the pair of spins is in 
 the singlet state, the rung energy~(\ref{eq:multiplets}) vanishes. When
 both pairs form a triplet, the rung energy is $-2J_\perp$  when the two
 triplets combine into a singlet, $-J_\perp$  when they combine into a
 triplet, $J_\perp$ when they combine into an $S=2$ quintuplet.\cite{arlego_4leg_2011}  When a
 small $J_\parallel \ll J_\perp$ is introduced, the ground state
 remains the singlet state formed of two triplets on the diagonals on
 the square. The lowest energy magnon band results from triplets
 generated by the pair of triplets on the diagonals of the square. Two
 magnon bands of higher energy are formed from one diagonal in the
 triplet state and the other diagonal in the singlet state.  Finally,
 a singlet excitation  from both diagonals in the singlet
 state, and a $S=2$ excitation formed from both diagonals in the
 triplet state can obtain\cite{arlego_4leg_2011}.

 In the opposite limit of $J_\parallel \ll J_\perp$ we consider the
 model using non-abelian\cite{tsvelik_field_theory} and abelian\cite{giamarchi_book_1d} bosonization. The first approach takes full advantage of the symmetries of the model, while the second approach gives a more detailed picture of the relevant observables. 
 
\section{Non-abelian bosonization approach}
\label{sec:bosonization}

Using non-abelian bosonization in the limit of $J_\perp=0$, the Hamiltonian of the four decoupled chains reads
\begin{eqnarray}
 \mathcal{H}_0=\sum_{p=1}^4  \frac{2\pi u}{3} \int dx (\mathbf{J}_{R,p}  \cdot \mathbf{J}_{R,p} + \mathbf{J}_{L,p}  \cdot \mathbf{J}_{L,p}),   
\end{eqnarray}
where  $u=\frac{\pi}2 J_\parallel a$ is the velocity of spin excitations, with $a$ the lattice spacing. The operators $J_{\nu,p}$  ($\nu=R,L$) are the $SU(2)_1$ currents\cite{tsvelik_field_theory}  of a Wess Zumino Novikov Witten (WZNW) model\cite{witten_wz,knizhnik_wz,difrancesco_book_conformal}. Each of these models has central charge $c=1$.   
The spin operators on chain $j$ are represented\cite{tsvelik_field_theory} as
\begin{equation}\label{eq:spin-bosonized}  
\mathbf{S}_{j,p} = \mathbf{J}_{R,p}(ja) +  \mathbf{J}_{R,p}(ja)  + \lambda (-)^j \mathbf{n}_p(ja), 
\end{equation}
where the current operators $\mathbf{J}_{\nu,p}$ of momentum $q\sim 0$ have scaling dimension $1$ while the staggered spin operators $\mathbf{n}_p$ of momentum $q \sim \frac \pi a$ are $SU(2)_1$ WZNW spin-1/2 primaries with scaling dimension $1/2$. The coefficient $\lambda$ in known quantitatively in XXZ spin-1/2 chains\cite{lukyanov_xxz_asymptotics,hikihara_xxz,takayoshi_2010}.  The most relevant contribution in the renormalization group sense is  given by the staggered operators in (\ref{eq:interchain}) and reads
\begin{equation}\label{eq:stag-int} 
\mathcal{H}_{int,b} =  \frac{J_\perp\lambda^2} a  \int dx (\mathbf{n}_1+\mathbf{n}_3) \cdot (\mathbf{n}_2+\mathbf{n}_4),   
\end{equation}
while the current operators contribute a marginal interaction
\begin{equation}\label{eq:curr-int} 
 \mathcal{H}_{int,b} =  \frac{J_\perp} a \int dx \sum_{\nu,\nu'}(\mathbf{J}_{\nu,1}+\mathbf{J}_{\nu,3}) \cdot (\mathbf{J}_{\nu',2}+\mathbf{J}_{\nu',4}),  
\end{equation}
to the full bosonized Hamiltonian $\mathcal{H}=\mathcal{H}_0+\mathcal{H}_{int,f}+\mathcal{H}_{int,b}$. The relevant interaction in Eq.~(\ref{eq:stag-int}) gives rise to a gap $\Delta \sim J_\perp \lambda^2$ in the excitation spectrum, and the marginal interaction in Eq.~(\ref{eq:curr-int}) can yield logarithmic corrections to $\Delta$. 
The form of the interaction in Eqs.~(\ref{eq:stag-int})--(\ref{eq:curr-int}) hints that a coset construction\cite{difrancesco_book_conformal} 
\begin{equation}\label{eq:coset-ising-su22} 
  SU(2)_1 \times SU(2)_1 \sim SU(2)_2 \times \mathrm{Ising} 
\end{equation}
can be used to rewrite the model in terms of operators belonging to 
\begin{equation}
  \mathrm{Ising}_{odd} \times \mathrm{Ising}_{even} \times SU(2)_{2,even} \times SU(2)_{2,odd},  
\end{equation}
where $odd$ indicates that the coset construction is applied to the operators with odd chain index, and $even$ that it is applied to operators with even chain index.
The magnetic degrees of freedom are described by the $SU(2)_2$ WZNW models, and the remaining non-magnetic degrees of freedom\cite{shelton_spin_ladders} by the Ising models. Combining together the magnetic degrees of freedom, a second coset construction\cite{georges1995,ferrero_2007} 
\begin{equation}\label{eq:coset-susy-su24} 
  SU(2)_{2,even} \times SU(2)_{2,odd} \sim \frac{SU(2)_{2} \times SU(2)_{2}}{SU(2)_4} \times SU(2)_4,   
\end{equation}
 yields the final representation
\begin{equation}
   \mathrm{Ising}_{odd} \times \times \mathrm{Ising}_{even} \times \frac{SU(2)_{2} \times SU(2)_{2}}{SU(2)_4} \times SU(2)_4,     
 \end{equation}
 in which the coset $A(2,2)=\frac{SU(2)_{2} \times SU(2)_{2}}{SU(2)_4}$ is of central charge $2\times \frac 3 2 -2=1$ and belongs to the $\mathcal{N}=1$ superconformal minimal series\cite{difrancesco_book_conformal}.  The conformal weights $h_{rs}$ of the primary operators $\phi_{rs}$ of the coset are given by\cite{difrancesco_book_conformal}
 \begin{equation}
   h_{rs}=\frac{(3r-2s)^2-1}{48} +\frac {1-(-)^{r-s}}{32}.  
 \end{equation}
 We have $h_{rs}=h_{4-r, 6-s}$ so we only need $s\le 3$ in the Kac table~\ref{tab:coset-weights}.

 \begin{table}
   \centering
   \begin{tabular}{ccc}
     Operator & Weight & Sector \\
     \hline 
     $\phi_{(1,1)R}$ & 0 & Neveu-Schwarz (NS) \\
     $\phi_{(2,1) R}$& $\frac 3 8$ & Ramond (R)\\
     $\phi_{(3,1)R}$ &  1 & NS \\
     $\phi_{(1,2)R}$ & $\frac 1 {16}$ & R \\
     $\phi_{(2,2)R}$ & $\frac 1 {16}$ & NS \\
     $\phi_{(3,2)R}$ & $\frac 9 {16}$ &  R \\
     $\phi_{(1,3)R}$ & $\frac 1 6$ & NS\\
     $\phi_{(2,3)R}$ & $\frac 1 {24}$ & R \\
   \end{tabular}
   \caption{Conformal weights of the right  moving scaling fields in the coset $A(2,2)=SU(2)_2\times SU(2)_2/SU(2)_4$ coset seen as an $\mathcal{N}=1$ superconformal field theory.}
   \label{tab:coset-weights}
 \end{table}
 
Now, if we consider the staggered spin operators, according to Eq.~(\ref{eq:coset-ising-su22}), the sum of two spin-1/2 primaries (odd or even) in $SU(2)_1$ can be written as the product of a spin-1/2 primary in $SU(2)_2$ times an Ising operator.\cite{difrancesco_book_conformal} Since a spin $j$ primary in $SU(2)_k$ has scaling dimension $2 \frac{j(j+1)}{k+2}$, the spin-1/2 operators in $SU(2)_1$ have dimension $1/2$ while the spin-1/2 operators in $SU(2)_2$ have dimension $3/8$. The Ising operator has dimension $1/8$ and can be taken as an Ising disorder operator giving
\begin{eqnarray}
  \mathbf{n}_1+\mathbf{n}_3 \sim \mu_{odd} \mathbf{N}_{odd}, \\
  \mathbf{n}_2+\mathbf{n}_4 \sim \mu_{even} \mathbf{N}_{even},
\end{eqnarray}
allowing us to rewrite the most relevant interaction as
\begin{equation}
\mathcal{H}_{int,b} =  \frac{J_\perp \lambda^2}{a}  \mu_{odd} \mu_{even}  \mathbf{N}_{odd} \cdot \mathbf{N}_{even}. 
\end{equation}
Now, let's consider $\mathbf{N}_{odd} \cdot \mathbf{N}_{even}$ of
scaling dimension $3/4$. Using the second coset construction, Eq.~(\ref{eq:coset-susy-su24}), we can
rewrite it as a sum of product of one operator of the superconformal theory by
one operator of the $SU(2)_4$ theory.  Since both operators are of spin
1/2, their product yields operators of spin 0 (identity) or of
spin 1 in $SU(2)_4$. The spin-1 primaries in $SU(2)_4$ have scaling
dimension $2/3$, so the operator in the superconformal theory must
be of dimension $3/4-2/3=1/12$. Looking up the Kac
table~\ref{tab:coset-weights}, it is identified as
$\Phi_{(23)}(z,\bar{z})=\phi_{(23)R}(z)\phi_{(23)L}(\bar{z}) $. The
operator multiplying the identity in $SU(2)_4$ has to be of dimension
$3/4$ and the operator in the superconformal theory with matching
dimension\cite{difrancesco_book_conformal} is
$\Phi_{(21)}(z,\bar{z})$. Thus, we can write
\begin{equation}\label{eq:final-coset} 
  \mathbf{N}_{odd} \cdot \mathbf{N}_{even} \sim  \Phi_{(21)}+ \Phi_{(23)} \Phi^{(1)}_{SU(2)_4},  
\end{equation}
where $\Phi^{(1)}$ is an $SU(2)$ invariant combination of spin-1
primaries in $SU(2)_4$. Now, it is known that there exists a conformal
embedding\cite{difrancesco_book_conformal} $SU(3)_1 \subset SU(2)_4$ such that the the three $SU(2)_4$ currents plus the five spin-2 primaries of
$SU(2)_4$ can be written as eight $SU(3)_1$ currents and the spin-1
primaries can be expressed using the $SU(3)_1$ primaries in the
fundamental representation of $SU(3)$. The
interaction~(\ref{eq:final-coset}) can thus be rewritten using only
$SU(3)_1$ operators\cite{konik2015}. This implies that the most 
relevant interactions, Eq.~(\ref{eq:stag-int}),
are giving rise to a gapful ground state in which
the symmetry is enlarged from $SU(2)$ to $SU(3)$. 
In particular, the
excited states above the ground state belong to irreducible 
representations of $SU(3)$.  Moreover, some operators transforming according to different  irreducible representations of $SU(2)$ can belong to the same irreducible representation of $SU(3)$ and thus exhibit identical correlation functions.
Another model having $\mathrm{SU(3)}$ symmetric low energy spectrum, albeit less realistic than the four-leg tube, is  a two-leg ladder made of two spin-1 chains described by the  Takhtajan-Babujian Hamiltonian\cite{takhtajan_spin_s,babujian_spin_s}, whose low energy excitations are described by the $\mathrm{SU(2)}_2$ Wess-Zumino-Novikov-Witten model\cite{tsvelik_field}, and coupled by an exchange interaction. The coset decomposition, Eq.~(\ref{eq:coset-susy-su24}), yields the interchain interaction (without Ising disorder fields) and a spectrum with $\mathrm{SU(3)}$ symmetry is obtained.  
Of course, the marginal current-current interaction in Eq.~(\ref{eq:curr-int})
involves only the $SU(2)_4$ current and none of the spin-2 primaries, and lowers the symmetry of the full model back to $SU(2)$. However, such
a marginal perturbation is expected from perturbation theory
to  give only corrections $O(J_\perp^2\ln(J_\parallel/J_\perp)/J_\parallel)$ to
the gaps\cite{shelton_spin_ladders}  to the excited states, so that for weak coupling, the degeneracy lifting in the spectrum is at a much lower scale than the spin gap $\Delta=O(J_\perp)$.
Beyond perturbation theory, the  correction from the marginal terms can be estimated by the following renormalization group argument.
If the gap to some excited state is $\Delta_n$ its dependence on the scale $\ell$ is given by
\begin{equation}
  \Delta_n = J_\parallel e^{-\ell} \delta_n\left(\frac{J_\perp}{J_\parallel} e^\ell,\frac{J_\perp}{J_\parallel + \mathcal{C} J_\perp \ell}\right),  
\end{equation}
where the dimensionless gap $\delta_n$ depends on the dimensionless relevant and marginal couplings and $\mathcal{C}=O(1)$ is a prefactor entering the marginal flow equation. Renormalizing to the scale $\ell^*=\ln(J_\parallel/|J_\perp|)$, we find that the gap behaves as
\begin{equation}
  \Delta_n = J_\perp\delta_n\left(1,\frac{J_\perp}{J_\parallel + \mathcal{C} J_\perp \ln(J_\parallel/|J_\perp|)}\right),  
\end{equation}
and since $ \ln(J_\parallel/|J_\perp|) \ll J_\parallel/|J_\perp|$, $\delta_n$ can be expanded as a Taylor series. We note that the logarithmic corrections have been resummed in the denominator, and the first correction is then $O(J_\perp^2/J_\parallel)\ll |J_\perp|$ provided $J_\perp \ll J_\parallel$. For $J_\perp/J_\parallel$ sufficiently small, the approximate $\mathrm{SU(3)}$ symmetry is preserved.   The emergent $\mathrm{SU(3)}$ symmetry can be contrasted with the one obtained by DSE in the two-leg Hubbard ladder at half-filling\cite{lin_so8}. In the latter case, the coupling constants are all marginally relevant, and under the renormalization group flow, they flow towards the line that corresponds to the $\mathrm{SO(8)}$ Gross-Neveu model. In our case, there are both marginal and relevant couplings. The initial values of the relevant couplings are already on the $\mathrm{SU(3)}$ symmetric manifold, and the marginal couplings are driving the flow away from the symmetric manifold. However, their growth under the renormalization group being slow, the renormalized low-energy Hamiltonian always remain close to a Hamiltonian with $\mathrm{SU(3)}$ symmetry.   
In fact, by adding a diagonal rung interaction 
\begin{eqnarray}
  && -\frac{J_\perp}2  \sum_{j} \left[(\mathbf{S}_{j,1}+\mathbf{S}_{j,3})\cdot (\mathbf{S}_{j+1,2}+\mathbf{S}_{j+1,4}) \right. \nonumber \\ && \left.+ (\mathbf{S}_{j,2}+\mathbf{S}_{j,4})\cdot (\mathbf{S}_{j+1,1}+\mathbf{S}_{j+1,2})\right],  
\end{eqnarray}
to the lattice Hamiltonian, Eq.~(\ref{eq:4leg-tube}),  the marginal interaction is entirely canceled\cite{shelton_spin_ladders} and the $\mathrm{SU(3)}$ breaking interactions are irrelevant.
In such a model, the $\mathrm{SU(3)}$ symmetry in the low energy spectrum is easier to characterize in exact diagonalizations\cite{noack_2005}. 
Another consequence of Eq.~(\ref{eq:final-coset}) is that since
$\mathrm{(Ising)^2}$
\cite{zuber_77,schroer_ising,ogilvie_ising,boyanovsky_ising}, the
superconformal $c=1$ theory\cite{kiritsis1988} and the $SU(3)_1$
theory\cite{assaraf_su(n)} admit abelian bosonization\cite{haldane_bosonisation,giamarchi_book_1d} representations
one can use abelian bosonization to
recover Eq.~(\ref{eq:final-coset}) and express all operators in terms of
boson fields. This will be the object of Sec.~\ref{sec:abelian}.
In the present section,
we recall briefly the results obtained in Ref.~\onlinecite{kiritsis1988}.
Both $\phi_{(23)R}$ and $\phi_{(21)R}$ belong to the Ramond sector, and their bosonized expression is\cite{kiritsis1988,georges1995}
\begin{eqnarray}
  \phi_{(23)R}(z)=e^{\frac{i \Phi_R(z)}{2\sqrt{3}}}, \\
  \phi_{(21)R}(z)=e^{\frac{i \sqrt{3} \Phi_R(z)}{2}}, 
\end{eqnarray}
for
\begin{equation}
  H_R=v \int \frac{dx}{4\pi} (\nabla \Phi_R)^2.  
\end{equation}

Similar expressions hold for the antiholomorphic fields with $\Phi_L$ in the place of $\Phi_R$. This leads to a bosonized representation
\begin{equation}\label{eq:nonabelian-interaction} 
   \mathbf{N}_{odd} \cdot \mathbf{N}_{even}  \sim \cos (\sqrt{3}\phi_c) + \cos \left(\frac{\phi_c}{\sqrt{3}}\right)  \Phi^{(1)}_{SU(2)_4},  
   \end{equation}
   where $\phi_c=(\Phi_R+\Phi_L)/2$. In
   Eq.~(\ref{eq:nonabelian-interaction}), $\Phi^{(1)}$ is a
   combination of left and right moving spin-1 primary fields that is
   invariant under a global $\mathrm{SU(2)}$ rotation.  To obtain
   expressions for the spin operators $\mathbf{N}_{odd}$ or
   $\mathbf{N}_{even}$ themselves, we note that they must be the
   product of an operator in $A(2,2)$ by a primary operator of spin
   $1/2$. Matching the scaling dimensions gives a dimension
   $3/8-1/4=1/8$. According to table~\ref{tab:coset-weights}, there are two possible
   operators $\Phi_{(1,2)}$ and $\Phi_{(2,2)}$ with the required dimension.
   Both of them are
   twisted fields that don't have a representation in terms of a
   boson field\cite{kiritsis1988}. Moreover, the spin-1/2 primary
   operators in $\mathrm{SU(2)}_4$ cannot be expressed\cite{difrancesco_book_conformal}  in terms of the
    operators of $\mathrm{SU(3)}_1$. For that reason, the
    $\mathrm{SU(3)}$ symmetry of the low energy theory
    is not apparent in  the spin-spin correlation functions. 
   However, if we take a tensor $N_{odd}^a N_{even}^b$ which is
   rewritten as the product of an operator in $A(2,2)$ by a spin-1
   primary in $\mathrm{SU(2)}_4$, that is a $\mathrm{SU(3)_1}$
   primary, its correlation functions can reflect the underlying
   $\mathrm{SU(3)}$ symmetry. This suggests to consider symmetric tensor products, that is quadrupolar (nematic) order parameters to detect the $\mathrm{SU(3)}$ symmetry of the model.  Such nematic correlations are accessible in experimental systems by Resonant Inelastic X-Ray scattering measurements.\cite{savary_2015}.
   To conclude that section, we note that an alternative coset
   representation applicable to our model is
   given by $\mathrm{SU(2)}_1^4 \sim \mathrm{SU(2)}_4\times G_4$ with
   $G_4=\Bbb{Z}_2\times \mathrm{TIM}\times \Bbb{Z}_3$ a tensor
   product of minimal models\cite{lecheminant_2012,capponi_2013}.
   While its leads the same conclusion concerning the
   $\mathrm{SU(3)}$ symmetry of the low-energy Hamiltonian, it treats
   the odd and even ladders in a less symmetrical way since it is built from
   successive tensor products $\mathrm{SU(2)}_n\times\mathrm{SU(2)}_1$.
   This forces us to choose first a pair of spin chains and apply the coset representation~(\ref{eq:coset-ising-su22}), then decide in which order the remaining two spin chains are used to form coset representations of the tricritical Ising model (TIM) and of the three-state clock model $\Bbb{Z}_3$. We thus end up with two non-equivalent representations for the non-magnetic degrees of freedom of our model. Such a representation would be in fact more convenient in a case where the rung exchange interaction has reflection symmetry only around one of the diagonals of the tube.  

   \section{Abelian bosonization}\label{sec:abelian} 
In abelian bosonization\cite{luther_chaine_xxz,haldane_xxzchain,nijs_equivalence,giamarchi_book_1d}, the decoupled chains have the Hamiltonian~(\ref{eq:bosonized-legs}) 
\begin{equation}
  \label{eq:bosonized-legs}
 \mathcal{H}_0=\sum_{j=1}^4 \int \frac{dx}{2\pi} u \left[(\pi \Pi_j)^2 +(\partial_x \phi_j)^2 \right], 
\end{equation}
where $[\phi_j(x),\Pi_k(x')]=i\delta_{jk} \delta(x-x')$ and $u=\frac{\pi} 2 J_\parallel a$. 
Meanwhile,  the $\mathrm{SU(2)}_1$ currents are  
\begin{eqnarray}
  \label{eq:curr-bosonized}
  J_{\nu,p}^+(x)&=&(J_\nu^x+iJ_\nu^y)(x)=\frac 1{2\pi a} e^{-i\sqrt{2}(\theta_p-r_\nu \phi_p)(x)},\label{eq:jrplus} \\
  J_{\nu,p}^z&=&\frac 1{2\pi \sqrt{2}} \left[ r_\nu \pi \Pi_p - \partial_x \phi_p\right],
                 \label{eq:jrz}
\end{eqnarray}
with $r_R=1$ and $r_L=-1$ and $\partial_x \theta_p=\pi \Pi_p$, and  the spin-1/2 primaries are 
\begin{eqnarray} 
n_p^+(x)&=& (n_p^x+in_p^y)(x)= e^{-i\sqrt{2}\theta_p(x)},\label{eq:nplus} \\
  n_p^z(x)&=& \sin \sqrt{2} \phi_p (x)\label{eq:nz}, \\
  \epsilon_p(x)&=& \cos \sqrt{2} \phi_p(x)\label{eq:dimerization},  
\end{eqnarray}
where $\epsilon(x)$ is the dimerization operator, such that $\mathbf{S}_{j,p}\cdot \mathbf{S}_{j+1,p} \sim \frac{1}4 \left[(\pi \Pi_P)^2 + (\partial_x \phi_p)^2\right] + (-)^j \bar{\lambda} \epsilon_p(ja)$.  The coefficient $\bar{\lambda}$ has been determined in the case of XXZ spin-1/2 chains\cite{orignac04_spingap,takayoshi_2010}. 
\subsection{Hamiltonian in abelian bosonization}

\subsubsection{Derivation of the low energy Hamiltonian}

Introducing\cite{schulz_spins,strong_spinchains} 
\begin{eqnarray}\label{eq:rotation-oe} 
  \theta_{o,r}&=& \frac 1 {\sqrt{2}} (\theta_1 + r \theta_3) \;  \phi_{o,r}= \frac 1 {\sqrt{2}} (\phi_1 + r \phi_3) \\
    \theta_{e,r}&=& \frac 1 {\sqrt{2}} (\theta_2 + r \theta_4) \;  \phi_{o,r}= \frac 1 {\sqrt{2}} (\phi_2 + r \phi_4),   
\end{eqnarray}
the Hamiltonian of the decoupled chains becomes
\begin{equation}
  \mathcal{H}_0 = \sum_{\nu=e,o\atop r=\pm}  \int \frac{dx}{2\pi} u \left[(\pi \Pi_{\nu,r})^2 +(\partial_x \phi_{\nu,r})^2 \right], 
\end{equation}
and can be rewritten in terms of Majorana fermions\cite{shelton_spin_ladders,nersesyan_biquadratic} as
\begin{equation}
  \label{eq:majorana-4leg}
  \mathcal{H}_0= -i\frac u 2 \sum_{\nu=e,o j=0,1,2,3} \int dx (\zeta_{R,\nu,j} \partial_x \zeta_{R,\nu,j} -\zeta_{L,\nu,j} \partial_x \zeta_{L,\nu,j}),   
\end{equation}
where we have defined ($\nu=e,o$) 
\begin{eqnarray}
  \frac 1 {\sqrt{2}} (\zeta_{R,\nu,1} + i \zeta_{R,\nu,2}) &=&\frac{e^{i(\theta_{\nu+}-\phi_{\nu+})}}{\sqrt{2\pi\alpha}} \eta_{\nu+}, \\
  \frac 1 {\sqrt{2}} (\zeta_{R,\nu,3} + i \zeta_{R,\nu,0}) &=&\frac{e^{i(\theta_{\nu-}-\phi_{\nu-})}}{\sqrt{2\pi\alpha}} \eta_{\nu-}, \\
  \frac 1 {\sqrt{2}} (\zeta_{L,\nu,1} + i \zeta_{L,\nu,2}) &=&\frac{e^{i(\theta_{\nu+}+\phi_{\nu+})}}{\sqrt{2\pi\alpha}} \eta_{\nu+}, \\
  \frac 1 {\sqrt{2}} (\zeta_{L,\nu,3} + i \zeta_{L,\nu,0}) &=&\frac{e^{i(\theta_{\nu-}+\phi_{\nu-})}}{\sqrt{2\pi\alpha}} \eta_{\nu-}, 
\end{eqnarray}
with $\{\eta_{\nu r},\eta_{\nu' r'}\}=2\delta_{\nu\nu'} \delta_{rr'}$ Majorana fermion operators that ensure anticommutation of fermions with different $\nu$ or $r$ indices\cite{haldane_bosonisation}. 
Introducing the corresponding Ising order and disorder parameters\cite{zuber_77,schroer_ising,ogilvie_ising,boyanovsky_ising,fabrizio_dsg,nersesyan01_ising}, the most relevant interaction becomes (see App.~\ref{app:2leg-ising} for details) 
\begin{eqnarray}\label{eq:interchain-ising}
  \mathcal{H}_{int,b}&=& \frac{J_\perp \lambda^2}a \int dx (\mathbf{n}_1+\mathbf{n}_3)\cdot (\mathbf{n}_2+\mathbf{n}_4)  \\ 
&&=-  \frac{J_\perp \lambda^2} a \int dx \mu_{e,0}  \mu_{o,0} \left[ \sum_{j=1}^3 \mu_{o,j} \mu_{e,j} \prod_{1\le k\le 3 \atop k \ne j} \sigma_{o,k}   \sigma_{e,k} \right]  \nonumber 
\end{eqnarray}

We now pair differently the Majorana fermion operators entering the Hamiltonian~(\ref{eq:majorana-4leg}) to form new Dirac fermions and define  new boson fields $\vartheta_{j},\varphi_{j}$  such that
\begin{eqnarray}
  \label{eq:varfields}
\psi_{R,j}=\frac 1 {\sqrt{2}} (\zeta_{R,e,j}+i\zeta_{R,o,j}) =\frac{e^{i\vartheta_{j}-i\varphi_{j}}}{\sqrt{2\pi \alpha}} \eta_j,\nonumber \\
\psi_{L,j}=\frac 1 {\sqrt{2}} (\zeta_{L,e,j}+i\zeta_{L,o,j}) =\frac{e^{i\vartheta_{j}+i\varphi_{j}}}{\sqrt{2\pi \alpha}} \eta_j.  
\end{eqnarray}
We can express products of Ising order and disorder operators in terms of the new fields using\cite{fabrizio_dsg,nersesyan01_ising}
\begin{eqnarray}
  \label{eq:new-ising}
 && \cos \varphi_{j} = \mu_{e,j} \mu _{o,j} \; \sin \varphi_j = i \sigma_{e,j } \sigma_{o,j}  \eta_{e,j} \eta_{o,j} ,  \\ 
 &&  \cos \vartheta_{j} = \sigma_{e,j}\mu _{o,j} i \eta_j \eta_{e,j} \; \sin \vartheta_j =  \mu_{e,j} \sigma_{o,j} i \eta_j \eta_{o,j},  
\end{eqnarray}
In Eqs.~(\ref{eq:varfields})--(\ref{eq:new-ising}), $\eta_j,\eta_{e/o,j}$ are Majorana fermion operators normalized by $\eta_j^2=\eta_{e/o,j}^2=1$. 
Using Eqs~(\ref{eq:new-ising}), we rewrite the interchain coupling in the form, 
\begin{eqnarray}
 \mathcal{H}_{int,b}&=& \frac{J_\perp \lambda^2}{a}\int dx \cos \varphi_0 \left[\cos (\varphi_1+\varphi_2 -\varphi_3) \right.\nonumber \\ && \left. + \cos (\varphi_3+\varphi_1 -\varphi_2)+ \cos (\varphi_2+\varphi_3 -\varphi_1) \right. \nonumber \\ && \left. - 3 \cos (\varphi_1+\varphi_2+\varphi_3)\right],  
\end{eqnarray}
while the Hamiltonian of the decoupled chains reads
  \begin{equation}
    \mathcal{H}_0 =\sum_{j=0}^3 \int \frac{dx}{2\pi} u \left[(\partial_x \vartheta_j)^2 +(\partial_x \varphi_j)^2 \right]. 
  \end{equation}
To make the $\mathrm{SU(3)}$ symmetry apparent,  we introduce the linear combinations of the boson fields\cite{assaraf_su(n),citro_su3}
\begin{equation}
  \label{eq:su3-basis}
  \left(\begin{array}{c}\varphi_c \\ \varphi_a \\ \varphi_b \end{array} \right) = \left(\begin{array}{ccc}\frac 1 {\sqrt{3}} &\frac 1 {\sqrt{3}} & \frac 1 {\sqrt{3}} \\  \frac 1 {\sqrt{2}} & - \frac 1 {\sqrt{2}} & 0 \\ \frac 1 {\sqrt{6}} & \frac 1 {\sqrt{6}} & - \frac 2 {\sqrt{6}} \end{array} \right)  \left(\begin{array}{c}\varphi_1\\ \varphi_2 \\ \varphi_3 \end{array} \right),  
\end{equation}
to obtain
\begin{eqnarray}\label{eq:su3-exchange} 
   \mathcal{H}_{int,b}&=& \frac{J_\perp \lambda^2}{a}\int dx \cos \varphi_0 \left[2 \cos \left(\frac{\varphi_c}{\sqrt{3}} -\sqrt{\frac 2 3} \varphi_b\right) \cos \sqrt{2} \varphi_a \right. \nonumber \\  && \left. + \cos \left(\frac{\varphi_c}{\sqrt{3}} +2 \sqrt{\frac 2 3} \varphi_b\right) - 3 \cos \sqrt{3} \varphi_c\right],   
  \end{eqnarray}
  with the Hamiltonian of the decoupled chains
  \begin{equation}\label{eq:su3-decoupled} 
    \mathcal{H}_0 =\sum_{\nu=0,a,b,c} \int \frac{dx}{2\pi} u \left[(\partial_x \vartheta_\nu)^2 +(\partial_x \varphi_\nu)^2 \right]. 
  \end{equation}
  In Eq.~(\ref{eq:su3-exchange}), the fields all have scaling dimension $1$, yielding a spin gap $\Delta \sim J_\perp \lambda^2$ as in planar ladders\cite{schulz_spins,shelton_spin_ladders,cabra_ladders,lecheminant_revue_1d}, and long range ordering for the fields $\varphi_{0,a,b,c}$. 
   The interchain interaction, Eq.~(\ref{eq:su3-exchange}),  is  minimized by $\langle \varphi_{0,a,b}\rangle =0$, and $\pm \langle \varphi_c\rangle/\sqrt{3}=\pi -\arccos(1/\sqrt{3})$. As a consequence,  exponentials of any dual field $\vartheta_{0,a,b,c}$ have autocorrelation functions decaying exponentially with distance.\cite{giamarchi_book_1d}   
\subsubsection{Symmetries of the low energy Hamiltonian}

  Eq.~(\ref{eq:su3-exchange}) is expressed in terms of the Dirac fermion operators (\ref{eq:varfields})in the form
  \begin{eqnarray}\label{eq:su3-fermionized} 
    \mathcal{H}_{int,b}&=&\frac{J_\perp \lambda^2}{a}\int dx \cos \varphi_0 \left[\sum_j (e^{i\sqrt{3}\varphi_c} e^{-2i \varphi_j} + H. c.) -3 \cos \sqrt{3} \varphi_c  \right], \nonumber \\
    &=& 2 \pi J_\perp \lambda^2\int dx \cos \varphi_0 \left[ i \sum_j  (e^{i\sqrt{3} \varphi_c} \psi^\dagger_{R,j} \psi_{L,j} - \text{H. c.} ) \right] \nonumber \\ &&   -\frac { 3 J_\perp \lambda^2} a \int dx \cos \varphi_0 \cos \sqrt{3} \varphi_c. 
  \end{eqnarray}
  According to Eq.~(\ref{eq:su3-basis}), the total fermion density is
  \begin{equation}
    -\frac 1 \pi \partial_x (\varphi_1+\varphi_2+\varphi_3) = -\frac{\sqrt{3}}{\pi} \partial_x \varphi_c,  
  \end{equation}
  so that  any $\mathrm{SU(3)}$ rotation $U^\dagger \psi_{r,j} U=U_{jj'} \psi_{rj'}$ with $r=R,L$ leaves invariant  $\varphi_c$. Then, since $\sum_j \psi^\dagger_{R,j} \psi_{L,j}$ is also  invariant,  the interchain interaction~(\ref{eq:su3-fermionized}) is invariant.    
An alternative derivation of Eq.~(\ref{eq:su3-fermionized}) that does not rely on Majorana fermions and Ising order and disorder operators is shown in App.~\ref{app:alternative}. Although it allows to establish the $\mathrm{SU(3)}$ symmetry of the interaction, it is less convenient to derive bosonized representations of observables. 
  
 In terms of conformal field theory,  in Eq.~(\ref{eq:su3-exchange}), the fields 
 $e^{i(\sqrt{2}\varphi_a\pm \sqrt{2/3} \varphi_b)}$ and $e^{i\sqrt{8/3}\varphi_b}$ have scaling dimension $2/3$ which matches\cite{difrancesco_book_conformal} the scaling dimension of $\mathrm{SU(2)_4}$ primary fields of spin 1. They correspond to the operator $\mathrm{Tr}(g)$ in the   $\mathrm{SU(3)_1}$ WZNW model. Using Eq.~(\ref{eq:final-coset}), the fields $e^{i \sqrt{3} \varphi_c} $ and $e^{i\varphi_c/\sqrt{3}}$ are respectively identified with the operators $\phi_{(21)}$ and $\phi_{(23)}$ of the superconformal coset in complete agreement with Ref.\cite{kiritsis1988} and Eq.~(\ref{eq:nonabelian-interaction}).

 Now, let us briefly discuss the other symmetries  of Eq.~(\ref{eq:su3-exchange}).  
 In Eq.~(\ref{eq:su3-exchange}), the sign of $J_\perp$ can be absorbed by making $\varphi_0 \to \varphi_0+\pi$ or $\varphi_c \to \varphi_c +\pi \sqrt{3}$. The interaction is a periodic function of the fields and even in $\varphi_0$ and $\varphi_a$. It is also invariant  under the simultaneous sign change of $\varphi_c$ and $\varphi_b$.  It has  periodicity under  translations  
   \begin{eqnarray}
     \varphi_c &\to&  \varphi_c +\frac{2\pi}{\sqrt{3}} n_c, \\
     \varphi_b &\to& \varphi_b -\frac \pi {\sqrt{6}} n_c + \pi \sqrt{\frac{3}{2} } n_b, \\
     \varphi_a &\to& \varphi_a +\frac{\pi}{\sqrt{2}}  (n_b + n_c + 2n_a),   
   \end{eqnarray}
   with $n_a,n_b,n_c$ integers. Finally, it is invariant under the $\frac {2\pi} 3$ rotation
   \begin{eqnarray}\label{eq:rotation} 
     \varphi_a &=& - \frac 1 2 \varphi'_a -\frac{\sqrt{3}}2 \varphi'_b, \\
     \varphi_b &=& \frac{\sqrt{3}}2 \varphi'_a -\frac 1 2 \varphi'_b,   \\
   \end{eqnarray}
   which amounts to a circular permutation of $\varphi_{1,2,3}$.

 \subsection{SU(3) currents and conserved quantities}
 Having derived the bosonized Hamiltonian of the 4-leg tube, we now turn to the generators of $\mathrm{SU(3)}$ symmetry. Their density and currents are given by the $\mathrm{SU(3)}$ right and left moving currents.  We will first discuss the $\mathrm{SU(2)}_4$ currents, then we will turn to the spin-2 primaries. 
 \subsubsection{SU(2) currents} 
  We first turn our attention to the $\mathrm{SU(2)_4}$currents. The sum of the right-moving currents in odd and even chains are expressed in terms of Majorana fermions as\cite{shelton_spin_ladders} 
   \begin{eqnarray}
  J_{R1}^a + J_{R3}^a &=& - \frac i 2 \epsilon_{abc} \zeta_{R,o,b} \zeta_{R,o,c} \\ 
 J_{R2}^a + J_{R4}^a&=&- \frac i 2 \epsilon_{abc} \zeta_{R,e,b} \zeta_{R,e,c} 
   \end{eqnarray}
   so we can rewrite their sum  using 
   \begin{equation}
     \Psi_R=\left(\begin{array}{c} \psi_{R,1} \\  \psi_{R,2} \\  \psi_{R,3}\end{array} \right),  
   \end{equation}
   in the form 
   \begin{eqnarray}
     \sum_{n=1}^4 J_{R,n}^x &=&  \Psi_R^\dagger \Lambda^7 \Psi_R,  \\
     \sum_{n=1}^4 J_{R,n}^y &=& -\Psi_R^\dagger \Lambda^5 \Psi_R, \\
      \sum_{n=1}^4 J_{R,n}^z &=&  \Psi_R^\dagger \Lambda^2 \Psi_R, 
   \end{eqnarray}
   where $\Lambda^{2,5,7}$ are Gell-Mann matrices\cite{itzykson-zuber,slansky_1981}. Similar relations hold for the left moving currents $J_{L,n}^{x,y,z}$. The matrices $(\Lambda^7,-\Lambda^5,\Lambda^2)$ generate a spin-1 $\mathrm{su(2)}$ subalgebra of the $\mathrm{su(3)}$ algebra\cite{slansky_1981} engendered by the full set of Gell-Mann matrices. 
With the unitary transformation
   \begin{equation}\label{eq:su3-rotation} 
    \left(\begin{array}{c} \psi_{R,1} \\  \psi_{R,2} \\  \psi_{R,3}\end{array}\right) = e^{i\frac\pi 4\Lambda_1}  e^{i\frac\pi 4 (\Lambda_3-\sqrt{3}\Lambda_8)} \left(\begin{array}{c} \bar{\psi}_{R,1} \\  \bar{\psi}_{R,-1} \\  \bar{\psi}_{R,0}\end{array}\right) ,   
   \end{equation}
   we can write
   \begin{eqnarray}
      \sum_{n=1}^4 J_{R,n}^x &=&  \bar{\Psi}^\dagger_R(x)\frac{\Lambda_5 - \Lambda_7  }{\sqrt{2}}  \bar{\Psi}_R(x) ,\\
      \sum_{n=1}^4 J_{R,n}^y &=&  - \bar{\Psi}^\dagger_R(x)   \frac{\Lambda_4 + \Lambda_6  }{\sqrt{2}}  \bar{\Psi}_R(x) ,\\
     \sum_{n=1}^4 J_{R,n}^z &=&  - \bar{\Psi}^\dagger_R(x)\Lambda_3   \bar{\Psi}_R(x) , 
   \end{eqnarray}
   and recover (up to a $\pi/2$ rotation around the $z$ axis) the expression of the spin currents  in terms of $\mathrm{SU(3)}_1$ operators\cite{citro00_spintube} obtained when considering the bilinear-biquadratic spin-1 chain\cite{golinelli_incommensurate} at the Uimin-Lai-Sutherland\cite{uimin,lai,sutherland} critical point. Bosonizing the $\bar{\Psi}$ fermions, and introducing fields $\bar{\varphi}_{a,b,c}$ and their duals $\bar{\vartheta}_{a,b,c}$  as in Eqs.~(\ref{eq:varfields})--(\ref{eq:su3-basis}) we find
   \begin{equation}\label{eq:barphi-a} 
     -\frac{1}{\pi \sqrt{2}} \left(\sum_{p=1}^4 \partial_x \phi_p\right) = -\frac{\sqrt{2}}{\pi} \partial_x \bar{\varphi}_a,   
     \end{equation}
     allowing us to relate the total magnetization with $\partial_x\bar{\varphi_a}$. Similarly, the total magnetization current is related with $\partial_x \bar{\vartheta}_a$.
     After performing the $\frac \pi 2$ rotation around the $z$-axis, we find the bosonized expression
     \begin{eqnarray}
       \sum_{n=1}^4 (J_{R,n}^x+i J_{R,n}^y)&=&\frac{e^{-i\frac{\bar{\vartheta}_a-\bar{\varphi}_a}{\sqrt{2}}}}{\pi\alpha\sqrt{2}}\left[e^{-i\sqrt{\frac 3 2}(\bar{\vartheta}_b-\bar{\varphi}_b)}\eta_1\eta_0 \right. \nonumber \\ && \left. +   e^{i\sqrt{\frac 3 2}(\bar{\vartheta}_b-\bar{\varphi}_b)}\eta_0 \eta_{-1}\right], 
     \end{eqnarray}
     which recovers the coset representation\cite{zamolodchikov_1985} $\mathrm{SU(2)}_4 \sim \mathrm{U(1)}\times \Bbb{Z}_4$ of the $\mathrm{SU(2)}_4$ currents, with $\mathrm{U(1)}$ a free $c=1$ bosonic theory  and $\Bbb{Z}_4$ the four-state clock model\cite{jose_planar_2d} with $c=1$.   The right moving parafermion field of dimension $3/4$ is given by
     \begin{eqnarray}
       \psi_{R,Z_4} \sim e^{-i\sqrt{\frac 3 2}(\bar{\vartheta}_b-\bar{\varphi}_b)}\eta_1\eta_0 +   e^{i\sqrt{\frac 3 2}(\bar{\vartheta}_b-\bar{\varphi}_b)}\eta_0 \eta_{-1}. 
     \end{eqnarray}
     Using that coset decomposition, we obtain\cite{zamolodchikov_1985} the spin-1 primary operators of $\mathrm{SU(2)_4}$ in the form
     \begin{eqnarray}
       \Phi_{[11]}^{(1)}  \sim e^{i\sqrt{2}\bar{\varphi}_a} \sigma_2, \\
       \Phi_{[00]}^{(1)}  \sim \varepsilon^{(1)}, \\
     \end{eqnarray}
     where $\sigma_2$ is the spin field of dimension $1/6$ and $\varepsilon^{(1)}$ is the thermal operator of the $\Bbb{Z}_4$ clock model. We can identify $\sigma_2 \sim \cos (\sqrt{2/3} \bar{\varphi}_a)$ and $ \varepsilon^{(1)}\sim \cos (\sqrt{8/3} \bar{\varphi}_a)$ by comparing with~(\ref{eq:su3-exchange}). In the ground state of the 4-leg tube, the $\Bbb{Z}_4$ degrees of freedom exhibit long range ordering. If we consider the spin-1/2 $\mathrm{SU(2)}_4$ primaries, we have
     \begin{eqnarray}
       \Phi_{[1/2,1/2]}^{(1/2)} &\sim& \sigma_1 e^{i\frac{\bar{\varphi}_a}{\sqrt{2}}}, \\
       \Phi_{[1/2,-1/2]}^{(1/2)} &\sim& \mu_1 e^{i\frac{\bar{\vartheta}_a}{\sqrt{2}}}, \\
     \end{eqnarray}
     where  $\sigma_1$ and $\mu_1$  are the spin field of dimension $1/8$ of the $\Bbb{Z}_4$ clock model and its dual.  Since $\varphi_a$ is ordered, the field $e^{i\frac{\bar{\vartheta}_a} {\sqrt{2}}}$ in the second line has short range order. The $\mathrm{SU(2)}$ symmetry then implies that $\sigma_1$ is also short range ordered, and $\mu_1$ must be long range ordered.
     
     \subsubsection{spin-2 primaries} 
The five remaining $\mathrm{SU(3)_1}$ currents are 
   \begin{equation}
     \Psi^\dagger_R \Lambda^{1,3,4,6,8} \Psi_R,  
   \end{equation}
   but substituting (\ref{eq:varfields}) in the above expression shows that it depends on products $i \zeta_{R,o,\alpha} \zeta_{R,e,\beta}$ in the original decoupled chains basis.  Hence, these operators are not local operators in the initial lattice model.
   However, since
   \begin{eqnarray}\label{eq:currdiff} 
     J^a_{R,1}-J^a_{R,3} &=& i \zeta_{R,o,0} \zeta_{R,o,a}, \\
     J^a_{R,2}-J^a_{R,4} &=& i \zeta_{R,e,0} \zeta_{R,e,a}, 
   \end{eqnarray}
   we have
   \begin{eqnarray}
 &&    (J^1_{R,1}-J^1_{R,3} ) (J^2_{R,2}-J^2_{R,4} )+ (J^2_{R,1}-J^2_{R,3} ) (J^1_{R,2}-J^1_{R,4} ) \nonumber \\ && = i \zeta_{R,o,0}\zeta_{R,e,0} \Psi^\dagger_R \Lambda^1 \Psi_{R}, \\
 &&    (J^1_{R,1}-J^1_{R,3} ) (J^1_{R,2}-J^1_{R,4} )- (J^2_{R,1}-J^2_{R,3} ) (J^2_{R,2}-J^2_{R,4} ) \nonumber \\ &&= i \zeta_{R,o,0}\zeta_{R,e,0} \Psi^\dagger_R \Lambda^3 \Psi_{R}, \\
  &&   (J^1_{R,1}-J^1_{R,3} ) (J^3_{R,2}-J^3_{R,4} )+ (J^3_{R,1}-J^3_{R,3} ) (J^1_{R,2}-J^1_{R,4} ) \nonumber \\ &&= i \zeta_{R,o,0}\zeta_{R,e,0} \Psi^\dagger_R \Lambda^4 \Psi_{R}, \\
   &&  (J^2_{R,1}-J^2_{R,3} ) (J^3_{R,2}-J^3_{R,4} )+ (J^3_{R,1}-J^3_{R,3} ) (J^2_{R,2}-J^2_{R,4} ) \nonumber \\ &&= i \zeta_{R,o,0}\zeta_{R,e,0} \Psi^\dagger_R \Lambda^6 \Psi_{R}, \\
     && (J^1_{R,1}-J^1_{R,3} ) (J^1_{R,2}-J^1_{R,4} )+(J^2_{R,1}-J^2_{R,3} ) (J^2_{R,2}-J^2_{R,4} )\nonumber \\ && -2   (J^3_{R,1}-J^3_{R,3} ) (J^3_{R,2}-J^3_{R,4} )\nonumber \\ &&= i\sqrt{6} \zeta_{R,o,0}\zeta_{R,e,0} \Psi^\dagger_R \Lambda^8 \Psi_{R},
   \end{eqnarray}
  showing that  tensor products of current differences are expressible with the $\mathrm{SU(2)}_4$ spin-2 primaries. 
   \subsubsection{Conserved quantities} 
  If we turn to globally conserved quantities,  the isospin $I_3=I_{3,R}+I_{3,L}$ in the $\mathrm{SU(3)}$ theory is given by
   \begin{eqnarray}
     \label{eq:isospin-fermion}
     I_{3}=\frac 1 2 \int dx \sum_\nu \bar{\Psi}^\dagger_\nu(x)\Lambda_3   \bar{\Psi}_\nu(x),  
   \end{eqnarray}
   and identifies with half the total spin of the lattice system.  The other two components of the spin also give rise to conserved quantities, but they don't commute with $I_3$. 
   But, the second conserved  quantity is the  hypercharge
   \begin{equation}\label{eq:hypercharge-fermion} 
     Y=\frac 1 {\sqrt{3}}   \int dx \sum_\nu \bar{\Psi}^\dagger_\nu(x) \Lambda^{8} \bar{\Psi}_\nu(x), 
   \end{equation}
   which is a nonlocal quantity in the original spin variables.
   Therefore, although
   the low energy excited states are classified by irreducible representations of
   $\mathrm{SU(3)}$ and possess both isospin $I_3$ and hypercharge $Y$, only
   the former can be determined from local observables of the lattice model.  
   In particular, the group $\mathrm{SU(3)}$ possesses
   two non-equivalent irreducible
   representations\cite{kaeding_su3_1995} of dimension 3, called $\mathbf{3}$ and
   $\mathbf{\bar{3}}$  with opposite isospins and hypercharges.
   But since only $I_3$ can be measured from the total spin,
   those representations appear as two $\mathrm{SU(2)}$ spin-1 triplets
   in the spectrum.  More generally, the irreducible representations of $\mathrm{SU(3)}$  decompose into direct sums of irreducible representations of $\mathrm{SU(2)}$ of given total spin. In the presence of $\mathrm{SU(3)}$ symmetry, degeneracies between states of different total spin are obtained.  For the sake of concreteness, let's consider two elementary examples.  If we take the tensor product\cite{kaeding_su3_1995} of $\mathrm{SU(3)}$ representations, 
   \begin{equation}
     \label{eq:tensor33b}
      \mathbf{3}\otimes\mathbf{\bar{3}}=\mathbf{1}\oplus \mathbf{8}, 
    \end{equation}
    seen as an  $\mathrm{SU(2)}$ tensor product $1 \otimes 1=0\oplus 1 \oplus 2$,  the one-dimensional representation of $\mathrm{SU(3)}$ identifies with the $\mathrm{SU(2)}$ spin singlet, while the eight dimensional representation is reducible into the  direct sum of $\mathrm{SU(2)}$ spin-1 and spin-2 representations. When the $\mathrm{SU(3)}$ symmetry is present, spin-1 and spin-2 states forming to the $\mathrm{8}$ representations are degenerate in energy.    
  If we now take the tensor product 
   \begin{equation}
     \label{eq:tensor33}
     \mathbf{3}\otimes\mathbf{3}=\mathbf{\bar{3}}\oplus \mathbf{6}, 
   \end{equation}
   interpreted in terms of $\mathrm{SU(2)}$ representations, we have again the tensor product
   $1 \otimes 1=0\oplus 1 \oplus 2$. Since the $\mathbf{\bar{3}}$ representation of  $\mathrm{SU(3)}$ has to be identified with the spin-1 representation of $\mathrm{SU(2)}$,  the $\mathbf{6}$ representation is reducible into  a sum of spin-2 and a spin-0  representation of $\mathrm{SU(2)}$. So when the $\mathrm{SU(3)}$ symmetry is present,  a degeneracy between spin-2 and spin-0 states is observed.
 So, even though the hypercharge cannot be measured from the local spin observables, the $\mathrm{SU(3)}$ symmetry manifests itself in the form of apparently accidental degeneracies in the spectrum. Moreover, considering the degeneracies of states containing two triplet excitations can reveal the representation of $\mathrm{SU(3)}$ to which the triplet belongs, and thus indirectly characterize their  hypercharge.    
    
\subsection{Excited states} 
   \subsubsection{Soliton and antisolitons}

In the semiclassical limit, excitations above the ground state take the form of solitons that interpolate between the different minimas of Eq.~(\ref{eq:su3-exchange}). The fields $\varphi_{0,a,b,c}(x)$ take different limits $\varphi_{0,a,b,c}(\pm \infty)$ as $x\to \pm \infty$ such that the potential Eq.~(\ref{eq:su3-exchange}) has the same limit for $x\to \pm \infty$. Introducing the notation
 \begin{equation}
   \Delta \varphi_\nu = \int_{-\infty}^{+\infty} dx \partial_x \varphi_\nu(x) = \varphi_\nu(+\infty) -  \varphi_\nu(-\infty),   
 \end{equation}  
  we define the  charge $Q$, isospin $I_3$ and hypercharge $Y$ of a soliton given by 
  \begin{eqnarray}
    \label{eq:charge} 
    Q=-\frac{\sqrt{3}} \pi  \Delta \varphi_c, \\
    \label{eq:isospin} 
    I_3=-\frac{1} {\pi\sqrt{2}}  \Delta \varphi_a, \\
    \label{eq:hypercharge}
    Y=-\frac{\sqrt{2}} {\pi\sqrt{3}} \Delta \varphi_b.  
 \end{eqnarray}
 \subsubsection{Magnetic solitons and antisolitons} 

Let us first consider the semiclassical limit of~(\ref{eq:su3-exchange}) and search for the quantum numbers of solitons and antisolitons. 
 If we consider solitons in which $\Delta \varphi_0=\varphi_0(+\infty)-\varphi_0(-\infty)=\pi$ equating the limits at $\pm\infty$ of the potential yields  
  \begin{eqnarray}
    \sqrt{3} \Delta \varphi_c = (2 n_c +1) \pi, \nonumber \\ 
    \frac{\Delta \varphi_c}{\sqrt{3}} -\sqrt{2} \Delta \varphi_a -\sqrt{\frac 2 3} \Delta \varphi_b = (2 n_1 +1) \pi, \nonumber \\
    \frac{\Delta \varphi_c}{\sqrt{3}} +\sqrt{2} \Delta \varphi_a -\sqrt{\frac 2 3} \Delta \varphi_b = (2 n_{-1} +1) \pi, \nonumber\\
     \frac{\Delta \varphi_c}{\sqrt{3}} + 2 \sqrt{\frac 2 3} \Delta \varphi_b = (2 n_{0} +1) \pi, 
  \end{eqnarray}
  where $n_{c,-1,0,1}$ are integers, to ensure that the potential has the same limit at $\pm \infty$. Combining the above equations yields $n_c=n_0+n_1+n_{-1}+1$ and
  \begin{eqnarray}
    I_3 = \frac 1 2 (n_1-n_{-1}), \nonumber \\
    Y=\frac{1}{3} (n_1+n_{-1}-2n_0).  
  \end{eqnarray}
  To minimize $Q$, we have to set  $n_0+n_1+n_{-1}=-2$ ($Q=1$) or  $n_0+n_1+n_{-1}=-1$ ($Q=-1)$. 
  In the first case, $n_k=0, n_{j\ne k}=-1$, we
  have  charge $Q=1$ and isospin and hypercharge
  $(I_3,Y) \in \{(-1/2,1/3),(1/2,1/3),(0,-2/3)\}$.  The rotation~(\ref{eq:rotation}) can be used to generate all of them starting for instance with the one of isospin $0$ and hypercharge $-2/3$. These solitons carry the same quantum numbers as the fermions $\psi_{R/L,j}$  ($j=1,2,3$) but they also carry the topological charge associated with $\varphi_0$, so the bosonized form of their creation operator contains a factor  $e^{-i (\vartheta_j \pm \vartheta_0)}$. Given their quantum numbers, the solitons transform in the $\mathbf{3}$ representation of $\mathrm{SU(3)}$. 
  
  In the second case, we must  set
  $n_k=-1, n_{j\ne k}=0$, for $k=-1,0,1$ to find antisolitons with
  $SU(3)$ isospin and hypercharge $(I_3,Y) \in \{(1/2,-1/3),(1/2,-1/3),(0,2/3)\}$.   They carry quantum numbers as the antifermions $\psi_{R/L,j}$ ($j=1,3$), as well as the topological charge associated with $\varphi_0$, so he bosonized expression of the antisoliton creation operator contains $e^{i \vartheta_j \pm \vartheta_0}$.  The antisolitons transform in the  $\mathbf{\bar{3}}$ representation of $SU(3)$.  
  In terms of spin, since $S^z=2 I_3$, the solitons and the antisolitons give rise to two degenerate branches of gapped spin-1 excitations. Eqs.~(\ref{eq:currdiff}) show that the Matsubara response functions of current differences 
contain contributions from solitons and antisolitons that give rise to sharp peaks in the dynamical structure factor after analytic continuation.  
 Topological excitations with different $Q,I_3,Y$ might also exist at the semiclassical level, and would correspond for instance to bound states of solitons and/or antisolitons (breathers).\cite{rajaraman_instanton}  However, it is unclear which of these bound states persist at the fully quantum level. In the case of  the integrable quantum sine-Gordon model\cite{dashen_sinegordon,zamolodchikov79_smatrices,babujian99_ff_sg_1}, it is known that the number of bound states depends on the Tomonaga-Luttinger exponent. As the Tomonaga-Luttinger exponent increases, the number of breathers decreases, and beyond a critical value, solitons and antisolitons do not form bound states. In our case, the interaction Eq.~(\ref{eq:su3-exchange}) does not seem to lead to an integrable model and the breather stability remains an open question.    We can only state that if breather excitations exist, they must organize in $\mathrm{SU(3)}$ multiplets.

\subsubsection{Trimerized $SU(3)$ spin chain}

\begin{figure}[h]
      \centering
      \includegraphics[width=9cm]{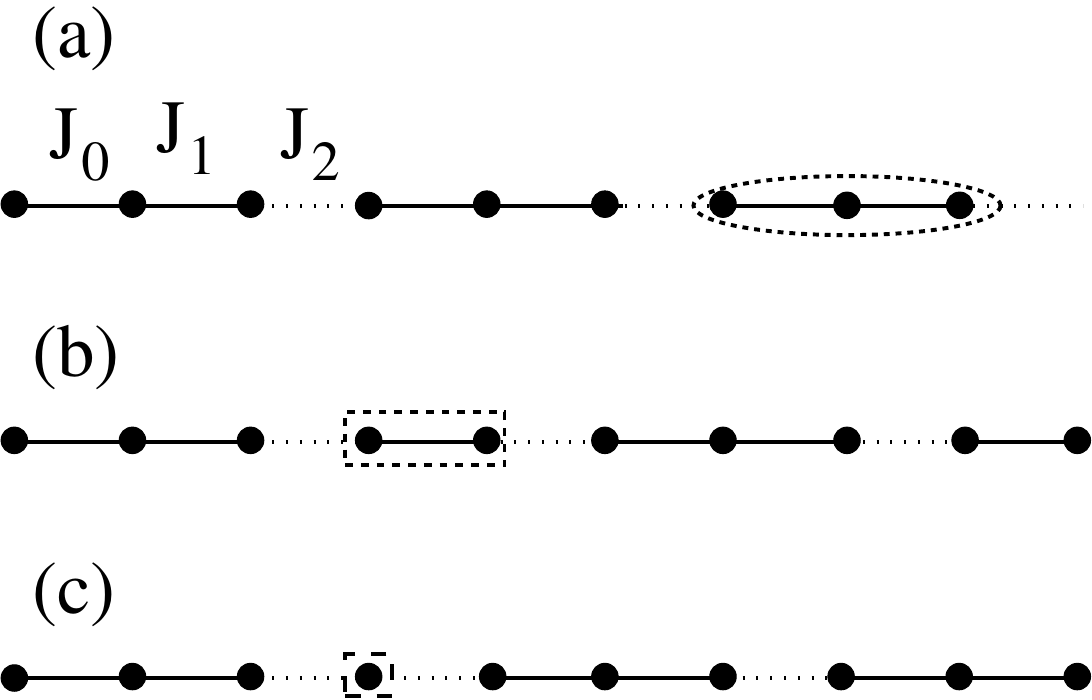} 
      \caption{(a) the exchange coupling in the trimerized SU(3) spin chain. $J_0=J_1$ are the string links, and $J_2$ is the weak link. The ellipse represents 3 $\mathrm{SU(3)}$ spin forming a single state. (b) Domain wall in which the trimerization pattern has been shifted by one lattice spacing. A SU(3) dimer (indicated by a rectangle) is formed. It allows the formation of an excitation belonging to the $\mathbf{\bar{3}}$ representation. (c) domain wall in which the pattern has been shifted by two lattice spacings. An isolated SU(3) spin is present. }
      \label{fig:patterns}
    \end{figure}

  To form a more accurate image of the magnetic solitons and antisolitons, we need to return to the original quantum Hamiltonian.  We will only assume that the fields $\varphi_0$ and $\varphi_c$ having long range order can be replaced by their expectation value in Eq.~(\ref{eq:su3-exchange}),
  the resulting low energy Hamiltonian reduces to the bosonized Hamiltonian of a trimerized $\mathrm{SU(3)}$ spin chain\cite{assaraf_su(n),citro00_spintube}, 
  \begin{eqnarray}
    \label{eq:su3-trimer}
    H&=&\sum_n  (J+\delta J_n)  \sum_{a=1}^8 \lambda_{n}^a \lambda_{n+1}^a,\\ 
    J_n&=&J+\delta J \left(e^{i\left[\frac {2\pi} 3 n -\frac{\langle \varphi_c\rangle}{\sqrt{3}}\right]} + e^{-i\left[\frac {2\pi} 3 n -\frac{\langle \varphi_c\rangle}{\sqrt{3}}\right]}\right),   
  \end{eqnarray}
  where the $SU(3)$ spins are in the $\mathbf{3}$ representation, $J$ is chosen\cite{uimin,lai,sutherland} to reproduce the excitation velocity $u$, and $\delta J\ll J$ is proportional to $J_\perp\langle \cos \varphi_0\rangle$.  In that improved approximation, only the fields carrying non-magnetic degrees of freedom are treated semiclassically. 
  For $\langle \varphi_c\rangle=0$ and $\delta J<0$,  the periodic pattern satisfies $0<J_{3n}<  J_{3n+2}=J_{3n+1}$, and we can consider as strong coupling fixed point a  trimerized chain as made of independent groups of 3 $\mathrm{SU(3)}$ spins that form a singlet in the ground state as shown on Fig.~\ref{fig:patterns}~(a) .
  For $\langle \varphi_c\rangle=0$ and $\delta J>0$, the periodic pattern satisfies $0<J_{3n+2}=J_{3n+1}<J_{3n}$, so we can take a strong coupling fixed point pairs of spins on the strong bond forming an effective spin in the $\mathbf{3^*}$ representation. We obtain a chain in which spins in the representation $\mathbf{3}$ and $\mathbf{\bar{3}}$ alternate and the spectrum is gapped\cite{barber_1989,klumper_1990}. We can picture the ground state as the spontaneous formation of singlet pairs with spins in the $\mathbf{3}$ and $\mathbf{\bar{3}}$ representation.  
    Let's discuss first the case of $\delta J<0$. 
    When solitons are present, $\langle \varphi_0\rangle$ is shifted by $\pi$ and $\langle \varphi_c\rangle/\sqrt{3}$ is shifted by $\pi/3$, so that the trimerization pattern shifts by two lattice spacings when moving from $-\infty$ to $+\infty$.
    A dimer defect is introduced somewhere along the chain, giving rise [see Fig.~\ref{fig:patterns}~(b)]  to a spin in the $\bar{3}$ representation. With $\langle \varphi_c\rangle/\sqrt{3}$ shifted by $-\pi$, a single $SU(3)$ spin in the $\mathbf{3}$ is present [see Fig.~\ref{fig:patterns}~(c)]. 
With $\delta J>0$, solitons  create defects $\mathbf{\bar{3}}-\mathbf{3}-\mathbf{3}-\mathbf{\bar{3}}$ that give rise to an unpaired spin in the $\mathbf{3}$ representation and antisolitons defects  $\mathbf{3}-\mathbf{\bar{3}}-\mathbf{\bar{3}}-\mathbf{3}$ that give rise to an unpaired spin in the $\mathbf{\bar{3}}$ representation. 
We can now return to the question of soliton/antisoliton bound  states.     
    If we consider pairs of  solitons in the $\mathbf{3}$ representation, of antisolitons in the $\mathbf{\bar{3}}$ representation, or  a soliton antisoliton pair, we need to consider the tensor products\cite{kaeding_su3_1995}
    \begin{eqnarray}
      \mathbf{3}\otimes \mathbf{3} &=& \mathbf{\bar{3}}\oplus \mathbf{6}, \\
      \mathbf{\bar{3}}\otimes \mathbf{\bar{3}} &=&  \mathbf{3} \oplus \mathbf{\bar{6}}, \\
      \mathbf{3}\otimes \mathbf{\bar{3}}  &=&  \mathbf{1}\oplus \mathbf{8}.  
    \end{eqnarray}
    In the simplest case, the only bound states of two solitons are antisolitons, the only bound states of two antisolitons are solitons, and there are no soliton-antisoliton bound states, so that solitons and antisolitons are the only excitations.
    The representations $\mathbf{6},\mathbf{\bar{6}},\mathbf{8}$ then correspond to excitations in the continuum formed of unbound soliton/antisoliton pairs.  In terms of $SU(2)$ representations, only two degenerate branches of gapped triplet excitations are present besides the continuum.
    
   In a slightly more complicated case,   soliton-antisolitons bound states (breathers)  in the $\mathbf{8}$ representation  are also present.  In terms of $\mathrm{SU(2)}$ representation, the $\mathbf{8}$ representation gives a gapped branch of spin-2 excitations degenerate with a gapped branch of spin-1 excitations.  To characterize the breather-soliton, breather-antisoliton and breather-breather bound states, we need the tensor products\cite{kaeding_su3_1995}
      \begin{eqnarray}
      \mathbf{3}\otimes \mathbf{8} &=&\mathbf{3}\oplus \mathbf{6}\oplus \mathbf{15}, \\
      \mathbf{\bar{3}}\otimes \mathbf{8} &=&    \mathbf{\bar{3}} \oplus \mathbf{\bar{6}}\oplus \mathbf{\bar{15}}, \\
      \mathbf{8}\otimes \mathbf{8}  &=&  \mathbf{1}\oplus \mathbf{8}\oplus \mathbf{8'}\oplus \mathbf{10}  \oplus \mathbf{\bar{10}}\oplus \mathbf{27}, 
      \end{eqnarray}
      indicating that the bound state of a soliton (respectively antisoliton) with a breather is a soliton  (respectively antisoliton), and bound states of breathers are breathers.
      In terms of the trimerized $\mathrm{SU(3)}$ chain, the excitations of a trimer are obtained by considering the tensor products
      \begin{eqnarray}
        \mathbf{3}\otimes \mathbf{3} &=& \mathbf{\bar{3}}\oplus \mathbf{6}, \\
        \mathbf{3}\otimes \mathbf{\bar{3}}  &=&  \mathbf{1}\oplus \mathbf{8}\\
        \mathbf{3}\otimes\mathbf{6}&=&\mathbf{8}\oplus\mathbf{10}                                                 \end{eqnarray}
 and would allow for both signs of $\delta J$ a delocalized excitation in the $\mathbf{8}$ representation of $\mathrm{SU(3)}$.       

 \subsubsection{Non-magnetic excitations}
      
Besides excitations possessing $\mathrm{SU(3)}$ spin and hypercharge, we can also have excitations involving only $\varphi_0$ and $\varphi_c$. For instance, when only $\varphi_c$ is varying the potential reduces to
  \begin{equation}
    J_\perp \lambda^2 \left[3 \cos (\varphi_c/\sqrt{3}) -3 \cos \sqrt{3} \varphi_c\right],  
  \end{equation}
  and it allows for short kinks interpolating from $\pi-\arccos(1/\sqrt{3})$ to $\pi + \arccos(1/\sqrt{3})$ and long kinks from $\arccos(1/\sqrt{3})-\pi$ to  $\pi-\arccos(1/\sqrt{3})$.  It is also possible to have kinks where $\Delta \langle \varphi_0\rangle =\pi$ and $\Delta \langle \varphi_c\rangle =\pm \pi \sqrt{3}$.    All those kinks are $SU(3)$ singlets and possess a non-integer charge $Q$. If they survive in the  quantum limit,
  they  give rise to  branches of gapped spin singlet  excitations. 

   \subsection{Observables}\label{sec:observables} 
   We would like to determine the observables that make the $\mathrm{SU(3)}$ symmetry of the model apparent. Since the spin-1/2 primaries in $\mathrm{SU(2)_4}$ are in the twisted sector\cite{difrancesco_book_conformal}, they cannot be realized with $SU(3)_1$ primaries.
   Thus, we need to consider operators containing the product of two spin-1/2 primaries that can be expressed in terms of $\mathrm{SU(2)_4}$ spin 1 primaries that are also $\mathrm{SU(3)_1}$ primaries.  
  Obvious candidates are the vector chiralities\cite{hikihara_2008} $(\mathbf{n}_1 \pm \mathbf{n}_3) \times (\mathbf{n}_2 \pm \mathbf{n}_3)$, and the nematic order parameter $2 Q^{ab}_{\pm \pm}  = (n_1^a\pm n_3^a) (n_2^b\pm n_4^b) + (n_1^b \pm n_3^b)(n_2^a \pm n_4^a) - \delta_{ab} (\mathbf{n}_1\pm \mathbf{n}_3)\cdot(\mathbf{n}_2\pm \mathbf{n}_4)/3$. We can also consider operators\cite{savary_2015} formed from the product of a dimerization operator by a spin operator such as $(\epsilon_1 \pm \epsilon_3) (n_2^a \pm n_4^a)+(\epsilon_2 \pm \epsilon_4) (n_1^a \pm n_3^a)$.
   
\subsubsection{Symmetric case}
\label{sec:symobs}
Let us first consider the case with both symmetric combinations, $(\mathbf{n}_1 + \mathbf{n}_3) \times (\mathbf{n}_2 + \mathbf{n}_4)$ and $2 Q^{ab}_{+ +}  = (n_1^a+ n_3^a) (n_2^b+ n_4^b) + (n_1^b + n_3^b)(n_2^a + n_4^a) - \delta_{ab} (\mathbf{n}_1+ \mathbf{n}_3)\cdot(\mathbf{n}_2 +\mathbf{n}_4)/3$ 
   Using the expression in terms of Ising order and disorder operators we find
   \begin{eqnarray}
     (n_2+n_4)^1(n_1+n_3)^2&=&(\mu_{e1} \sigma_{o1})(\sigma_{e2}\mu_{o2})  (\sigma_{e3}\sigma_{o3}) (\mu_{e0} \mu_{o0}) \nonumber \\ && \times (i\eta_{e2} \eta_{e3})   (i\eta_{o3} \eta_{o1}),\\
     &=&-i\eta_1 \eta_2 \sin \vartheta_1 \cos \vartheta_2 \sin \varphi_3 \cos \varphi_0,  \nonumber 
   \end{eqnarray}
   and similarly, exchanging $e$ and $o$ indices, 
   \begin{eqnarray}
      (n_2+n_4)^2(n_1+n_3)^1&=&(\mu_{o1} \sigma_{e1})(\sigma_{o2}\mu_{e2})  (\sigma_{e3}\sigma_{o3}) (\mu_{e0} \mu_{o0}) \nonumber \\ && \times (i\eta_{o2} \eta_{o3})   (i\eta_{e3} \eta_{e1}),\\
     &=&+ i\eta_1 \eta_2 \cos \vartheta_1 \sin \vartheta_2 \sin \varphi_3 \cos \varphi_0, \nonumber   
   \end{eqnarray}
   yielding
   \begin{eqnarray}
     \label{eq:symtens-od} 
   &&  Q^{12}_{++} = \sin(\vartheta_2 -\vartheta_1) \sin \varphi_3 \cos \varphi_0 i \eta_1 \eta_2,\\ 
    && \lbrack (\mathbf{n}_2+\mathbf{n}_4)\times (\mathbf{n}_1+\mathbf{n}_3)\rbrack^3 =  -\sin(\vartheta_2 +\vartheta_1) \sin \varphi_3 \cos \varphi_0 i \eta_1 \eta_2.   \nonumber \\         
   \end{eqnarray}
   The other components are obtained by circular permutations. For the diagonal components of the nematic order parameter, we find
   \begin{eqnarray}
     \label{eq:symtens-diag}
     Q^{11}_{++}-Q^{22}_{++} &=& -\sin(\varphi_2-\varphi_1) \sin \varphi_3 \cos \varphi_0, \\
     Q^{33}_{++} &=& -\cos \varphi_0\left[\cos (\varphi_2+\varphi_3 -\varphi_1) + \cos (\varphi_3+\varphi_1 -\varphi_2) \right. \nonumber \\ && \left. -2  \cos (\varphi_1+\varphi_2 -\varphi_3)\right]  
   \end{eqnarray}
   Now, let's write $Q^{ab}_{++}$ in terms of fermion operators. Using $\varphi_3=\sqrt{3} \varphi_c - \varphi_1 -\varphi_2$, we can show that
   \begin{eqnarray}\label{eq:symtens-gellmann} 
    && Q^{12}_{++} \sim  i \cos \varphi_0\left[ e^{-i \sqrt{3} \varphi_c} \Psi^\dagger_R \Lambda_1 \Psi_L -  \text{H. c.} \right],  \\
    && Q^{23} _{++} \sim  i \cos \varphi_0\left[ e^{-i \sqrt{3} \varphi_c} \Psi^\dagger_R \Lambda_6 \Psi_L -   \text{H. c.} \right],  \\
    && Q^{13}_{++} \sim i \cos \varphi_0\left[ e^{-i \sqrt{3} \varphi_c} \Psi^\dagger_R \Lambda_4 \Psi_L -   \text{H. c.} \right], \\
     && Q^{11}_{++}- Q^{12} _{++} \sim i \cos \varphi_0\left[ e^{-i \sqrt{3} \varphi_c} \Psi^\dagger_R \Lambda_3 \Psi_L -  \text{H. c.} \right], \\
   &&  Q^{33} _{++} \sim i \cos \varphi_0\left[ e^{-i \sqrt{3} \varphi_c} \Psi^\dagger_R \Lambda_8 \Psi_L -  \text{H. c.} \right],
   \end{eqnarray}
   showing that the nematic order parameter transforms according to the \textbf{8} representation of $\mathrm{SU(3)}$.
   Now, if we turn to
   \begin{eqnarray}
     && (n_2+n_4)^1 (\epsilon_1+\epsilon_3) +  (n_1+n_3)^1 (\epsilon_2+\epsilon_4)  \nonumber \\
     && = -\cos (\vartheta_2-\vartheta_3) \cos \varphi_1 \cos \varphi_0 i \eta_2 \eta_3,   
   \end{eqnarray}
   and similar expressions obtained by circular permutations, we find
   \begin{eqnarray}\label{eq:dimvec-gellmann} 
    && (n_2+n_4)^1 (\epsilon_1+\epsilon_3) +  (n_1+n_3)^1 (\epsilon_2+\epsilon_4)  \nonumber \\ && \sim \cos \varphi_0\left[ e^{-i \sqrt{3} \varphi_c} \Psi^\dagger_R \Lambda_7 \Psi_L +  \text{H.c.} \right], \\
  &&   (n_2+n_4)^2 (\epsilon_1+\epsilon_3) +  (n_1+n_3)^2 (\epsilon_2+\epsilon_4)   \nonumber \\ && \sim\cos \varphi_0\left[ e^{-i \sqrt{3} \varphi_c} \Psi^\dagger_R \Lambda_5 \Psi_L +  \text{H.c.}  \right], \\
&&      (n_2+n_4)^3 (\epsilon_1+\epsilon_3) +  (n_1+n_3)^3 (\epsilon_2+\epsilon_4)  \nonumber \\ && \sim  \cos \varphi_0\left[ e^{-i \sqrt{3} \varphi_c} \Psi^\dagger_R \Lambda_2 \Psi_L +  \text{H.c.} \right], 
   \end{eqnarray}
   showing that these operators also transform according to the  \textbf{8} representation of $\mathrm{SU(3)}$. If we consider their correlation functions, since $\cos \varphi_0$, $\sin \varphi_3$ and $\cos \varphi_3$ are all long range ordered, their exponential decay is determined by the one of $e^{i(\vartheta_i-\vartheta_j)}$. 
   As a result, they must present the same correlation correlation length as $Q^{ab}_{++} $, and at long distance, the correlation functions are proportional to each other. The difference in amplitude  result from the different expectation values $\langle \cos \varphi_3 \rangle \ne \langle \sin \varphi_3\rangle$ and the different prefactors $\lambda$ and $\bar{\lambda}$.  This proportionality is a first sign of the hidden $\mathrm{SU(3)}$ symmetry of the model.
   Moreover, in the case in which  stable breathers belonging to the  \textbf{8} representation exist, an excited state containing a single breather will present a non-vanishing matrix element with the ground state for one of the 8 operators we have just identified. Calling $q$ that operator, 
   the Fourier transform of its ground state correlator $\langle \{q(x,t),q(0,0)\}\rangle$ contains a contribution
   \begin{equation}
     |\langle B,k|q|0\rangle|^2 \delta (\omega -\sqrt{(uk)^2 + m_B^2}),  
   \end{equation}
   separate from any continuum. The dynamical structure factors of the operators $Q_{++}^{ab}$ and $(\mathbf{n_2}+\mathbf{n_4})(\epsilon_1+\epsilon_3)+ (\mathbf{n_1}+\mathbf{n_2})(\epsilon_2+\epsilon_4)$ then show sharp peaks associated with the breathers.
   In the lattice model, the operators to consider are
   \begin{eqnarray}\label{eq:nematic-lattice} 
     \tensor{Q}_{++}(n) &\sim& \frac 1 2 \left[(\mathbf{S}_{n,1}+\mathbf{S}_{n,3}-\mathbf{S}_{n-1,1}-\mathbf{S}_{n-1,3})\right. \nonumber \\ && \left.\otimes (\mathbf{S}_{n,2}+\mathbf{S}_{n,4}-\mathbf{S}_{n-1,2}-\mathbf{S}_{n-1,4}) \right. \nonumber \\ && \left.+  (\mathbf{S}_{n,2}+\mathbf{S}_{n,4}-\mathbf{S}_{n-1,2}-\mathbf{S}_{n-1,4})\right. \nonumber \\ && \left.\otimes (\mathbf{S}_{n,1}+\mathbf{S}_{n,3}-\mathbf{S}_{n-1,1}-\mathbf{S}_{n-1,3}) \right. \nonumber \\ && \left. - \frac 2 3  (\mathbf{S}_{n,1}+\mathbf{S}_{n,3}-\mathbf{S}_{n-1,1}-\mathbf{S}_{n-1,3})\right. \nonumber \\ && \left. \cdot   (\mathbf{S}_{n,2}+\mathbf{S}_{n,4}-\mathbf{S}_{n-1,2}-\mathbf{S}_{n-1,4}) \openone \right], 
   \end{eqnarray}
   for the ferroquadrupolar order parameter, 
where $\mathbf{S}_{n,p}-\mathbf{S}_{n-1,p}$ is used to filter out the contribution from $\mathbf{J}_{R,p}+\mathbf{J}_{L,p}$ and retain only $\mathbf{n}_p$. Similarly, for the vector operator $ (\epsilon_1+\epsilon_3)(\mathbf{n}_2+\mathbf{n}_4) +  (\epsilon_2+\epsilon_4)(\mathbf{n}_1+\mathbf{n}_3) $, the lattice expression is 
   \begin{eqnarray}\label{eq:dimvec-lattice}                                     
     && [\mathbf{S}_{n,1} \cdot (\mathbf{S}_{n+1,1}-\mathbf{S}_{n-1,1})+  \mathbf{S}_{n,3} \cdot (\mathbf{S}_{n+1,3}-\mathbf{S}_{n-1,3})] \nonumber \\ && \times (\mathbf{S}_{n,2}+\mathbf{S}_{n,4}-\mathbf{S}_{n-1,2}-\mathbf{S}_{n-1,4}) \nonumber \\ 
     &&+ [ \mathbf{S}_{n,2} \cdot (\mathbf{S}_{n+1,2}-\mathbf{S}_{n-1,2})+  \mathbf{S}_{n,4} \cdot (\mathbf{S}_{n+1,4}-\mathbf{S}_{n-1,4})] \nonumber \\ && \times (\mathbf{S}_{n,1}+\mathbf{S}_{n,3}-\mathbf{S}_{n-1,1}-\mathbf{S}_{n-1,3}). 
   \end{eqnarray}
   Turning to  $(\mathbf{n}_1+\mathbf{n}_3)\times(\mathbf{n}_2+\mathbf{n}_4)$ we find
   \begin{eqnarray}
   &&  (n_1+n_3)^1 (n_2+n_4)^2 - (n_1+n_3)^2 (n_2+n_4)^1\nonumber \\ &=& \sin (\vartheta_1+\vartheta_2) \sin \varphi_3 \cos \varphi_0 i \eta_1 \eta_2, \\
 &&    (n_1+n_3)^2 (n_2+n_4)^3 - (n_1+n_3)^3 (n_2+n_4)^2\nonumber \\ &=& \sin (\vartheta_2+\vartheta_3) \sin \varphi_1 i \cos \varphi_0 \eta_2 \eta_3, \\
 &&     (n_1+n_3)^3 (n_2+n_4)^1 - (n_1+n_3)^1 (n_2+n_4)^3\nonumber \\ &=& \sin (\vartheta_1+\vartheta_3) \sin \varphi_2  \cos \varphi_0i \eta_3 \eta_1, 
   \end{eqnarray}
   allowing us to rewrite
   \begin{eqnarray}
  &&   [(\mathbf{n}_1+\mathbf{n}_3)\times(\mathbf{n}_2+\mathbf{n}_4)]_j = \sin (\sqrt{3} \vartheta_c- \vartheta_j) \sin \varphi_j  \cos \varphi_0\frac i 2  \epsilon_{jkl}\eta_j \eta_l, \nonumber \\
     &\sim& \frac 1 4 \left[e^{i\sqrt{3} \vartheta_c} (\psi^\dagger_{R j} - \psi^\dagger_{L j} ) -  \text{H. c.} \right]   \cos \varphi_0
   \end{eqnarray}
   showing that these operators transform in the $\mathbf{3}$ and $\mathbf{\bar{3}}$ representations of $\mathrm{SU(3)}$. However, since they don't shift $\varphi_0$,  their matrix elements between the ground state and states containing a soliton or an antisoliton vanish.  On the lattice, those operators can be written
   \begin{eqnarray}
     \label{eq:lattice-cross}
     (\mathbf{S}_{n+1,1}+ \mathbf{S}_{n+1,3}-\mathbf{S}_{n,1}- \mathbf{S}_{n,3})\times(\mathbf{S}_{n+1,2}+ \mathbf{S}_{n+1,4}-\mathbf{S}_{n,2}- \mathbf{S}_{n,4}). \nonumber \\ 
   \end{eqnarray}
   The other symmetric operators are $\tensor{Q}_{--}=(\mathbf{n}_1-\mathbf{n}_3)\otimes (\mathbf{n}_2-\mathbf{n}_4)$,   $(\mathbf{n}_2-\mathbf{n}_4) (\epsilon_1-\epsilon_3) +  (\mathbf{n}_1-\mathbf{n}_3) (\epsilon_2-\epsilon_4)$   and  $(\mathbf{n}_1-\mathbf{n}_3)\times(\mathbf{n}_2-\mathbf{n}_4)$. The expression of $Q_{--}$ is deduced from the one of $Q_{++}$ by the duality transformation (see App.~\ref{app:2leg-ising})  $\mu_{o,j} \leftrightarrow \sigma_{o,j}$ and  $\mu_{e,j} \leftrightarrow \sigma_{e,j}$.  
   As a result, its  bosonized expression is given by the change of variable $\varphi_j \to \frac \pi 2 -\varphi_j$ and $\vartheta_j \to \frac \pi 2 -\vartheta_j$.  Under such duality,
\begin{eqnarray}
  \psi_{Rj} \to \psi_{Rj}^\dagger, \\
  \psi_{Lj} \to - \psi_{Lj}^\dagger,
\end{eqnarray}
and $\sqrt{3} \varphi_c \to \frac{3\pi}{2} -\sqrt{3}\varphi_c$. We then obtain from Eq.~(\ref{eq:symtens-gellmann})
\begin{eqnarray}
  \label{eq:symtensmm}
   && Q^{12}_{--} \sim   \sin \varphi_0\left[ e^{-i \sqrt{3} \varphi_c} \Psi^\dagger_R \Lambda_1 \Psi_L+\text{H. c.} \right],  \\
    && Q^{23} _{--} \sim   \sin \varphi_0\left[ e^{-i \sqrt{3} \varphi_c} \Psi^\dagger_R \Lambda_6 \Psi_L +  \text{H. c.} \right],  \\
    && Q^{13}_{--} \sim  \sin \varphi_0\left[ e^{-i \sqrt{3} \varphi_c} \Psi^\dagger_R \Lambda_4 \Psi_L + \text{H. c.}  \right], \\
    && Q^{11}_{--}- Q^{12} _{--} \sim \sin \varphi_0\left[ e^{-i \sqrt{3} \varphi_c} \Psi^\dagger_R \Lambda_3 \Psi_L +  \text{H. c.} \right],\\
    && Q^{33} _{--} \sim  \sin \varphi_0\left[ e^{-i \sqrt{3} \varphi_c} \Psi^\dagger_R \Lambda_8 \Psi_L +  \text{H. c.} \right],
\end{eqnarray}
Applying the same argument to (\ref{eq:dimvec-gellmann}), we obtain
\begin{eqnarray}
   && (n_2-n_4)^1 (\epsilon_1-\epsilon_3) +  (n_1-n_3)^1 (\epsilon_2-\epsilon_4) \nonumber \\  &\sim&  i \sin \varphi_0\left[ e^{-i \sqrt{3} \varphi_c} \Psi^\dagger_R \Lambda_7 \Psi_L -  e^{-i \sqrt{3} \varphi_c} \Psi^\dagger_L \Lambda_7 \Psi_R\right], \\
    && (n_2-n_4)^2 (\epsilon_1-\epsilon_3) +  (n_1-n_3)^2 (\epsilon_2-\epsilon_4) \nonumber \\ &\sim& i \sin \varphi_0\left[ e^{-i \sqrt{3} \varphi_c} \Psi^\dagger_R \Lambda_5 \Psi_L -  e^{-i \sqrt{3} \varphi_c} \Psi^\dagger_L \Lambda_5 \Psi_R\right], \\
     &&  (n_2+n_4)^3 (\epsilon_1+\epsilon_3) +  (n_1+n_3)^3 (\epsilon_2+\epsilon_4)  \nonumber \\&\sim& i \sin \varphi_0\left[ e^{-i \sqrt{3} \varphi_c} \Psi^\dagger_R \Lambda_2 \Psi_L -  e^{-i \sqrt{3} \varphi_c} \Psi^\dagger_L \Lambda_2 \Psi_R\right], 
\end{eqnarray}
so both $\tensor{Q}_{--}$ and $(\mathbf{n}_2-\mathbf{n}_4) (\epsilon_1-\epsilon_3) +  (\mathbf{n}_1-\mathbf{n}_3) (\epsilon_2-\epsilon_4)$  transform in the $\mathbf{8}$ representation of $\mathrm{SU(3)}$.  Because of the presence of the factor $\sin \varphi_0$, the correlation function of operators $Q^{ab}_{--}$ is shorter than the one of the operators $Q_{ab}^{++}$.   We also find that
\begin{equation}
   [(\mathbf{n}_1-\mathbf{n}_3)\times(\mathbf{n}_2-\mathbf{n}_4)]_j \sim \frac 1 4 \left[e^{i\sqrt{3} \vartheta_c} (\psi^\dagger_{R j} + \psi^\dagger_{L j} ) + \text{H. c.} \right]   \sin \varphi_0, 
\end{equation}

so $(\mathbf{n}_1-\mathbf{n}_3)\times(\mathbf{n}_2-\mathbf{n}_4)$ is a linear combination operators transforming in the $\mathbf{3}$ and $\mathbf{\bar{3}}$ representations. Lattice expressions can be obtained on the model of Eqs.~(\ref{eq:nematic-lattice})--~(\ref{eq:dimvec-lattice}) and Eq.~(\ref{eq:lattice-cross}).

 \subsubsection{Asymmetric case}
\label{sec:asym-obs}
We now turn to combinations of operators that are symmetric on one pair of legs and antisymmetric on the other pair. We begin with operators symmetric on even legs, and antisymmetric on the odd legs. 
For the quadrupole operator,  $Q_{-+}^{ab}=(n_1-n_3)^a(n_2+n_4)^b+  (n_1-n_3)^b(n_2+n_4)^a$, we find
\begin{eqnarray}
 && Q_{-+}^{12} = \sin (\varphi_1-\varphi_2) \cos \vartheta_3\sin \vartheta_0 i\eta_3 \eta_0,  \\ 
 &&  Q_{-+}^{23} =\sin (\varphi_2-\varphi_3) \cos \vartheta_1\sin \vartheta_0 i\eta_1 \eta_0, \\ 
 && Q_{-+}^{31} =\sin (\varphi_3-\varphi_1) \cos \vartheta_2\sin \vartheta_0 i\eta_2 \eta_0,  \\
 && Q_{-+}^{11}-Q_{-+}^{22}= \sin(\vartheta_1 -\vartheta_2) \cos \vartheta_3 \sin \vartheta_0 \eta_1 \eta_2\eta_3\eta_0,   \\
&&  Q_{-+}^{11}+Q_{-+}^{22}-2Q_{-+}^{33}=\frac 1 2 \left[\sin (\vartheta_1+\vartheta_3-\vartheta_2) \right. \\ && \left.+ \sin (\vartheta_2+\vartheta_3-\vartheta_1) -2  \sin (\vartheta_1+\vartheta_2-\vartheta_3) \right] \sin \vartheta_0 \eta_1 \eta_2\eta_3\eta_0,  \nonumber \\ 
 && \mathrm{Tr}(Q_{-+}) =\frac 1 4 \left[3 \sin (\vartheta_1+\vartheta_2+\vartheta_3) + \sin (\vartheta_1+\vartheta_2-\vartheta_3) \right. \nonumber \\ && \left.+ \sin (\vartheta_3+\vartheta_1-\vartheta_2)+  \sin (\vartheta_2+\vartheta_3-\vartheta_1)\right] \sin \vartheta_0 \eta_1 \eta_2\eta_3\eta_0,  \nonumber \\    
\end{eqnarray}
which can be rewritten
\begin{eqnarray}
 && Q_{-+}^{12} \sim \sin \vartheta_0 \left[ e^{-i\sqrt{3} \vartheta_c } \Psi_R \Lambda_1 \Psi_L + \mathrm{H. c.}\right] \eta_1 \eta_2\eta_3\eta_0,\\
 && Q_{-+}^{23} \sim \sin \vartheta_0 \left[ e^{-i\sqrt{3} \vartheta_c } \Psi_R \Lambda_6 \Psi_L + \mathrm{H. c.}\right] \eta_1 \eta_2\eta_3\eta_0, \\
&&  Q_{-+}^{31} \sim \sin \vartheta_0 \left[ e^{-i\sqrt{3} \vartheta_c } \Psi_R \Lambda_4 \Psi_L + \mathrm{H. c.}\right] \eta_1 \eta_2\eta_3\eta_0, \\
&&  Q_{-+}^{11}-Q_{-+}^{22}\sim \sin \vartheta_0 \left[ e^{-i\sqrt{3} \vartheta_c } \Psi_R \Lambda_3 \Psi_L + \mathrm{H. c.}\right] \eta_1 \eta_2\eta_3\eta_0, \nonumber \\
 && Q_{-+}^{11}+Q_{-+}^{22}-2Q_{-+}^{33}\sim \sin \vartheta_0 \left[ e^{-i\sqrt{3} \vartheta_c } \Psi_R \Lambda_8 \Psi_L + \mathrm{H. c.}\right] \nonumber \\ && \times \eta_1 \eta_2\eta_3\eta_0, \\
 && \mathrm{Tr}(Q_{-+}) \sim \sin \vartheta_0 \left[ e^{-i\sqrt{3} \vartheta_c } (C+ \Psi_R  \Psi_L) + \mathrm{H. c.}\right] \eta_1 \eta_2\eta_3\eta_0, \nonumber \\
\end{eqnarray}
showing (see App.~\ref{app:bilinear}) that $Q^{ab}_{-+}$ is a linear combination of operators transforming in the $\mathbf{6}$ and $\mathbf{\bar{6}}$ representations of $\mathrm{SU(3)}$. In contrast with the symmetric case,  all operators have the same prefactor $\lambda^2$ and there are no differences in expectation values. As a result, all the operators have the same auto correlation function except $\mathrm{Tr}(Q_{-+})$ that has an extra contribution from the correlator of $e^{-i\sqrt{3} \theta_c }$. That contribution does not produce cross correlation with $   e^{-i\sqrt{3} \theta_c } \Psi_R  \Psi_L$ thanks to the unbroken $\mathrm{U(1)}$ symmetry $\vartheta_c \to \vartheta_c+\alpha$. Moreover, $e^{-i\sqrt{3} \theta_c }$ creates an excitation with larger charge $Q$ that the operators transforming in the $\mathbf{6},\mathbf{\bar{6}}$ representations. In turn, this implies the creation of a larger number of solitons and a faster exponential decay for the correlator of $e^{-i\sqrt{3} \theta_c }$.  Asymptotically, the six operators show the same correlation function, revealing the $\mathrm{SU(3)}$ symmetry. 
If we turn to $(\mathbf{n}_1-\mathbf{n}_3)\times(\mathbf{n}_2+\mathbf{n}_4)$,
\begin{eqnarray}
  [(\mathbf{n}_1-\mathbf{n}_3)\times(\mathbf{n}_2+\mathbf{n}_4)]^1 &=& \cos \vartheta_1 \sin(\varphi_2+\varphi_3) \sin \vartheta_0 i \eta_1 \eta_0, \nonumber \\
 \\
  \lbrack(\mathbf{n}_1-\mathbf{n}_3)\times(\mathbf{n}_2+\mathbf{n}_4)\rbrack^2&=& \cos \vartheta_2 \sin(\varphi_1+\varphi_3) \sin \vartheta_0 i \eta_2 \eta_0, \nonumber \\
  \\
  \lbrack(\mathbf{n}_1-\mathbf{n}_3)\times(\mathbf{n}_2+\mathbf{n}_4)\rbrack^3&=& \cos \vartheta_3 \sin(\varphi_1+\varphi_2) \sin \vartheta_0 i \eta_3 \eta_0, \nonumber \\
\end{eqnarray}
we can rewrite
\begin{eqnarray}
   [(\mathbf{n}_1-\mathbf{n}_3)\times(\mathbf{n}_2+\mathbf{n}_4)]^j&\sim& \frac 1 4 \sin\vartheta_0 \left[e^{i\sqrt{3}\varphi_c} (\psi_{Rj} + \psi^\dagger_{Lj}) + \mathrm{H.c.}\right],  \nonumber \\
\end{eqnarray}
showing that $(\mathbf{n}_1-\mathbf{n}_3)\times(\mathbf{n}_2+\mathbf{n}_4)$ is a linear combination of operators transforming in the $\mathbf{3}$ and $\mathbf{\bar{3}}$ representations of $\mathrm{SU(3)}$. We note that the operator carries the same quantum numbers as the solitons or antisolitons, implying that they will give rise to sharp peaks in its dynamical structure factor.
If we turn our attention to $(\epsilon_1-\epsilon_3)(\mathbf{n}_2+\mathbf{n}_4)-(\mathbf{n}_1-\mathbf{n}_3)(\epsilon_2+\epsilon_4)$, we find
\begin{eqnarray}
 && (\epsilon_1-\epsilon_3)({n}_2+{n}_4)^1-({n}_1-{n}_3)(\epsilon_2+\epsilon_4)^1 \nonumber \\ &\sim & \sin \vartheta_0 \left[ e^{-i \sqrt{3} \vartheta_c}\Psi_R \Lambda_7 \Psi_L +\mathrm{H. c.}\right], \\
  &&(\epsilon_1-\epsilon_3)({n}_2+{n}_4)^2-({n}_1-{n}_3)(\epsilon_2+\epsilon_4)^2 \nonumber \\ &\sim & \sin \vartheta_0 \left[ e^{-i \sqrt{3} \vartheta_c}\Psi_R \Lambda_5 \Psi_L +\mathrm{H. c.}\right], \\
 &&  (\epsilon_1-\epsilon_3)({n}_2+{n}_4)^3-({n}_1-{n}_3)(\epsilon_2+\epsilon_4)^3 \nonumber \\ &\sim & \sin \vartheta_0 \left[ e^{-i \sqrt{3} \vartheta_c}\Psi_R \Lambda_2 \Psi_L +\mathrm{H. c.}\right], 
\end{eqnarray}
showing that
$(\epsilon_1-\epsilon_3)(\mathbf{n}_2+\mathbf{n}_4)-(\mathbf{n}_1-\mathbf{n}_3)(\epsilon_2+\epsilon_4)$
is a linear combination of operators transforming in the $\mathbf{3}$
and $\mathbf{\bar{3}}$ representation.  We can also consider
$Q_{-+}^{ab} =(n_1+n_3)^a (n_2-n_4)^b + (n_1+n_3)^a (n_2-n_4)^b $ and
the vector product
$(\mathbf{n}_1+\mathbf{n}_3)\times (\mathbf{n}_2-\mathbf{n}_4)$. As
before, their bosonized expressions are obtained from those of $\tensor{Q}_{+-}$ and
$(\mathbf{n}_1-\mathbf{n}_3)\times (\mathbf{n}_2+\mathbf{n}_4)$ by the
duality transformation $\vartheta_j \to \frac \pi 2 -\vartheta_j$ and
$\varphi_j \to \frac \pi 2 -\varphi_j$.  In the end, the operators
$Q_{-+}^{ab}$ are also linear combination of operators in the
$\mathbf{6}$ and $\mathbf{\bar{6}}$ representation, while the
components
$(\mathbf{n}_1+\mathbf{n}_3)\times (\mathbf{n}_2-\mathbf{n}_4)$ are
also linear combination of operators in the $\mathbf{3}$ and
$\mathbf{\bar{3}}$ representation.

\section{Conclusion}
\label{sec:ccl}
We have found that the field theory describing the low energy
excitations of the four leg spin tube in the limit of weak rung
exchange has an enlarged $\mathrm{SU(3)}$ symmetry, broken down to
$\mathrm{SU(2)}$ only by marginal perturbations. By adding diagonal
interactions, the marginal perturbations can be canceled, enhancing
the $\mathrm{SU(3)}$ symmetry of the spectrum at low energy. The
spectrum of the low energy theory organizes in multiplets of
$\mathrm{SU(3)}$ classified by isospin and
hypercharge\cite{itzykson-zuber}. While the isospin is directly
related with the total spin, the hypercharge is non-local in the spin
operators of the tube. The $\mathrm{SU(3)}$ symmetry is thus revealed
by apparently accidental degeneracies of the spectrum
when it is decomposed into the expected $\mathrm{SU(2)}$ spin multiplets.  In
particular, we have identified two degenerate $\mathrm{SU(2)}$
triplets that correspond to the fundamental and conjugate
representations of $\mathrm{SU(3)}$. Such degeneracies should be
detectable in exact diagonalization
studies.\cite{noack_2005,sandvik_computational_2010} We have shown that the
triplet excitations would give rise to coherent peaks in the dynamical
spin structure factor near zero momentum.  Moreover, ferroquadrupolar
(or nematic) correlations can reveal the enlarged $\mathrm{SU(3)}$
symmetry. Such correlations functions are accessible with
Density Matrix Renormalization
Group\cite{white_dmrg_letter,white_dmrg,schollwock_2005,hallberg_2006}
or Quantum Monte Carlo\cite{sandvik_computational_2010,fehske_2008}
The same dynamical symmetry enlargement should be observed in a
two-leg spin-1 ladder with biquadratic
interactions\cite{golinelli_incommensurate} along the legs when the
biquadratic interactions are tuned to the
Takhtajan-Babujian\cite{takhtajan_spin_s,babujian_spin_s}
point. Although this is a less realistic model,  it
is less computationally expensive for numerical simulations.
Concerning the spectrum of the model, open questions remain concerning
first the presence of spin singlet gapped excitations resulting from
the non-magnetic modes, and second the existence of
soliton-antisoliton bound states. Since the low-energy theory does not
seem to be integrable, these questions will have to be addressed by
other non-perturbative methods such as the truncated conformal space
approximation.\cite{james_2018} 
Another open issue is the nature of
edge states in a semi-infinite four leg spin
tube.\cite{lecheminant02_magnet} If the rung interactions are
ferromagnetic, one would expect spin-1 edge states\cite{ng_edges}
similar to those of the spin-2 chain. If the open boundary conditions
are compatible with the bulk $\mathrm{SU(3)}$ symmetry, those spin-1
edge states could turn out to be in a $\mathbf{3}$ or a
$\mathbf{\bar{3}}$ representation of $\mathrm{SU(3)}$.  
Beyond the case of spin systems,  an emergent $\mathrm{SU(3)}$ symmetry should also be present in the four leg Hubbard tube at half-filling as a consequence of spin-charge separation. Upon doping, the spin gap is robust, and the $\mathrm{SU(3)}$ emergent symmetry should be observable in the four-leg Hubbard tube\cite{jiang_2020,chung_2020,ehlers_2018} and the four-leg t-J tube\cite{jiang_2018}.

\appendix
\section{Expression of staggered spin components of a pair of spin-1/2 chains in terms of Ising order and disorder operators}\label{app:2leg-ising} 
Here we summarize the derivation in\cite{shelton_spin_ladders,nersesyan_biquadratic}. For the sake of definiteness, we treat the case of the chains with odd index. Analogous relations are obtained for the chains of even index. 
Considering the addition of two $SU(2)_1$ currents, we find 
\begin{eqnarray}\label{eq:jrplus-odd}
  J_{R,1}^+  + J_{R,3}^+ = \frac 1 {2\pi a} \left[  e^{-i\sqrt{2}(\theta_1-  \phi_1)(x)}  + e^{-i\sqrt{2}(\theta_3-  \phi_3)(x)},\right]  \\
\end{eqnarray}
Introducing
\begin{eqnarray}
  \phi_{o,+} =\frac 1 {\sqrt{2}} (\phi_1+\phi_3) \\
  \phi_{o,-} =\frac 1 {\sqrt{2}} (\phi_1-\phi_3) 
\end{eqnarray}
and the corresponding dual variables, we rewrite
\begin{eqnarray}
    J_{R,1}^+  + J_{R,3}^+ &=& \frac 1 {2\pi a} \left[  e^{-i (\theta_{o,+} -\phi_{o,+}) -i (\theta_{o,-} -\phi_{o,-}) }  \right. \nonumber \\ && \left. + e^{-i (\theta_{o,+} -\phi_{o,+}) + i (\theta_{o,-} -\phi_{o,-}) }\right],
\end{eqnarray}
and introduce fermion operators
\begin{eqnarray}
  \psi_{\nu,o,+} = \frac 1 {\sqrt{2\pi a}} e^{i (\theta_{o,+} -r_\nu \phi_{o,+})}\eta_{o,+}\\ 
  \psi_{\nu,o,-} = \frac 1 {\sqrt{2\pi a}} e^{i (\theta_{o,-} -r_\nu \phi_{o,-})}\eta_{o,-}\\ 
\end{eqnarray}
Allowing to rewrite the currents
\begin{eqnarray}\label{eq:curr-sum} 
  J^+_{R,1}+J^+_{R,3} = \psi^\dagger_{R,o,+} \eta_{o,+}\eta_{o,-}  (\psi^\dagger_{R,o,-} + \psi_{R,o,-}),  
\end{eqnarray}
The operator $i \eta_{o,+}\eta_{o,-} $ is hermitian and satisfies $(i \eta_{o,+}\eta_{o,-})^2 =1$. It can be diagonalized\cite{delft_bosonization} with eigenvalues $\pm 1$, allowing us to replace $  \eta_{o,+}\eta_{o,-}   $ with $\pm i$ in Eq.~(\ref{eq:curr-sum}). Picking $\eta_{o,+}\eta_{o,-}=-i$ , and introducing Majorana fermion operators
\begin{eqnarray}
  \psi_{R,o,+}=\frac{1}{\sqrt{2}}(\zeta_{R,o,2}+i \zeta_{R,o,1}), \\
   \psi_{R,o,-}=\frac{1}{\sqrt{2}}(\zeta_{R,o,3}+i \zeta_{R,o,0}), 
\end{eqnarray}
the currents are finally rewritten in the form
\begin{eqnarray}
  J^a_{R,1}+J_{R,3}^a = - \frac i 2  \epsilon_{abc} \zeta_{R,o,b} \zeta_{R,o,c}.  
\end{eqnarray}
We also have
\begin{eqnarray}
  \label{eq:primary-sum}
  n_{1}^++n_3^+ &=& 2 e^{i\theta_{o,+}} \cos \theta_{o,-}, \\ 
  n_1^3 + n_3^3 &=& 2 \sin \phi_{o,+} \cos \phi_{o,-}, \\
 \epsilon_1+\epsilon_3&=& 2 \cos \phi_{0,+} \cos \phi_{o,-}                   
\end{eqnarray}
and using\cite{nersesyan01_ising} we can write
\begin{eqnarray}
  \label{eq:2Ising-mapping}
  \cos \phi_{o,+}&=& \mu_{o,1} \mu_{o,2} \;   \cos \phi_{o,-} = \mu_{o,3} \mu_{o,0}, \\
  \sin\phi_{o,+}&=& \sigma_{o,1} \sigma_{o,2} i\eta_{o,1} \eta_{o,2} \;   \sin \phi_{o,-} = \sigma_{o,3} \sigma_{o,0}  i\eta_{o,3} \eta_{o,0}, \\
  \cos \theta_{o,+}&=& \sigma_{o,1} \mu_{o,2} \eta_{o,1}\;   \cos \theta_{o,-} = \sigma_{o,3} \mu_{o,0} \eta_{o,3}, \\
  \sin\theta_{o,+}&=& \mu_{o,1} \sigma_{o,2}  \eta_{o,2} \;   \sin \theta_{o,-} = \mu_{o,3} \sigma_{o,0}  \eta_{o,0},
\end{eqnarray}
where $\sigma_{o,j}$ and $\mu_{o,j}$ are Ising order and disorder operators, and  $\eta_{o,j} (j=0,1,2,3) $ are Majorana fermion operators with $\eta_{o,j}^2=1$. 
The $SU(2)_1$ primary operators are rewritten
\begin{eqnarray}\label{eq:su2-2-Ising} 
  n_1^1+n_3^1&=&\mu_{o,1} \sigma_{o,2} \sigma_{o,3} \mu_{o,0} (i\eta_{o,2} \eta_{o,3}), \\
  n_1^2+n_3^2&=& \sigma_{o,1}\mu_{o,2} \sigma_{o,3} \mu_{o,0} (i\eta_{o,3} \eta_{o,1}), \\
  n_1^3+n_3^3&=& \sigma_{o,1}\sigma_{o,2} \mu_{o,3} \mu_{o,0} (i\eta_{o,1}\eta_{o,2}), \\
  \epsilon_1+\epsilon_3 &=& \mu_{o,1}\mu_{o,2}\mu_{o,3}\mu_{o,0}. 
\end{eqnarray}
For differences of $\mathrm{SU(2)_1}$ primaries, we have
\begin{eqnarray}
  \label{eq:primary-diff}
  n_{1}^+-n_3^+ &=& 2 i e^{i\theta_{o,+}} \sin \theta_{o,-}, \\ 
  n_1^3 - n_3^3 &=& 2 \cos \phi_{o,+} \sin \phi_{o,-}, \\
 \epsilon_1-\epsilon_3&=& 2 \sin \phi_{0,+} \sin \phi_{o,-},                    
\end{eqnarray}
leading to
\begin{eqnarray}
  \label{eq:su2-2-dual}
  n_1^1-n_3^1&=&\sigma_{o,1} \mu_{o,2} \mu_{o,3} \sigma_{o,0} (i\eta_{o,1} \eta_{o,0}), \\
  n_1^2-n_3^2&=& \mu_{o,1}\sigma_{o,2} \mu_{o,3} \sigma_{o,0} (i\eta_{o,2} \eta_{o,0}), \\
  n_1^3-n_3^3&=& \mu_{o,1}\mu_{o,2} \sigma_{o,3} \sigma_{o,0} (i\eta_{o,3}\eta_{o,0}), \\
  \epsilon_1-\epsilon_3 &=& \sigma_{o,1}\sigma_{o,2}\sigma_{o,3}\sigma_{o,0}\eta_{o,1} \eta_{o,2}\eta_{o,3} \eta_{o,0}. 
\end{eqnarray}
The differences are obtained from the sums by the duality transformation $\mu \leftrightarrow \sigma$. 

With the help of Eq.~(\ref{eq:su2-2-Ising}),  the most relevant operator reads
\begin{eqnarray}
\mathcal{H}_{int,b}&=&  \frac{J_\perp \lambda^2} a \mu_{e,0}  \mu_{o,0} \left[ \mu_{o,1} \mu_{e,1} \sigma_{o,2}   \sigma_{e,2}   \sigma_{o,3}   \sigma_{e,3} \eta_{o,2}\eta_{e,2} \eta_{o,3}\eta_{e,3}  \right. \nonumber \\ && +\left.       \mu_{o,2} \mu_{e,2} \sigma_{o,3}   \sigma_{e,3}   \sigma_{o,1}   \sigma_{e,1}  \eta_{o,3}\eta_{e,3} \eta_{o,1}\eta_{e,1} \right. \nonumber \\ && +\left.  \mu_{o,3} \mu_{e,3} \sigma_{o,1}   \sigma_{e,1}   \sigma_{o,2}   \sigma_{e,2}\eta_{o,1}\eta_{e,1} \eta_{o,2}\eta_{e,2} \right].  
\end{eqnarray}
The products of Majorana fermion operators $\eta_{o,j}\eta_{e,j} $ commute among themselves, making them simultaneously diagonalizable. Since
\begin{equation}
   \eta_{o,2}\eta_{e,2} \eta_{o,3}\eta_{e,3}\eta_{o,3}\eta_{e,3} \eta_{o,1}\eta_{e,1} \eta_{o,1}\eta_{e,1} \eta_{o,2}\eta_{e,2} =-1
\end{equation}
the product of eigenvalues has to be $-1$. The most symmetrical choice is to take the eigenvalue $-1$ for all products of four Majorana fermions. This gives Eq.~(\ref{eq:interchain-ising}).

\section{Alternative derivation of the interchain interaction}
\label{app:alternative}

Using Eqs.~(\ref{eq:nplus})--~(\ref{eq:nz}),  and the fields defined in Eq.~(\ref{eq:rotation-oe}) we first write the interaction
\begin{eqnarray}
&&  (\mathbf{n}_1+\mathbf{n}_3)\cdot (\mathbf{n}_2+\mathbf{n}_3) = 2 \cos (\theta_{e+}-\theta_{o+})  \nonumber \\ && \times \left[\cos (\theta_{e-}+\theta_{o-})+\cos (\theta_{e-}-\theta_{o-}) \right] \nonumber \\ && + \left[\cos (\phi_{e+}-\phi_{o+})- \cos (\phi_{e+}+\phi_{o+})\right] \nonumber \\ && \times\left[\cos (\phi_{e-}-\phi_{o-})+\cos (\phi_{e-}+\phi_{o-})\right], 
\end{eqnarray}
then we introduce the fields
\begin{equation}
  \left(
    \begin{array}{c}
      \phi_c \\ \phi_s \\ \phi_f \\ \phi_{sf} 
    \end{array}
\right) = \left(
  \begin{array}{rrrr}
    \frac 1 2 & \frac 1 2 & \frac 1 2 & \frac 1 2 \\
    \frac 1 2 & \frac 1 2 & -\frac 1 2 & -\frac 1 2 \\
    \frac 1 2 & -\frac 1 2 & \frac 1 2 & -\frac 1 2 \\
    \frac 1 2 & -\frac 1 2 & -\frac 1 2 & \frac 1 2 
  \end{array}
\right) \left(
  \begin{array}{c}
    \phi_1 \\
    \phi_2 \\
    \phi_3 \\
    \phi_4 
  \end{array}
\right),  
\end{equation}
and rewrite
\begin{eqnarray}
  && (\mathbf{n}_1+\mathbf{n}_3)\cdot (\mathbf{n}_2+\mathbf{n}_3) = 2 \cos (\sqrt{2}\theta_f) \left[\cos (\sqrt{2}\theta_s)+\cos (\sqrt{2}\theta_{sf}) \right] \nonumber \\ && + \left[\cos (\sqrt{2}\phi_f)- \cos (\sqrt{2}\phi_c)\right]\left[\cos (\sqrt{2}\phi_s)+\cos (\sqrt{2} \phi_{sf}) \right]. 
\end{eqnarray}
Noting that the operators $\cos \sqrt{2} \theta_\nu$, $\sin \sqrt{2} \theta_\nu$ and $\sin \sqrt{2} \phi_\nu$ transform as the components of a vector under $\mathrm{SU(2)}$ rotation, we make a unitary transformation such that
\begin{eqnarray}
  \mathcal{U}^\dagger \cos \sqrt{2} \theta_\nu \mathcal{U} &= &\sin \sqrt{2}\phi_\nu, \\
  \mathcal{U}^\dagger \sin \sqrt{2} \theta_\nu \mathcal{U} &= &\sin \sqrt{2}\theta_\nu, \\
    \mathcal{U}^\dagger\sin \sqrt{2}\phi_\nu  \mathcal{U} &= & - \cos \sqrt{2} \theta_\nu,
\end{eqnarray}
for $\nu=s,f,sf$ to obtain
\begin{eqnarray}
&&   \mathcal{U}^\dagger  (\mathbf{n}_1+\mathbf{n}_3)\cdot (\mathbf{n}_2+\mathbf{n}_3) \mathcal{U} = 2 \sin (\sqrt{2}\phi_f) \left[\sin (\sqrt{2}\phi_s)+\sin (\sqrt{2}\phi_{sf}) \right] \nonumber \\ && + \left[\cos (\sqrt{2}\phi_f)- \cos (\sqrt{2}\phi_c)\right]\left[\cos (\sqrt{2}\phi_s)+\cos (\sqrt{2} \phi_{sf}) \right]. 
\end{eqnarray}
Now, with the fields
\begin{eqnarray}
  && \varphi_0 =\frac{\phi_s -\phi_{sf}}{\sqrt{2}} \; \varphi_1 = -\frac{\phi_s +\phi_{sf}}{\sqrt{2}}  \\
  && \varphi_2 =\frac{\phi_f+\phi_c}{\sqrt{2}} \; \varphi_3 =\frac{\phi_f-\phi_c}{\sqrt{2}} 
\end{eqnarray}
we recover the form
\begin{eqnarray}
&&  \mathcal{U}^\dagger  (\mathbf{n}_1+\mathbf{n}_3)\cdot   (\mathbf{n}_2+\mathbf{n}_4)\mathcal{U} =\cos \varphi_0 \left[\cos (\varphi_1+\varphi_2 -\varphi_3) \right.\nonumber \\ && \left. + \cos (\varphi_3+\varphi_1 -\varphi_2)+ \cos (\varphi_2+\varphi_3 -\varphi_1) - 3 \cos (\varphi_1+\varphi_2+\varphi_3)\right], \nonumber \\   
\end{eqnarray}
and with a change of variables analogous to Eq.~(\ref{eq:su3-basis}),
\begin{equation}\label{eq:su3-alt} 
  \left(\begin{array}{c}\varphi_c \\ \varphi_a \\ \varphi_b \end{array} \right) = \left(\begin{array}{ccc}\frac 1 {\sqrt{3}} &\frac 1 {\sqrt{3}} & \frac 1 {\sqrt{3}} \\  \frac 1 {\sqrt{2}} & - \frac 1 {\sqrt{2}} & 0 \\ \frac 1 {\sqrt{6}} & \frac 1 {\sqrt{6}} & - \frac 2 {\sqrt{6}} \end{array} \right)  \left(\begin{array}{c}\varphi_2\\ \varphi_3 \\ \varphi_1 \end{array} \right),  
\end{equation}
we recover the form Eq.~(\ref{eq:su3-exchange}) for the interaction. Note that in Eq.~(\ref{eq:su3-alt}), we have made a circular permutation of $\varphi_{1,2,3}$ compared with Eq.~(\ref{eq:su3-basis}). The reason is that such choice of variables gives us
\begin{equation}
  -\frac{\sqrt{2}} \pi \partial_x \varphi_a = -\frac{\sqrt{2}}\pi \partial_x \phi_c,  
\end{equation}
and this implies that $\partial_x \varphi_a$ is proportional to the magnetization density.
While the approach in this appendix is convenient to establish the $SU(3)$ symmetry in the low energy Hamiltonian, it is impractical to derive bosonized expressions
of the $\mathrm{SU(2)_4}$ currents $\sum_{n} J_{R/L,n}^{x,y}$. The reason is that the transformation under the $SU(2)$
rotation of the operators $e^{i\pm (\theta_\nu\pm \phi_\nu)/\sqrt{2}}$ is ambiguous. 
Indeed, the expression $e^{i(\theta_\sigma-\phi_\sigma)/\sqrt{2}}$  appears both in the bosonized representation of the spin up annihilation operator and  in the bosonized representation of the spin down creation operator\cite{giamarchi_book_1d}. However,  those operators transform differently under $\mathrm{SU(2)}$ rotation. To have a well defined transformation under $\mathrm{SU(2)}$ rotation we must specify if we are considering $e^{i(\theta_\sigma-\phi_\sigma)/\sqrt{2}}\eta_{\uparrow}$  or $e^{i(\theta_\sigma-\phi_\sigma)/\sqrt{2}}\eta_{\downarrow}$.
\section{Transformation of fermion bilinears}
\label{app:bilinear}
If we consider a fermion bilinear
\begin{equation}
  \sum_{\alpha,\beta} \psi_{R,\alpha} M_{\alpha\beta} \psi_{L,\beta},  
\end{equation} 
since we can always write $M=(M+{}^tM)/2 + (M-{}^tM)/2$, we can without loss of generality consider the cases $M={}^tM$ and $M=-{}^tM$ separately.
In the first case,  the fermion bilinear can be written
\begin{eqnarray}
   \frac 1 2  \sum_{\alpha,\beta} M_{\alpha\beta}  (\psi_{R,\alpha} \psi_{L,\beta} + \psi_{R,\beta} \psi_{L,\alpha} ),  
\end{eqnarray}
and under a $\mathrm{SU(3)}$ rotation $\psi_{\nu,\alpha}=\sum_{\alpha'} \tilde{\psi}_{\nu\alpha'}$, it becomes
\begin{eqnarray}
  \frac 1 2  \sum_{\alpha,\beta, \alpha',\beta'} M_{\alpha\beta}U_{\alpha\alpha'} U_{\beta\beta'} ( \tilde{\psi}_{R,\alpha'} \tilde{\psi}_{L,\beta'}  + \tilde{\psi}_{R,\beta'} \tilde{\psi}_{L,\alpha'} ),\nonumber \\
  =  \frac 1 2  \sum_{\alpha',\beta'} ({}tUMU)_{\alpha'\beta'}   ( \tilde{\psi}_{R,\alpha'} \tilde{\psi}_{L,\beta'}  + \tilde{\psi}_{R,\beta'} \tilde{\psi}_{L,\alpha'} ). 
\end{eqnarray}
The matrix $M$ is transformed in the new symmetric matrix ${}^tU M U$. Symmetric matrices in $\mathrm{M}_3(\Bbb{C})$ are a 6-dimensional vector space, showing that the fermion bilinear are in the $\mathbf{6}$ representation of $\mathrm{SU(3)}$. The Gell-Mann matrices $\Lambda_1,\Lambda_3,\Lambda_4,\Lambda_6,\Lambda_8$ and the identity matrix span the space of symmetric matrices. 
With $M$ antisymmetric, the fermion bilinear is now written
\begin{eqnarray}
   \frac 1 2  \sum_{\alpha,\beta} M_{\alpha\beta}  (\psi_{R,\alpha} \psi_{L,\beta} - \psi_{R,\beta} \psi_{L,\alpha} ),  
\end{eqnarray}
and  $M$ now transforms in the antisymmetric matrix ${}^tU M U$.  Antisymmetric matrices in $\mathrm{M}_3(\Bbb{C})$ are a 3-dimensional vector space, spanned by $\Lambda_2,\Lambda_5,\Lambda_7$. Moreover, since $\mathrm{det} U=1$, antisymmetric combinations of $U_{\alpha\alpha'} U_{\beta\beta'}$ are combining into $U^{-1}$. The three dimensional representation corresponds to $\mathbf{\bar{3}}$.

\begin{acknowledgments}
  I thank Sylvain Capponi, Bruce Normand for discussions and suggestions and Philippe Lecheminant for comments on the manuscript. 
\end{acknowledgments}


\begin{thebibliography}{101}%
\makeatletter
\providecommand \@ifxundefined [1]{%
 \@ifx{#1\undefined}
}%
\providecommand \@ifnum [1]{%
 \ifnum #1\expandafter \@firstoftwo
 \else \expandafter \@secondoftwo
 \fi
}%
\providecommand \@ifx [1]{%
 \ifx #1\expandafter \@firstoftwo
 \else \expandafter \@secondoftwo
 \fi
}%
\providecommand \natexlab [1]{#1}%
\providecommand \enquote  [1]{``#1''}%
\providecommand \bibnamefont  [1]{#1}%
\providecommand \bibfnamefont [1]{#1}%
\providecommand \citenamefont [1]{#1}%
\providecommand \href@noop [0]{\@secondoftwo}%
\providecommand \href [0]{\begingroup \@sanitize@url \@href}%
\providecommand \@href[1]{\@@startlink{#1}\@@href}%
\providecommand \@@href[1]{\endgroup#1\@@endlink}%
\providecommand \@sanitize@url [0]{\catcode `\\12\catcode `\$12\catcode
  `\&12\catcode `\#12\catcode `\^12\catcode `\_12\catcode `\%12\relax}%
\providecommand \@@startlink[1]{}%
\providecommand \@@endlink[0]{}%
\providecommand \url  [0]{\begingroup\@sanitize@url \@url }%
\providecommand \@url [1]{\endgroup\@href {#1}{\urlprefix }}%
\providecommand \urlprefix  [0]{URL }%
\providecommand \Eprint [0]{\href }%
\providecommand \doibase [0]{https://doi.org/}%
\providecommand \selectlanguage [0]{\@gobble}%
\providecommand \bibinfo  [0]{\@secondoftwo}%
\providecommand \bibfield  [0]{\@secondoftwo}%
\providecommand \translation [1]{[#1]}%
\providecommand \BibitemOpen [0]{}%
\providecommand \bibitemStop [0]{}%
\providecommand \bibitemNoStop [0]{.\EOS\space}%
\providecommand \EOS [0]{\spacefactor3000\relax}%
\providecommand \BibitemShut  [1]{\csname bibitem#1\endcsname}%
\let\auto@bib@innerbib\@empty
\bibitem [{\citenamefont {Gomes}(2016)}]{gomes_2016}%
  \BibitemOpen
  \bibfield  {author} {\bibinfo {author} {\bibfnamefont {P.~R.~S.}\
  \bibnamefont {Gomes}},\ }\bibfield  {title} {\bibinfo {title} {Aspects of
  {Emergent} {Symmetries}},\ }\href {https://doi.org/10.1142/S0217751X1630009X}
  {\bibfield  {journal} {\bibinfo  {journal} {International Journal of Modern
  Physics A}\ }\textbf {\bibinfo {volume} {31}},\ \bibinfo {pages} {1630009}
  (\bibinfo {year} {2016})},\ \bibinfo {note} {arXiv:1510.04492 [cond-mat,
  physics:hep-th]}\BibitemShut {NoStop}%
\bibitem [{\citenamefont {{Di Francesco}}\ \emph {et~al.}(1997)\citenamefont
  {{Di Francesco}}, \citenamefont {Mathieu},\ and\ \citenamefont
  {Senechal}}]{difrancesco_book_conformal}%
  \BibitemOpen
  \bibfield  {author} {\bibinfo {author} {\bibfnamefont {P.}~\bibnamefont {{Di
  Francesco}}}, \bibinfo {author} {\bibfnamefont {P.}~\bibnamefont {Mathieu}},\
  and\ \bibinfo {author} {\bibfnamefont {D.}~\bibnamefont {Senechal}},\
  }\href@noop {} {\emph {\bibinfo {title} {Conformal Field Theory}}}\ (\bibinfo
   {publisher} {Springer-Verlag},\ \bibinfo {address} {Berlin},\ \bibinfo
  {year} {1997})\BibitemShut {NoStop}%
\bibitem [{\citenamefont {Witten}(1984)}]{witten_wz}%
  \BibitemOpen
  \bibfield  {author} {\bibinfo {author} {\bibfnamefont {E.}~\bibnamefont
  {Witten}},\ }\href@noop {} {\bibfield  {journal} {\bibinfo  {journal}
  {Commun. Math. Phys.}\ }\textbf {\bibinfo {volume} {92}},\ \bibinfo {pages}
  {455} (\bibinfo {year} {1984})}\BibitemShut {NoStop}%
\bibitem [{\citenamefont {Chen}\ \emph {et~al.}(2015)\citenamefont {Chen},
  \citenamefont {Xue}, \citenamefont {McCulloch}, \citenamefont {Chung},
  \citenamefont {Huang},\ and\ \citenamefont {Yip}}]{chen_2015}%
  \BibitemOpen
  \bibfield  {author} {\bibinfo {author} {\bibfnamefont {P.}~\bibnamefont
  {Chen}}, \bibinfo {author} {\bibfnamefont {Z.-L.}\ \bibnamefont {Xue}},
  \bibinfo {author} {\bibfnamefont {I.}~\bibnamefont {McCulloch}}, \bibinfo
  {author} {\bibfnamefont {M.-C.}\ \bibnamefont {Chung}}, \bibinfo {author}
  {\bibfnamefont {C.-C.}\ \bibnamefont {Huang}},\ and\ \bibinfo {author}
  {\bibfnamefont {S.-K.}\ \bibnamefont {Yip}},\ }\bibfield  {title} {\bibinfo
  {title} {Quantum {Critical} {Spin}-2 {Chain} with {Emergent} {SU}(3)
  {Symmetry}},\ }\href {https://doi.org/10.1103/PhysRevLett.114.145301}
  {\bibfield  {journal} {\bibinfo  {journal} {Physical Review Letters}\
  }\textbf {\bibinfo {volume} {114}},\ \bibinfo {pages} {145301} (\bibinfo
  {year} {2015})}\BibitemShut {NoStop}%
\bibitem [{\citenamefont {Li}\ \emph {et~al.}(2022)\citenamefont {Li},
  \citenamefont {Quito}, \citenamefont {Miranda}, \citenamefont {Pereira},
  \citenamefont {Affleck},\ and\ \citenamefont {Lopes}}]{li_2022}%
  \BibitemOpen
  \bibfield  {author} {\bibinfo {author} {\bibfnamefont {C.}~\bibnamefont
  {Li}}, \bibinfo {author} {\bibfnamefont {V.~L.}\ \bibnamefont {Quito}},
  \bibinfo {author} {\bibfnamefont {E.}~\bibnamefont {Miranda}}, \bibinfo
  {author} {\bibfnamefont {R.}~\bibnamefont {Pereira}}, \bibinfo {author}
  {\bibfnamefont {I.}~\bibnamefont {Affleck}},\ and\ \bibinfo {author}
  {\bibfnamefont {P.~L.~S.}\ \bibnamefont {Lopes}},\ }\bibfield  {title}
  {\bibinfo {title} {The case of {SU}(3) criticality in spin-2 chains},\ }\href
  {https://doi.org/10.1103/PhysRevB.105.085140} {\bibfield  {journal} {\bibinfo
   {journal} {Physical Review B}\ }\textbf {\bibinfo {volume} {105}},\ \bibinfo
  {pages} {085140} (\bibinfo {year} {2022})},\ \bibinfo {note} {arXiv:
  2108.10329}\BibitemShut {NoStop}%
\bibitem [{\citenamefont {Nersesyan}\ and\ \citenamefont
  {Tsvelik}(1997{\natexlab{a}})}]{nersesyan_biquad}%
  \BibitemOpen
  \bibfield  {author} {\bibinfo {author} {\bibfnamefont {A.}~\bibnamefont
  {Nersesyan}}\ and\ \bibinfo {author} {\bibfnamefont {A.~M.}\ \bibnamefont
  {Tsvelik}},\ }\bibfield  {title} {\bibinfo {title} {One-dimensional
  spin-liquid without magnon excitations},\ }\href@noop {} {\bibfield
  {journal} {\bibinfo  {journal} {Phys. Rev. Lett.}\ }\textbf {\bibinfo
  {volume} {78}},\ \bibinfo {pages} {3939} (\bibinfo {year}
  {1997}{\natexlab{a}})},\ \bibinfo {note} {ibid. , \textbf{79}, E
  1171}\BibitemShut {NoStop}%
\bibitem [{\citenamefont {Lin}\ \emph {et~al.}(1998)\citenamefont {Lin},
  \citenamefont {Balents},\ and\ \citenamefont {Fisher}}]{lin_so8}%
  \BibitemOpen
  \bibfield  {author} {\bibinfo {author} {\bibfnamefont {H.}~\bibnamefont
  {Lin}}, \bibinfo {author} {\bibfnamefont {L.}~\bibnamefont {Balents}},\ and\
  \bibinfo {author} {\bibfnamefont {M.~P.~A.}\ \bibnamefont {Fisher}},\
  }\href@noop {} {\bibfield  {journal} {\bibinfo  {journal} {Phys. Rev. B}\
  }\textbf {\bibinfo {volume} {58}},\ \bibinfo {pages} {1794} (\bibinfo {year}
  {1998})}\BibitemShut {NoStop}%
\bibitem [{\citenamefont {Gross}\ and\ \citenamefont
  {Neveu}(1974)}]{gross_neveu}%
  \BibitemOpen
  \bibfield  {author} {\bibinfo {author} {\bibfnamefont {D.~J.}\ \bibnamefont
  {Gross}}\ and\ \bibinfo {author} {\bibfnamefont {A.}~\bibnamefont {Neveu}},\
  }\bibfield  {title} {\bibinfo {title} {Dynamical symmetry breaking in
  asymptotically free field theories},\ }\href@noop {} {\bibfield  {journal}
  {\bibinfo  {journal} {Phys. Rev. D}\ }\textbf {\bibinfo {volume} {10}},\
  \bibinfo {pages} {3235} (\bibinfo {year} {1974})}\BibitemShut {NoStop}%
\bibitem [{\citenamefont {Konik}\ \emph {et~al.}(2000)\citenamefont {Konik},
  \citenamefont {Lesage}, \citenamefont {Ludwig},\ and\ \citenamefont
  {Saleur}}]{konik_exact_commensurate_ladder}%
  \BibitemOpen
  \bibfield  {author} {\bibinfo {author} {\bibfnamefont {R.}~\bibnamefont
  {Konik}}, \bibinfo {author} {\bibfnamefont {F.}~\bibnamefont {Lesage}},
  \bibinfo {author} {\bibfnamefont {A.~W.~W.}\ \bibnamefont {Ludwig}},\ and\
  \bibinfo {author} {\bibfnamefont {H.}~\bibnamefont {Saleur}},\ }\href@noop {}
  {\bibfield  {journal} {\bibinfo  {journal} {Phys. Rev. B}\ }\textbf {\bibinfo
  {volume} {61}},\ \bibinfo {pages} {4983} (\bibinfo {year}
  {2000})}\BibitemShut {NoStop}%
\bibitem [{\citenamefont {Essler}\ and\ \citenamefont
  {Konik}(2004)}]{essler04_condmat_exact_review}%
  \BibitemOpen
  \bibfield  {author} {\bibinfo {author} {\bibfnamefont {F.~H.~L.}\
  \bibnamefont {Essler}}\ and\ \bibinfo {author} {\bibfnamefont {R.~M.}\
  \bibnamefont {Konik}},\ }\bibfield  {title} {\bibinfo {title} {Applications
  of massive integrable quantum field theories to problems in condensed matter
  physics},\ }in\ \href@noop {} {\emph {\bibinfo {booktitle} {From {Fields} to
  {Strings}: {Circumnavigating} {Theoretical} {Physics}: {Ian} {Kogan}
  {Memorial} {Collection}}}},\ Vol.\ \bibinfo {volume} {Part 2: From Fields to
  Strings -- Condensed Matter},\ \bibinfo {editor} {edited by\ \bibinfo
  {editor} {\bibnamefont {{Misha Shifman}}}, \bibinfo {editor} {\bibnamefont
  {{Arkady Vainshtein}}},\ and\ \bibinfo {editor} {\bibnamefont {{John
  Wheater}}}}\ (\bibinfo  {publisher} {World Scientific},\ \bibinfo {address}
  {Singapore},\ \bibinfo {year} {2004})\ p.\ \bibinfo {pages} {684},\ \bibinfo
  {note} {cond-mat/0412421}\BibitemShut {NoStop}%
\bibitem [{\citenamefont {Schulz}(1998)}]{schulz_son}%
  \BibitemOpen
  \bibfield  {author} {\bibinfo {author} {\bibfnamefont {H.}~\bibnamefont
  {Schulz}},\ }\bibfield  {title} {\bibinfo {title} {So(n) symmetries in the
  two--chain model of correlated fermions}} (\bibinfo {year} {1998}),\ \bibinfo
  {note} {cond-mat/9808167}\BibitemShut {NoStop}%
\bibitem [{\citenamefont {Essler}\ and\ \citenamefont
  {Konik}(2007)}]{essler_2007}%
  \BibitemOpen
  \bibfield  {author} {\bibinfo {author} {\bibfnamefont {F.~H.~L.}\
  \bibnamefont {Essler}}\ and\ \bibinfo {author} {\bibfnamefont
  {R.}~\bibnamefont {Konik}},\ }\bibfield  {title} {\bibinfo {title} {Dynamical
  {Spin} {Response} of {Doped} {Two}-{Leg} {Hubbard}-like {Ladders}},\ }\href
  {https://doi.org/10.1103/PhysRevB.75.144403} {\bibfield  {journal} {\bibinfo
  {journal} {Physical Review B}\ }\textbf {\bibinfo {volume} {75}},\ \bibinfo
  {pages} {144403} (\bibinfo {year} {2007})},\ \bibinfo {note}
  {arXiv:cond-mat/0607783}\BibitemShut {NoStop}%
\bibitem [{\citenamefont {Assaraf}\ \emph {et~al.}(2004)\citenamefont
  {Assaraf}, \citenamefont {Azaria}, \citenamefont {Boulat}, \citenamefont
  {Caffarel},\ and\ \citenamefont {Lecheminant}}]{assaraf_2004}%
  \BibitemOpen
  \bibfield  {author} {\bibinfo {author} {\bibfnamefont {R.}~\bibnamefont
  {Assaraf}}, \bibinfo {author} {\bibfnamefont {P.}~\bibnamefont {Azaria}},
  \bibinfo {author} {\bibfnamefont {E.}~\bibnamefont {Boulat}}, \bibinfo
  {author} {\bibfnamefont {M.}~\bibnamefont {Caffarel}},\ and\ \bibinfo
  {author} {\bibfnamefont {P.}~\bibnamefont {Lecheminant}},\ }\bibfield
  {title} {\bibinfo {title} {Dynamical {Symmetry} {Enlargement} versus
  {Spin}-{Charge} {Decoupling} in the {One}-{Dimensional} {SU}(4) {Hubbard}
  {Model}},\ }\href {https://doi.org/10.1103/PhysRevLett.93.016407} {\bibfield
  {journal} {\bibinfo  {journal} {Physical Review Letters}\ }\textbf {\bibinfo
  {volume} {93}},\ \bibinfo {pages} {016407} (\bibinfo {year}
  {2004})}\BibitemShut {NoStop}%
\bibitem [{\citenamefont {Bunder}\ and\ \citenamefont
  {Lin}(2007)}]{bunder_dynamical_2007}%
  \BibitemOpen
  \bibfield  {author} {\bibinfo {author} {\bibfnamefont {J.~E.}\ \bibnamefont
  {Bunder}}\ and\ \bibinfo {author} {\bibfnamefont {H.-H.}\ \bibnamefont
  {Lin}},\ }\bibfield  {title} {\bibinfo {title} {Dynamical symmetry
  enlargement in metallic zigzag carbon nanotubes},\ }\href
  {https://doi.org/10.1103/PhysRevB.75.075418} {\bibfield  {journal} {\bibinfo
  {journal} {Physical Review B}\ }\textbf {\bibinfo {volume} {75}},\ \bibinfo
  {pages} {075418} (\bibinfo {year} {2007})},\ \bibinfo {note} {arXiv:0901.4097
  [cond-mat]}\BibitemShut {NoStop}%
\bibitem [{\citenamefont {Zamolodchikov}\ and\ \citenamefont
  {Zamolodchikov}(1979)}]{zamolodchikov79_smatrices}%
  \BibitemOpen
  \bibfield  {author} {\bibinfo {author} {\bibfnamefont {A.~B.}\ \bibnamefont
  {Zamolodchikov}}\ and\ \bibinfo {author} {\bibfnamefont {A.~B.}\ \bibnamefont
  {Zamolodchikov}},\ }\bibfield  {title} {\bibinfo {title} {Factorized s
  matrices in two dimensions},\ }\href@noop {} {\bibfield  {journal} {\bibinfo
  {journal} {Ann. Phys. (N. Y.)}\ }\textbf {\bibinfo {volume} {120}},\ \bibinfo
  {pages} {253} (\bibinfo {year} {1979})}\BibitemShut {NoStop}%
\bibitem [{\citenamefont {Karowski}\ and\ \citenamefont
  {Thun}(1981)}]{karowski81_gross_neveu}%
  \BibitemOpen
  \bibfield  {author} {\bibinfo {author} {\bibfnamefont {M.}~\bibnamefont
  {Karowski}}\ and\ \bibinfo {author} {\bibfnamefont {H.~J.}\ \bibnamefont
  {Thun}},\ }\bibfield  {title} {\bibinfo {title} {Complete s-matrix of the
  o(2n) gross-neveu model},\ }\href@noop {} {\bibfield  {journal} {\bibinfo
  {journal} {Nucl. Phys. B}\ }\textbf {\bibinfo {volume} {190[FS3]}},\ \bibinfo
  {pages} {61} (\bibinfo {year} {1981})}\BibitemShut {NoStop}%
\bibitem [{\citenamefont {Karowski}\ and\ \citenamefont
  {Wiesz}(1978)}]{karowski_ff}%
  \BibitemOpen
  \bibfield  {author} {\bibinfo {author} {\bibfnamefont {M.}~\bibnamefont
  {Karowski}}\ and\ \bibinfo {author} {\bibfnamefont {P.}~\bibnamefont
  {Wiesz}},\ }\bibfield  {title} {\bibinfo {title} {Exact form factors in
  (1+1)dimensional field theoretic models with soliton behavior},\ }\href@noop
  {} {\bibfield  {journal} {\bibinfo  {journal} {Nucl. Phys. B}\ }\textbf
  {\bibinfo {volume} {139}},\ \bibinfo {pages} {455} (\bibinfo {year}
  {1978})}\BibitemShut {NoStop}%
\bibitem [{\citenamefont {Cabra}\ \emph {et~al.}(1998)\citenamefont {Cabra},
  \citenamefont {Honecker},\ and\ \citenamefont {Pujol}}]{cabra_ladders}%
  \BibitemOpen
  \bibfield  {author} {\bibinfo {author} {\bibfnamefont {D.}~\bibnamefont
  {Cabra}}, \bibinfo {author} {\bibfnamefont {A.}~\bibnamefont {Honecker}},\
  and\ \bibinfo {author} {\bibfnamefont {P.}~\bibnamefont {Pujol}},\ }\bibfield
   {title} {\bibinfo {title} {Magnetization plateaux in n-leg spin ladders},\
  }\href@noop {} {\bibfield  {journal} {\bibinfo  {journal} {Phys. Rev. B}\
  }\textbf {\bibinfo {volume} {58}},\ \bibinfo {pages} {6241} (\bibinfo {year}
  {1998})},\ \bibinfo {note} {cond-mat/9802035}\BibitemShut {NoStop}%
\bibitem [{\citenamefont {Kawano}\ and\ \citenamefont
  {Takahashi}(1997)}]{kawano_3legs}%
  \BibitemOpen
  \bibfield  {author} {\bibinfo {author} {\bibfnamefont {K.}~\bibnamefont
  {Kawano}}\ and\ \bibinfo {author} {\bibfnamefont {M.}~\bibnamefont
  {Takahashi}},\ }\href@noop {} {\bibfield  {journal} {\bibinfo  {journal} {J.
  Phys. Soc. Jpn.}\ }\textbf {\bibinfo {volume} {66}},\ \bibinfo {pages} {4001}
  (\bibinfo {year} {1997})}\BibitemShut {NoStop}%
\bibitem [{\citenamefont {Tandon}\ \emph {et~al.}(1999)\citenamefont {Tandon},
  \citenamefont {Lal}, \citenamefont {Pati}, \citenamefont {Ramasesha},\ and\
  \citenamefont {Sen}}]{tandon_3legs}%
  \BibitemOpen
  \bibfield  {author} {\bibinfo {author} {\bibfnamefont {K.}~\bibnamefont
  {Tandon}}, \bibinfo {author} {\bibfnamefont {S.}~\bibnamefont {Lal}},
  \bibinfo {author} {\bibfnamefont {S.~K.}\ \bibnamefont {Pati}}, \bibinfo
  {author} {\bibfnamefont {S.}~\bibnamefont {Ramasesha}},\ and\ \bibinfo
  {author} {\bibfnamefont {D.}~\bibnamefont {Sen}},\ }\href@noop {} {\bibfield
  {journal} {\bibinfo  {journal} {Phys. Rev. B}\ }\textbf {\bibinfo {volume}
  {59}},\ \bibinfo {pages} {396} (\bibinfo {year} {1999})},\ \bibinfo {note}
  {cond-mat/9806111}\BibitemShut {NoStop}%
\bibitem [{\citenamefont {Arlego}\ and\ \citenamefont
  {Brenig}(2011)}]{arlego_4leg_2011}%
  \BibitemOpen
  \bibfield  {author} {\bibinfo {author} {\bibfnamefont {M.}~\bibnamefont
  {Arlego}}\ and\ \bibinfo {author} {\bibfnamefont {W.}~\bibnamefont
  {Brenig}},\ }\bibfield  {title} {\bibinfo {title} {Series {Expansion}
  {Analysis} of a {Frustrated} {Four}-{Spin}-{Tube}},\ }\href
  {https://doi.org/10.1103/PhysRevB.84.134426} {\bibfield  {journal} {\bibinfo
  {journal} {Physical Review B}\ }\textbf {\bibinfo {volume} {84}},\ \bibinfo
  {pages} {134426} (\bibinfo {year} {2011})},\ \bibinfo {note} {arXiv:1106.2101
  [cond-mat]}\BibitemShut {NoStop}%
\bibitem [{\citenamefont {Dagotto}\ and\ \citenamefont
  {Rice}(1996)}]{dagotto_ladder_review}%
  \BibitemOpen
  \bibfield  {author} {\bibinfo {author} {\bibfnamefont {E.}~\bibnamefont
  {Dagotto}}\ and\ \bibinfo {author} {\bibfnamefont {T.~M.}\ \bibnamefont
  {Rice}},\ }\href@noop {} {\bibfield  {journal} {\bibinfo  {journal}
  {Science}\ }\textbf {\bibinfo {volume} {271}},\ \bibinfo {pages} {5249}
  (\bibinfo {year} {1996})}\BibitemShut {NoStop}%
\bibitem [{\citenamefont {Dagotto}(1999)}]{dagotto_supra_ladder_review}%
  \BibitemOpen
  \bibfield  {author} {\bibinfo {author} {\bibfnamefont {E.}~\bibnamefont
  {Dagotto}},\ }\bibfield  {title} {\bibinfo {title} {Experiments on ladders
  reveal a complex interplay between a spin-gapped normal state and
  superconductivity},\ }\href@noop {} {\bibfield  {journal} {\bibinfo
  {journal} {Rep. Prog. Phys.}\ }\textbf {\bibinfo {volume} {62}},\ \bibinfo
  {pages} {1525} (\bibinfo {year} {1999})}\BibitemShut {NoStop}%
\bibitem [{\citenamefont {Gavilano}\ \emph {et~al.}(2003)\citenamefont
  {Gavilano}, \citenamefont {Rau}, \citenamefont {Mushkolaj}, \citenamefont
  {Ott}, \citenamefont {Millet},\ and\ \citenamefont
  {Mila}}]{gavilano03_spintube}%
  \BibitemOpen
  \bibfield  {author} {\bibinfo {author} {\bibfnamefont {J.~L.}\ \bibnamefont
  {Gavilano}}, \bibinfo {author} {\bibfnamefont {D.}~\bibnamefont {Rau}},
  \bibinfo {author} {\bibfnamefont {S.}~\bibnamefont {Mushkolaj}}, \bibinfo
  {author} {\bibfnamefont {H.~R.}\ \bibnamefont {Ott}}, \bibinfo {author}
  {\bibfnamefont {P.}~\bibnamefont {Millet}},\ and\ \bibinfo {author}
  {\bibfnamefont {F.}~\bibnamefont {Mila}},\ }\bibfield  {title} {\bibinfo
  {title} {Low-dimensional spin s=1/2 system at the quantum critical limit:
  $\mathrm{Na_2V_3O_7}$},\ }\href@noop {} {\bibfield  {journal} {\bibinfo
  {journal} {Phys. Rev. Lett.}\ }\textbf {\bibinfo {volume} {90}},\ \bibinfo
  {pages} {167202} (\bibinfo {year} {2003})}\BibitemShut {NoStop}%
\bibitem [{\citenamefont {Manaka}\ \emph {et~al.}(2009)\citenamefont {Manaka},
  \citenamefont {Hirai}, \citenamefont {Hachigo}, \citenamefont {Mitsunaga},
  \citenamefont {Ito},\ and\ \citenamefont {Terada}}]{manaka2009}%
  \BibitemOpen
  \bibfield  {author} {\bibinfo {author} {\bibfnamefont {H.}~\bibnamefont
  {Manaka}}, \bibinfo {author} {\bibfnamefont {Y.}~\bibnamefont {Hirai}},
  \bibinfo {author} {\bibfnamefont {Y.}~\bibnamefont {Hachigo}}, \bibinfo
  {author} {\bibfnamefont {M.}~\bibnamefont {Mitsunaga}}, \bibinfo {author}
  {\bibfnamefont {M.}~\bibnamefont {Ito}},\ and\ \bibinfo {author}
  {\bibfnamefont {N.}~\bibnamefont {Terada}},\ }\bibfield  {title} {\bibinfo
  {title} {Spin-{Liquid} {State} {Study} of {Equilateral} {Triangle} {S}=3/2
  {Spin} {Tubes} {Formed} in {CsCrF4}},\ }\href
  {https://doi.org/10.1143/JPSJ.78.093701} {\bibfield  {journal} {\bibinfo
  {journal} {Journal of the Physical Society of Japan}\ }\textbf {\bibinfo
  {volume} {78}},\ \bibinfo {pages} {093701} (\bibinfo {year}
  {2009})}\BibitemShut {NoStop}%
\bibitem [{\citenamefont {Garlea}\ \emph {et~al.}(2008)\citenamefont {Garlea},
  \citenamefont {Zheludev}, \citenamefont {Regnault}, \citenamefont {Chung},
  \citenamefont {Qiu}, \citenamefont {Boehm}, \citenamefont {Habicht},\ and\
  \citenamefont {Meissner}}]{garlea2008}%
  \BibitemOpen
  \bibfield  {author} {\bibinfo {author} {\bibfnamefont {V.~O.}\ \bibnamefont
  {Garlea}}, \bibinfo {author} {\bibfnamefont {A.}~\bibnamefont {Zheludev}},
  \bibinfo {author} {\bibfnamefont {L.-P.}\ \bibnamefont {Regnault}}, \bibinfo
  {author} {\bibfnamefont {J.-H.}\ \bibnamefont {Chung}}, \bibinfo {author}
  {\bibfnamefont {Y.}~\bibnamefont {Qiu}}, \bibinfo {author} {\bibfnamefont
  {M.}~\bibnamefont {Boehm}}, \bibinfo {author} {\bibfnamefont
  {K.}~\bibnamefont {Habicht}},\ and\ \bibinfo {author} {\bibfnamefont
  {M.}~\bibnamefont {Meissner}},\ }\bibfield  {title} {\bibinfo {title}
  {Excitations in a {Four}-{Leg} {Antiferromagnetic} {Heisenberg} {Spin}
  {Tube}},\ }\href {https://doi.org/10.1103/PhysRevLett.100.037206} {\bibfield
  {journal} {\bibinfo  {journal} {Physical Review Letters}\ }\textbf {\bibinfo
  {volume} {100}},\ \bibinfo {pages} {037206} (\bibinfo {year}
  {2008})}\BibitemShut {NoStop}%
\bibitem [{\citenamefont {{Zheludev}}\ \emph {et~al.}(2008)\citenamefont
  {{Zheludev}}, \citenamefont {{Garlea}}, \citenamefont {{Regnault}},
  \citenamefont {{Manaka}}, \citenamefont {{Tsvelik}},\ and\ \citenamefont
  {{Chung}}}]{zheludev_universality}%
  \BibitemOpen
  \bibfield  {author} {\bibinfo {author} {\bibfnamefont {A.}~\bibnamefont
  {{Zheludev}}}, \bibinfo {author} {\bibfnamefont {V.~O.}\ \bibnamefont
  {{Garlea}}}, \bibinfo {author} {\bibfnamefont {L.-P.}\ \bibnamefont
  {{Regnault}}}, \bibinfo {author} {\bibfnamefont {H.}~\bibnamefont
  {{Manaka}}}, \bibinfo {author} {\bibfnamefont {A.}~\bibnamefont
  {{Tsvelik}}},\ and\ \bibinfo {author} {\bibfnamefont {J.-H.}\ \bibnamefont
  {{Chung}}},\ }\bibfield  {title} {\bibinfo {title} {{Extended Universal
  Finite-T Renormalization of Excitations in a Class of One-Dimensional Quantum
  Magnets}},\ }\href {https://doi.org/10.1103/PhysRevLett.100.157204}
  {\bibfield  {journal} {\bibinfo  {journal} {Physical Review Letters}\
  }\textbf {\bibinfo {volume} {100}},\ \bibinfo {pages} {157204} (\bibinfo
  {year} {2008})}\BibitemShut {NoStop}%
\bibitem [{\citenamefont {Garlea}\ \emph {et~al.}(2009)\citenamefont {Garlea},
  \citenamefont {Zheludev}, \citenamefont {Habicht}, \citenamefont {Meissner},
  \citenamefont {Grenier}, \citenamefont {Regnault},\ and\ \citenamefont
  {Ressouche}}]{garlea_2009}%
  \BibitemOpen
  \bibfield  {author} {\bibinfo {author} {\bibfnamefont {V.~O.}\ \bibnamefont
  {Garlea}}, \bibinfo {author} {\bibfnamefont {A.}~\bibnamefont {Zheludev}},
  \bibinfo {author} {\bibfnamefont {K.}~\bibnamefont {Habicht}}, \bibinfo
  {author} {\bibfnamefont {M.}~\bibnamefont {Meissner}}, \bibinfo {author}
  {\bibfnamefont {B.}~\bibnamefont {Grenier}}, \bibinfo {author} {\bibfnamefont
  {L.-P.}\ \bibnamefont {Regnault}},\ and\ \bibinfo {author} {\bibfnamefont
  {E.}~\bibnamefont {Ressouche}},\ }\bibfield  {title} {\bibinfo {title}
  {Dimensional crossover in a spin-liquid-to-helimagnet quantum phase
  transition},\ }\href {https://doi.org/10.1103/PhysRevB.79.060404} {\bibfield
  {journal} {\bibinfo  {journal} {Physical Review B}\ }\textbf {\bibinfo
  {volume} {79}},\ \bibinfo {pages} {060404} (\bibinfo {year} {2009})},\
  \bibinfo {note} {publisher: American Physical Society}\BibitemShut {NoStop}%
\bibitem [{\citenamefont {Schrettle}\ \emph {et~al.}(2013)\citenamefont
  {Schrettle}, \citenamefont {Krohns}, \citenamefont {Lunkenheimer},
  \citenamefont {Loidl}, \citenamefont {Wulf}, \citenamefont {Yankova},\ and\
  \citenamefont {Zheludev}}]{schrettle_2013}%
  \BibitemOpen
  \bibfield  {author} {\bibinfo {author} {\bibfnamefont {F.}~\bibnamefont
  {Schrettle}}, \bibinfo {author} {\bibfnamefont {S.}~\bibnamefont {Krohns}},
  \bibinfo {author} {\bibfnamefont {P.}~\bibnamefont {Lunkenheimer}}, \bibinfo
  {author} {\bibfnamefont {A.}~\bibnamefont {Loidl}}, \bibinfo {author}
  {\bibfnamefont {E.}~\bibnamefont {Wulf}}, \bibinfo {author} {\bibfnamefont
  {T.}~\bibnamefont {Yankova}},\ and\ \bibinfo {author} {\bibfnamefont
  {A.}~\bibnamefont {Zheludev}},\ }\bibfield  {title} {\bibinfo {title}
  {Magnetic-field induced multiferroicity in a quantum critical frustrated spin
  liquid},\ }\href {https://doi.org/10.1103/PhysRevB.87.121105} {\bibfield
  {journal} {\bibinfo  {journal} {Physical Review B}\ }\textbf {\bibinfo
  {volume} {87}},\ \bibinfo {pages} {121105} (\bibinfo {year} {2013})},\
  \bibinfo {note} {arXiv:1203.3127 [cond-mat]}\BibitemShut {NoStop}%
\bibitem [{\citenamefont {Glazkov}(2020)}]{glazkov_2020}%
  \BibitemOpen
  \bibfield  {author} {\bibinfo {author} {\bibfnamefont {V.~N.}\ \bibnamefont
  {Glazkov}},\ }\bibfield  {title} {\bibinfo {title} {Magnetic {Resonance} in
  {Collective} {Paramagnets} with {Gapped} {Excitation} {Spectrum}},\ }\href
  {https://doi.org/10.1134/S1063776120070067} {\bibfield  {journal} {\bibinfo
  {journal} {Journal of Experimental and Theoretical Physics}\ }\textbf
  {\bibinfo {volume} {131}},\ \bibinfo {pages} {46} (\bibinfo {year} {2020})},\
  \bibinfo {note} {arXiv: 2007.04885}\BibitemShut {NoStop}%
\bibitem [{\citenamefont {Lecheminant}(2005)}]{lecheminant_revue_1d}%
  \BibitemOpen
  \bibfield  {author} {\bibinfo {author} {\bibfnamefont {P.}~\bibnamefont
  {Lecheminant}},\ }\bibfield  {title} {\bibinfo {title} {One-dimensional
  quantum spin liquids},\ }in\ \href@noop {} {\emph {\bibinfo {booktitle}
  {Frustrated spin systems}}},\ \bibinfo {editor} {edited by\ \bibinfo {editor}
  {\bibfnamefont {H.~T.}\ \bibnamefont {Diep}}}\ (\bibinfo  {publisher} {World
  Scientific},\ \bibinfo {address} {Singapore},\ \bibinfo {year} {2005})\ p.\
  \bibinfo {pages} {307},\ \bibinfo {note} {and references therein}\BibitemShut
  {NoStop}%
\bibitem [{\citenamefont {Haldane}(1983)}]{haldane_gap}%
  \BibitemOpen
  \bibfield  {author} {\bibinfo {author} {\bibfnamefont {F.~D.~M.}\
  \bibnamefont {Haldane}},\ }\href@noop {} {\bibfield  {journal} {\bibinfo
  {journal} {Phys. Rev. Lett.}\ }\textbf {\bibinfo {volume} {50}},\ \bibinfo
  {pages} {1153} (\bibinfo {year} {1983})}\BibitemShut {NoStop}%
\bibitem [{\citenamefont {Affleck}(1988)}]{affleck_houches}%
  \BibitemOpen
  \bibfield  {author} {\bibinfo {author} {\bibfnamefont {I.}~\bibnamefont
  {Affleck}},\ }\bibfield  {title} {\bibinfo {title} {Field theory methods and
  quantum critical phenomena},\ }in\ \href@noop {} {\emph {\bibinfo {booktitle}
  {Fields, Strings and Critical Phenomena}}},\ \bibinfo {editor} {edited by\
  \bibinfo {editor} {\bibfnamefont {E.}~\bibnamefont {Brezin}}\ and\ \bibinfo
  {editor} {\bibfnamefont {J.}~\bibnamefont {Zinn-Justin}}}\ (\bibinfo
  {publisher} {Elsevier Science},\ \bibinfo {address} {Amsterdam},\ \bibinfo
  {year} {1988})\ p.\ \bibinfo {pages} {563}\BibitemShut {NoStop}%
\bibitem [{\citenamefont {Sierra}(1996)}]{sierra_nchains}%
  \BibitemOpen
  \bibfield  {author} {\bibinfo {author} {\bibfnamefont {G.}~\bibnamefont
  {Sierra}},\ }\bibfield  {title} {\bibinfo {title} {The non linear sigma model
  and spin ladders},\ }\href@noop {} {\bibfield  {journal} {\bibinfo  {journal}
  {J. Phys. A}\ }\textbf {\bibinfo {volume} {29}},\ \bibinfo {pages} {3299}
  (\bibinfo {year} {1996})},\ \Eprint {https://arxiv.org/abs/cond-mat/9512007}
  {cond-mat/9512007} \BibitemShut {NoStop}%
\bibitem [{\citenamefont {Cabra}\ and\ \citenamefont
  {Pujol}(2004)}]{cabra_field-theoretical_2004}%
  \BibitemOpen
  \bibfield  {author} {\bibinfo {author} {\bibfnamefont {D.~C.}\ \bibnamefont
  {Cabra}}\ and\ \bibinfo {author} {\bibfnamefont {P.}~\bibnamefont {Pujol}},\
  }\bibfield  {title} {\bibinfo {title} {Field-theoretical methods in quantum
  magnetism},\ }in\ \href {https://doi.org/10.1007/BFb0119596} {\emph {\bibinfo
  {booktitle} {Quantum {Magnetism}}}},\ \bibinfo {series} {Lecture {Notes} in
  {Physics}}, Vol.\ \bibinfo {volume} {645},\ \bibinfo {editor} {edited by\
  \bibinfo {editor} {\bibfnamefont {U.}~\bibnamefont {Schollwöck}}, \bibinfo
  {editor} {\bibfnamefont {J.}~\bibnamefont {Richter}}, \bibinfo {editor}
  {\bibfnamefont {D.~J.~J.}\ \bibnamefont {Farnell}},\ and\ \bibinfo {editor}
  {\bibfnamefont {R.~F.}\ \bibnamefont {Bishop}}}\ (\bibinfo  {publisher}
  {Springer Berlin Heidelberg},\ \bibinfo {address} {Berlin, Heidelberg},\
  \bibinfo {year} {2004})\ pp.\ \bibinfo {pages} {253--305}\BibitemShut
  {NoStop}%
\bibitem [{\citenamefont {Totsuka}\ and\ \citenamefont
  {Suzuki}(1996)}]{totsuka_3legs}%
  \BibitemOpen
  \bibfield  {author} {\bibinfo {author} {\bibfnamefont {K.}~\bibnamefont
  {Totsuka}}\ and\ \bibinfo {author} {\bibfnamefont {M.}~\bibnamefont
  {Suzuki}},\ }\bibfield  {title} {\bibinfo {title} {Phase diagram of three leg
  spin ladder},\ }\href@noop {} {\bibfield  {journal} {\bibinfo  {journal} {J.
  Phys. A}\ }\textbf {\bibinfo {volume} {29}},\ \bibinfo {pages} {3559}
  (\bibinfo {year} {1996})}\BibitemShut {NoStop}%
\bibitem [{\citenamefont {Sakai}\ \emph {et~al.}(2010)\citenamefont {Sakai},
  \citenamefont {Sato}, \citenamefont {Okamoto}, \citenamefont {Okunishi},\
  and\ \citenamefont {Itoi}}]{sakai_2010}%
  \BibitemOpen
  \bibfield  {author} {\bibinfo {author} {\bibfnamefont {T.}~\bibnamefont
  {Sakai}}, \bibinfo {author} {\bibfnamefont {M.}~\bibnamefont {Sato}},
  \bibinfo {author} {\bibfnamefont {K.}~\bibnamefont {Okamoto}}, \bibinfo
  {author} {\bibfnamefont {K.}~\bibnamefont {Okunishi}},\ and\ \bibinfo
  {author} {\bibfnamefont {C.}~\bibnamefont {Itoi}},\ }\bibfield  {title}
  {\bibinfo {title} {Quantum spin nanotubes—frustration, competing orders and
  criticalities},\ }\href {https://doi.org/10.1088/0953-8984/22/40/403201}
  {\bibfield  {journal} {\bibinfo  {journal} {Journal of Physics: Condensed
  Matter}\ }\textbf {\bibinfo {volume} {22}},\ \bibinfo {pages} {403201}
  (\bibinfo {year} {2010})},\ \bibinfo {note} {arXiv:1007.5102
  [cond-mat]}\BibitemShut {NoStop}%
\bibitem [{\citenamefont {Giamarchi}(2004)}]{giamarchi_book_1d}%
  \BibitemOpen
  \bibfield  {author} {\bibinfo {author} {\bibfnamefont {T.}~\bibnamefont
  {Giamarchi}},\ }\href@noop {} {\emph {\bibinfo {title} {Quantum Physics in
  One Dimension}}}\ (\bibinfo  {publisher} {Oxford University Press},\ \bibinfo
  {address} {Oxford},\ \bibinfo {year} {2004})\BibitemShut {NoStop}%
\bibitem [{\citenamefont {Hikihara}\ \emph {et~al.}(2008)\citenamefont
  {Hikihara}, \citenamefont {Kecke}, \citenamefont {Momoi},\ and\ \citenamefont
  {Furusaki}}]{hikihara_2008}%
  \BibitemOpen
  \bibfield  {author} {\bibinfo {author} {\bibfnamefont {T.}~\bibnamefont
  {Hikihara}}, \bibinfo {author} {\bibfnamefont {L.}~\bibnamefont {Kecke}},
  \bibinfo {author} {\bibfnamefont {T.}~\bibnamefont {Momoi}},\ and\ \bibinfo
  {author} {\bibfnamefont {A.}~\bibnamefont {Furusaki}},\ }\bibfield  {title}
  {\bibinfo {title} {Vector chiral and multipolar orders in the spin-1/2
  frustrated ferromagnetic chain in magnetic field},\ }\href
  {https://doi.org/10.1103/PhysRevB.78.144404} {\bibfield  {journal} {\bibinfo
  {journal} {Physical Review B}\ }\textbf {\bibinfo {volume} {78}},\ \bibinfo
  {pages} {144404} (\bibinfo {year} {2008})},\ \bibinfo {note} {arXiv:0807.0858
  [cond-mat]}\BibitemShut {NoStop}%
\bibitem [{\citenamefont {Tsvelik}(1995)}]{tsvelik_field_theory}%
  \BibitemOpen
  \bibfield  {author} {\bibinfo {author} {\bibfnamefont {A.~M.}\ \bibnamefont
  {Tsvelik}},\ }\href@noop {} {\emph {\bibinfo {title} {Quantum Field Theory in
  Condensed Matter Physics}}}\ (\bibinfo  {publisher} {Cambridge University
  Press},\ \bibinfo {address} {Cambridge},\ \bibinfo {year} {1995})\BibitemShut
  {NoStop}%
\bibitem [{\citenamefont {Knizhnik}\ and\ \citenamefont
  {Zamolodchikov}(1984)}]{knizhnik_wz}%
  \BibitemOpen
  \bibfield  {author} {\bibinfo {author} {\bibfnamefont {V.~G.}\ \bibnamefont
  {Knizhnik}}\ and\ \bibinfo {author} {\bibfnamefont {A.~B.}\ \bibnamefont
  {Zamolodchikov}},\ }\href@noop {} {\bibfield  {journal} {\bibinfo  {journal}
  {Nucl. Phys. B}\ }\textbf {\bibinfo {volume} {247}},\ \bibinfo {pages} {83}
  (\bibinfo {year} {1984})}\BibitemShut {NoStop}%
\bibitem [{\citenamefont {Lukyanov}\ and\ \citenamefont
  {Terras}(2003)}]{lukyanov_xxz_asymptotics}%
  \BibitemOpen
  \bibfield  {author} {\bibinfo {author} {\bibfnamefont {S.}~\bibnamefont
  {Lukyanov}}\ and\ \bibinfo {author} {\bibfnamefont {V.}~\bibnamefont
  {Terras}},\ }\bibfield  {title} {\bibinfo {title} {Long-distance asymptotics
  of spin-spin correlation functions for the xxz spin chain},\ }\href@noop {}
  {\bibfield  {journal} {\bibinfo  {journal} {Nucl. Phys. B}\ }\textbf
  {\bibinfo {volume} {654}},\ \bibinfo {pages} {323} (\bibinfo {year}
  {2003})},\ \bibinfo {note} {hep-th/0206093}\BibitemShut {NoStop}%
\bibitem [{\citenamefont {Hikihara}\ and\ \citenamefont
  {Furusaki}(1998)}]{hikihara_xxz}%
  \BibitemOpen
  \bibfield  {author} {\bibinfo {author} {\bibfnamefont {T.}~\bibnamefont
  {Hikihara}}\ and\ \bibinfo {author} {\bibfnamefont {A.}~\bibnamefont
  {Furusaki}},\ }\bibfield  {title} {\bibinfo {title} {Correlation amplitude
  for the $s=1/2$ xxz spin chain in the critical region: Numerical
  renormalization-group study of an open chain},\ }\href@noop {} {\bibfield
  {journal} {\bibinfo  {journal} {Phys. Rev. B}\ }\textbf {\bibinfo {volume}
  {58}},\ \bibinfo {pages} {R853} (\bibinfo {year} {1998})}\BibitemShut
  {NoStop}%
\bibitem [{\citenamefont {Takayoshi}\ and\ \citenamefont
  {Sato}(2010)}]{takayoshi_2010}%
  \BibitemOpen
  \bibfield  {author} {\bibinfo {author} {\bibfnamefont {S.}~\bibnamefont
  {Takayoshi}}\ and\ \bibinfo {author} {\bibfnamefont {M.}~\bibnamefont
  {Sato}},\ }\bibfield  {title} {\bibinfo {title} {Coefficients of bosonized
  dimer operators in spin-1/2 {XXZ} chains and their applications},\ }\href
  {https://doi.org/10.1103/PhysRevB.82.214420} {\bibfield  {journal} {\bibinfo
  {journal} {Physical Review B}\ }\textbf {\bibinfo {volume} {82}},\ \bibinfo
  {pages} {214420} (\bibinfo {year} {2010})}\BibitemShut {NoStop}%
\bibitem [{\citenamefont {Shelton}\ \emph {et~al.}(1996)\citenamefont
  {Shelton}, \citenamefont {Nersesyan},\ and\ \citenamefont
  {Tsvelik}}]{shelton_spin_ladders}%
  \BibitemOpen
  \bibfield  {author} {\bibinfo {author} {\bibfnamefont {D.~G.}\ \bibnamefont
  {Shelton}}, \bibinfo {author} {\bibfnamefont {A.~A.}\ \bibnamefont
  {Nersesyan}},\ and\ \bibinfo {author} {\bibfnamefont {A.~M.}\ \bibnamefont
  {Tsvelik}},\ }\href@noop {} {\bibfield  {journal} {\bibinfo  {journal} {Phys.
  Rev. B}\ }\textbf {\bibinfo {volume} {53}},\ \bibinfo {pages} {8521}
  (\bibinfo {year} {1996})}\BibitemShut {NoStop}%
\bibitem [{\citenamefont {Georges}\ and\ \citenamefont
  {Sengupta}(1995)}]{georges1995}%
  \BibitemOpen
  \bibfield  {author} {\bibinfo {author} {\bibfnamefont {A.}~\bibnamefont
  {Georges}}\ and\ \bibinfo {author} {\bibfnamefont {A.~M.}\ \bibnamefont
  {Sengupta}},\ }\bibfield  {title} {\bibinfo {title} {Solution of the
  two-impurity, two-channel {Kondo} model},\ }\href
  {https://doi.org/10.1103/PhysRevLett.74.2808} {\bibfield  {journal} {\bibinfo
   {journal} {Physical Review Letters}\ }\textbf {\bibinfo {volume} {74}},\
  \bibinfo {pages} {2808} (\bibinfo {year} {1995})}\BibitemShut {NoStop}%
\bibitem [{\citenamefont {Ferrero}\ \emph {et~al.}(2007)\citenamefont
  {Ferrero}, \citenamefont {Leo}, \citenamefont {Lecheminant},\ and\
  \citenamefont {Fabrizio}}]{ferrero_2007}%
  \BibitemOpen
  \bibfield  {author} {\bibinfo {author} {\bibfnamefont {M.}~\bibnamefont
  {Ferrero}}, \bibinfo {author} {\bibfnamefont {L.~D.}\ \bibnamefont {Leo}},
  \bibinfo {author} {\bibfnamefont {P.}~\bibnamefont {Lecheminant}},\ and\
  \bibinfo {author} {\bibfnamefont {M.}~\bibnamefont {Fabrizio}},\ }\bibfield
  {title} {\bibinfo {title} {Strong correlations in a nutshell},\ }\href
  {https://doi.org/10.1088/0953-8984/19/43/433201} {\bibfield  {journal}
  {\bibinfo  {journal} {Journal of Physics: Condensed Matter}\ }\textbf
  {\bibinfo {volume} {19}},\ \bibinfo {pages} {433201} (\bibinfo {year}
  {2007})}\BibitemShut {NoStop}%
\bibitem [{\citenamefont {Konik}\ \emph {et~al.}(2015)\citenamefont {Konik},
  \citenamefont {Palmai}, \citenamefont {Takacs},\ and\ \citenamefont
  {Tsvelik}}]{konik2015}%
  \BibitemOpen
  \bibfield  {author} {\bibinfo {author} {\bibfnamefont {R.~M.}\ \bibnamefont
  {Konik}}, \bibinfo {author} {\bibfnamefont {T.}~\bibnamefont {Palmai}},
  \bibinfo {author} {\bibfnamefont {G.}~\bibnamefont {Takacs}},\ and\ \bibinfo
  {author} {\bibfnamefont {A.~M.}\ \bibnamefont {Tsvelik}},\ }\bibfield
  {title} {\bibinfo {title} {Studying the {Perturbed}
  {Wess}-{Zumino}-{Novikov}-{Witten} {SU}(2)k {Theory} {Using} the {Truncated}
  {Conformal} {Spectrum} {Approach}},\ }\href
  {https://doi.org/10.1016/j.nuclphysb.2015.08.016} {\bibfield  {journal}
  {\bibinfo  {journal} {Nuclear Physics B}\ }\textbf {\bibinfo {volume}
  {899}},\ \bibinfo {pages} {547} (\bibinfo {year} {2015})},\ \bibinfo {note}
  {arXiv: 1505.03860}\BibitemShut {NoStop}%
\bibitem [{\citenamefont {Takhtajan}(1982)}]{takhtajan_spin_s}%
  \BibitemOpen
  \bibfield  {author} {\bibinfo {author} {\bibfnamefont {L.}~\bibnamefont
  {Takhtajan}},\ }\bibfield  {title} {\bibinfo {title} {??},\ }\href@noop {}
  {\bibfield  {journal} {\bibinfo  {journal} {Phys. Lett. A}\ }\textbf
  {\bibinfo {volume} {87}},\ \bibinfo {pages} {479} (\bibinfo {year}
  {1982})}\BibitemShut {NoStop}%
\bibitem [{\citenamefont {Babujian}(1982)}]{babujian_spin_s}%
  \BibitemOpen
  \bibfield  {author} {\bibinfo {author} {\bibfnamefont {J.}~\bibnamefont
  {Babujian}},\ }\bibfield  {title} {\bibinfo {title} {??},\ }\href@noop {}
  {\bibfield  {journal} {\bibinfo  {journal} {Phys. Lett. A}\ }\textbf
  {\bibinfo {volume} {90}},\ \bibinfo {pages} {479} (\bibinfo {year}
  {1982})}\BibitemShut {NoStop}%
\bibitem [{\citenamefont {Tsvelik}(1990)}]{tsvelik_field}%
  \BibitemOpen
  \bibfield  {author} {\bibinfo {author} {\bibfnamefont {A.~M.}\ \bibnamefont
  {Tsvelik}},\ }\href@noop {} {\bibfield  {journal} {\bibinfo  {journal} {Phys.
  Rev. B}\ }\textbf {\bibinfo {volume} {42}},\ \bibinfo {pages} {10499}
  (\bibinfo {year} {1990})}\BibitemShut {NoStop}%
\bibitem [{\citenamefont {Noack}\ and\ \citenamefont
  {Manmana}(2005)}]{noack_2005}%
  \BibitemOpen
  \bibfield  {author} {\bibinfo {author} {\bibfnamefont {R.~M.}\ \bibnamefont
  {Noack}}\ and\ \bibinfo {author} {\bibfnamefont {S.~R.}\ \bibnamefont
  {Manmana}},\ }\bibfield  {title} {\bibinfo {title} {Diagonalization‐ and
  {Numerical} {Renormalization}‐{Group}‐{Based} {Methods} for {Interacting}
  {Quantum} {Systems}},\ }in\ \href@noop {} {\emph {\bibinfo {booktitle} {{AIP}
  {Conf}. {Proc}.}}},\ Vol.\ \bibinfo {volume} {789},\ \bibinfo {editor}
  {edited by\ \bibinfo {editor} {\bibnamefont {{Avella, A.}}}\ and\ \bibinfo
  {editor} {\bibnamefont {{Mancini, F.}}}}\ (\bibinfo  {publisher} {AIP
  Publishing},\ \bibinfo {address} {Melville, NY},\ \bibinfo {year} {2005})\
  p.~\bibinfo {pages} {93}\BibitemShut {NoStop}%
\bibitem [{\citenamefont {Zuber}\ and\ \citenamefont
  {Itzykson}(1977)}]{zuber_77}%
  \BibitemOpen
  \bibfield  {author} {\bibinfo {author} {\bibfnamefont {J.~B.}\ \bibnamefont
  {Zuber}}\ and\ \bibinfo {author} {\bibfnamefont {C.}~\bibnamefont
  {Itzykson}},\ }\href@noop {} {\bibfield  {journal} {\bibinfo  {journal}
  {Phys. Rev. D}\ }\textbf {\bibinfo {volume} {15}},\ \bibinfo {pages} {2875}
  (\bibinfo {year} {1977})}\BibitemShut {NoStop}%
\bibitem [{\citenamefont {Schroer}\ and\ \citenamefont
  {Truong}(1978)}]{schroer_ising}%
  \BibitemOpen
  \bibfield  {author} {\bibinfo {author} {\bibfnamefont {B.}~\bibnamefont
  {Schroer}}\ and\ \bibinfo {author} {\bibfnamefont {T.~T.}\ \bibnamefont
  {Truong}},\ }\bibfield  {title} {\bibinfo {title} {The order/disorder quantum
  field operators associated with the two-dimewnsional ising model in the
  continuum limit},\ }\href@noop {} {\bibfield  {journal} {\bibinfo  {journal}
  {Nucl. Phys. B}\ }\textbf {\bibinfo {volume} {144}},\ \bibinfo {pages} {80}
  (\bibinfo {year} {1978})}\BibitemShut {NoStop}%
\bibitem [{\citenamefont {Ogilvie}(1981)}]{ogilvie_ising}%
  \BibitemOpen
  \bibfield  {author} {\bibinfo {author} {\bibfnamefont {M.}~\bibnamefont
  {Ogilvie}},\ }\href@noop {} {\bibfield  {journal} {\bibinfo  {journal} {Ann.
  Phys. (N. Y.)}\ }\textbf {\bibinfo {volume} {136}},\ \bibinfo {pages} {273}
  (\bibinfo {year} {1981})}\BibitemShut {NoStop}%
\bibitem [{\citenamefont {Boyanovsky}(1989)}]{boyanovsky_ising}%
  \BibitemOpen
  \bibfield  {author} {\bibinfo {author} {\bibfnamefont {D.}~\bibnamefont
  {Boyanovsky}},\ }\href@noop {} {\bibfield  {journal} {\bibinfo  {journal}
  {Phys. Rev. B}\ }\textbf {\bibinfo {volume} {39}},\ \bibinfo {pages} {6744}
  (\bibinfo {year} {1989})}\BibitemShut {NoStop}%
\bibitem [{\citenamefont {Kiritsis}(1988)}]{kiritsis1988}%
  \BibitemOpen
  \bibfield  {author} {\bibinfo {author} {\bibfnamefont {E.~B.}\ \bibnamefont
  {Kiritsis}},\ }\bibfield  {title} {\bibinfo {title} {The c=2/3 minimal n=1
  superconformal system and its realisation in the critical o(2) gaussian
  model},\ }\href {https://doi.org/10.1088/0305-4470/21/2/011} {\bibfield
  {journal} {\bibinfo  {journal} {Journal of Physics A: Mathematical and
  General}\ }\textbf {\bibinfo {volume} {21}},\ \bibinfo {pages} {297}
  (\bibinfo {year} {1988})}\BibitemShut {NoStop}%
\bibitem [{\citenamefont {Assaraf}\ \emph {et~al.}(1999)\citenamefont
  {Assaraf}, \citenamefont {Azaria}, \citenamefont {Caffarel},\ and\
  \citenamefont {Lecheminant}}]{assaraf_su(n)}%
  \BibitemOpen
  \bibfield  {author} {\bibinfo {author} {\bibfnamefont {R.}~\bibnamefont
  {Assaraf}}, \bibinfo {author} {\bibfnamefont {P.}~\bibnamefont {Azaria}},
  \bibinfo {author} {\bibfnamefont {M.}~\bibnamefont {Caffarel}},\ and\
  \bibinfo {author} {\bibfnamefont {P.}~\bibnamefont {Lecheminant}},\
  }\bibfield  {title} {\bibinfo {title} {Metal-insulator transition in the
  one-dimensional su(n) hubbard model},\ }\href@noop {} {\bibfield  {journal}
  {\bibinfo  {journal} {Phys. Rev. B}\ }\textbf {\bibinfo {volume} {60}},\
  \bibinfo {pages} {2299} (\bibinfo {year} {1999})},\ \bibinfo {note}
  {cond-mat/9903057}\BibitemShut {NoStop}%
\bibitem [{\citenamefont {Haldane}(1981)}]{haldane_bosonisation}%
  \BibitemOpen
  \bibfield  {author} {\bibinfo {author} {\bibfnamefont {F.~D.~M.}\
  \bibnamefont {Haldane}},\ }\href@noop {} {\bibfield  {journal} {\bibinfo
  {journal} {J. Phys. C}\ }\textbf {\bibinfo {volume} {14}},\ \bibinfo {pages}
  {2585} (\bibinfo {year} {1981})}\BibitemShut {NoStop}%
\bibitem [{\citenamefont {Savary}\ and\ \citenamefont
  {Senthil}(2015)}]{savary_2015}%
  \BibitemOpen
  \bibfield  {author} {\bibinfo {author} {\bibfnamefont {L.}~\bibnamefont
  {Savary}}\ and\ \bibinfo {author} {\bibfnamefont {T.}~\bibnamefont
  {Senthil}},\ }\href {https://doi.org/10.48550/arXiv.1506.04752} {\bibinfo
  {title} {Probing {Hidden} {Orders} with {Resonant} {Inelastic} {X}-{Ray}
  {Scattering}}} (\bibinfo {year} {2015}),\ \bibinfo {note} {arXiv:1506.04752
  [cond-mat]}\BibitemShut {NoStop}%
\bibitem [{\citenamefont {Lecheminant}\ and\ \citenamefont
  {Nonne}(2012)}]{lecheminant_2012}%
  \BibitemOpen
  \bibfield  {author} {\bibinfo {author} {\bibfnamefont {P.}~\bibnamefont
  {Lecheminant}}\ and\ \bibinfo {author} {\bibfnamefont {H.}~\bibnamefont
  {Nonne}},\ }\bibfield  {title} {\bibinfo {title} {Exotic quantum criticality
  in one-dimensional coupled dipolar bosons tubes},\ }\href
  {https://doi.org/10.1103/PhysRevB.85.195121} {\bibfield  {journal} {\bibinfo
  {journal} {Physical Review B}\ }\textbf {\bibinfo {volume} {85}},\ \bibinfo
  {pages} {195121} (\bibinfo {year} {2012})}\BibitemShut {NoStop}%
\bibitem [{\citenamefont {Capponi}\ \emph {et~al.}(2013)\citenamefont
  {Capponi}, \citenamefont {Lecheminant},\ and\ \citenamefont
  {Moliner}}]{capponi_2013}%
  \BibitemOpen
  \bibfield  {author} {\bibinfo {author} {\bibfnamefont {S.}~\bibnamefont
  {Capponi}}, \bibinfo {author} {\bibfnamefont {P.}~\bibnamefont
  {Lecheminant}},\ and\ \bibinfo {author} {\bibfnamefont {M.}~\bibnamefont
  {Moliner}},\ }\bibfield  {title} {\bibinfo {title} {Quantum phase transitions
  in multileg spin ladders with ring exchange},\ }\href
  {https://doi.org/10.1103/PhysRevB.88.075132} {\bibfield  {journal} {\bibinfo
  {journal} {Physical Review B}\ }\textbf {\bibinfo {volume} {88}},\ \bibinfo
  {pages} {075132} (\bibinfo {year} {2013})}\BibitemShut {NoStop}%
\bibitem [{\citenamefont {Luther}\ and\ \citenamefont
  {Peschel}(1975)}]{luther_chaine_xxz}%
  \BibitemOpen
  \bibfield  {author} {\bibinfo {author} {\bibfnamefont {A.}~\bibnamefont
  {Luther}}\ and\ \bibinfo {author} {\bibfnamefont {I.}~\bibnamefont
  {Peschel}},\ }\href@noop {} {\bibfield  {journal} {\bibinfo  {journal} {Phys.
  Rev. B}\ }\textbf {\bibinfo {volume} {12}},\ \bibinfo {pages} {3908}
  (\bibinfo {year} {1975})}\BibitemShut {NoStop}%
\bibitem [{\citenamefont {Haldane}(1980)}]{haldane_xxzchain}%
  \BibitemOpen
  \bibfield  {author} {\bibinfo {author} {\bibfnamefont {F.~D.~M.}\
  \bibnamefont {Haldane}},\ }\href@noop {} {\bibfield  {journal} {\bibinfo
  {journal} {Phys. Rev. Lett.}\ }\textbf {\bibinfo {volume} {45}},\ \bibinfo
  {pages} {1358} (\bibinfo {year} {1980})}\BibitemShut {NoStop}%
\bibitem [{\citenamefont {den Nijs}(1981)}]{nijs_equivalence}%
  \BibitemOpen
  \bibfield  {author} {\bibinfo {author} {\bibfnamefont {M.~P.~M.}\
  \bibnamefont {den Nijs}},\ }\href@noop {} {\bibfield  {journal} {\bibinfo
  {journal} {Phys. Rev. B}\ }\textbf {\bibinfo {volume} {23}},\ \bibinfo
  {pages} {6111} (\bibinfo {year} {1981})}\BibitemShut {NoStop}%
\bibitem [{\citenamefont {Orignac}(2004)}]{orignac04_spingap}%
  \BibitemOpen
  \bibfield  {author} {\bibinfo {author} {\bibfnamefont {E.}~\bibnamefont
  {Orignac}},\ }\bibfield  {title} {\bibinfo {title} {Quantitative expression
  of the spin gap via bosonization for a dimerized spin-1/2 chain},\
  }\href@noop {} {\bibfield  {journal} {\bibinfo  {journal} {Eur. Phys. J. B}\
  }\textbf {\bibinfo {volume} {39}},\ \bibinfo {pages} {335} (\bibinfo {year}
  {2004})},\ \Eprint {https://arxiv.org/abs/cond-mat/0403175}
  {cond-mat/0403175} \BibitemShut {NoStop}%
\bibitem [{\citenamefont {Schulz}(1986)}]{schulz_spins}%
  \BibitemOpen
  \bibfield  {author} {\bibinfo {author} {\bibfnamefont {H.~J.}\ \bibnamefont
  {Schulz}},\ }\href@noop {} {\bibfield  {journal} {\bibinfo  {journal} {Phys.
  Rev. B}\ }\textbf {\bibinfo {volume} {34}},\ \bibinfo {pages} {6372}
  (\bibinfo {year} {1986})}\BibitemShut {NoStop}%
\bibitem [{\citenamefont {Strong}\ and\ \citenamefont
  {Millis}(1992)}]{strong_spinchains}%
  \BibitemOpen
  \bibfield  {author} {\bibinfo {author} {\bibfnamefont {S.~P.}\ \bibnamefont
  {Strong}}\ and\ \bibinfo {author} {\bibfnamefont {A.~J.}\ \bibnamefont
  {Millis}},\ }\href@noop {} {\bibfield  {journal} {\bibinfo  {journal} {Phys.
  Rev. Lett.}\ }\textbf {\bibinfo {volume} {69}},\ \bibinfo {pages} {2419}
  (\bibinfo {year} {1992})}\BibitemShut {NoStop}%
\bibitem [{\citenamefont {Nersesyan}\ and\ \citenamefont
  {Tsvelik}(1997{\natexlab{b}})}]{nersesyan_biquadratic}%
  \BibitemOpen
  \bibfield  {author} {\bibinfo {author} {\bibfnamefont {A.}~\bibnamefont
  {Nersesyan}}\ and\ \bibinfo {author} {\bibfnamefont {A.~M.}\ \bibnamefont
  {Tsvelik}},\ }\bibfield  {title} {\bibinfo {title} {One-dimensional
  spin-liquid without magnon excitations},\ }\href@noop {} {\bibfield
  {journal} {\bibinfo  {journal} {Phys. Rev. Lett.}\ }\textbf {\bibinfo
  {volume} {78}},\ \bibinfo {pages} {3939} (\bibinfo {year}
  {1997}{\natexlab{b}})},\ \bibinfo {note} {ibid. , \textbf{79}, E
  1171}\BibitemShut {NoStop}%
\bibitem [{\citenamefont {Fabrizio}\ \emph {et~al.}(2000)\citenamefont
  {Fabrizio}, \citenamefont {Gogolin},\ and\ \citenamefont
  {Nersesyan}}]{fabrizio_dsg}%
  \BibitemOpen
  \bibfield  {author} {\bibinfo {author} {\bibfnamefont {M.}~\bibnamefont
  {Fabrizio}}, \bibinfo {author} {\bibfnamefont {A.~O.}\ \bibnamefont
  {Gogolin}},\ and\ \bibinfo {author} {\bibfnamefont {A.~A.}\ \bibnamefont
  {Nersesyan}},\ }\bibfield  {title} {\bibinfo {title} {Critical properties of
  the double-frequency sine-gordon model with applications},\ }\href@noop {}
  {\bibfield  {journal} {\bibinfo  {journal} {Nucl. Phys. B}\ }\textbf
  {\bibinfo {volume} {580}},\ \bibinfo {pages} {647} (\bibinfo {year}
  {2000})}\BibitemShut {NoStop}%
\bibitem [{\citenamefont {Nersesyan}(2001)}]{nersesyan01_ising}%
  \BibitemOpen
  \bibfield  {author} {\bibinfo {author} {\bibfnamefont {A.~A.}\ \bibnamefont
  {Nersesyan}},\ }\bibfield  {title} {\bibinfo {title} {Ising-model description
  of quantum critical points in 1d electron and spin systems},\ }in\ \href@noop
  {} {\emph {\bibinfo {booktitle} {New theoretical approaches to Strongly
  Correlated Systems}}},\ \bibinfo {series} {Nato Science Series II},
  Vol.~\bibinfo {volume} {23},\ \bibinfo {editor} {edited by\ \bibinfo {editor}
  {\bibfnamefont {A.~M.}\ \bibnamefont {Tsvelik}}}\ (\bibinfo  {publisher}
  {Kluwer},\ \bibinfo {address} {Dordrecht, Netherlands},\ \bibinfo {year}
  {2001})\ Chap.~\bibinfo {chapter} {4}, p.~\bibinfo {pages} {89}\BibitemShut
  {NoStop}%
\bibitem [{\citenamefont {Citro}\ \emph
  {et~al.}(2000{\natexlab{a}})\citenamefont {Citro}, \citenamefont {Orignac},
  \citenamefont {Andrei}, \citenamefont {Itoi},\ and\ \citenamefont
  {Qin}}]{citro_su3}%
  \BibitemOpen
  \bibfield  {author} {\bibinfo {author} {\bibfnamefont {R.}~\bibnamefont
  {Citro}}, \bibinfo {author} {\bibfnamefont {E.}~\bibnamefont {Orignac}},
  \bibinfo {author} {\bibfnamefont {N.}~\bibnamefont {Andrei}}, \bibinfo
  {author} {\bibfnamefont {C.}~\bibnamefont {Itoi}},\ and\ \bibinfo {author}
  {\bibfnamefont {S.}~\bibnamefont {Qin}},\ }\href@noop {} {\bibinfo {title}
  {Critical behaviour of a spin-tube model in a magnetic field}} (\bibinfo
  {year} {2000}{\natexlab{a}}),\ \bibinfo {note} {cond-mat/9904371}\BibitemShut
  {NoStop}%
\bibitem [{\citenamefont {Itzykson}\ and\ \citenamefont
  {Zuber}(1980)}]{itzykson-zuber}%
  \BibitemOpen
  \bibfield  {author} {\bibinfo {author} {\bibfnamefont {C.}~\bibnamefont
  {Itzykson}}\ and\ \bibinfo {author} {\bibfnamefont {J.~B.}\ \bibnamefont
  {Zuber}},\ }\href@noop {} {\emph {\bibinfo {title} {Quantum Field Theory}}}\
  (\bibinfo  {publisher} {Mc Graw Hill},\ \bibinfo {address} {New-York},\
  \bibinfo {year} {1980})\BibitemShut {NoStop}%
\bibitem [{\citenamefont {Slansky}(1981)}]{slansky_1981}%
  \BibitemOpen
  \bibfield  {author} {\bibinfo {author} {\bibfnamefont {R.}~\bibnamefont
  {Slansky}},\ }\bibfield  {title} {\bibinfo {title} {Group theory for unified
  model building},\ }\href {https://doi.org/10.1016/0370-1573(81)90092-2}
  {\bibfield  {journal} {\bibinfo  {journal} {Physics Reports}\ }\textbf
  {\bibinfo {volume} {79}},\ \bibinfo {pages} {1} (\bibinfo {year}
  {1981})}\BibitemShut {NoStop}%
\bibitem [{\citenamefont {Citro}\ \emph
  {et~al.}(2000{\natexlab{b}})\citenamefont {Citro}, \citenamefont {Orignac},
  \citenamefont {Andrei}, \citenamefont {Itoi},\ and\ \citenamefont
  {Qin}}]{citro00_spintube}%
  \BibitemOpen
  \bibfield  {author} {\bibinfo {author} {\bibfnamefont {R.}~\bibnamefont
  {Citro}}, \bibinfo {author} {\bibfnamefont {E.}~\bibnamefont {Orignac}},
  \bibinfo {author} {\bibfnamefont {N.}~\bibnamefont {Andrei}}, \bibinfo
  {author} {\bibfnamefont {C.}~\bibnamefont {Itoi}},\ and\ \bibinfo {author}
  {\bibfnamefont {S.}~\bibnamefont {Qin}},\ }\bibfield  {title} {\bibinfo
  {title} {Critical behaviour of a spin-tube model in a magnetic field},\
  }\href@noop {} {\bibfield  {journal} {\bibinfo  {journal} {J. Phys.: Condens.
  Matter}\ }\textbf {\bibinfo {volume} {12}},\ \bibinfo {pages} {3041}
  (\bibinfo {year} {2000}{\natexlab{b}})},\ \Eprint
  {https://arxiv.org/abs/cond-mat/9904371} {cond-mat/9904371} \BibitemShut
  {NoStop}%
\bibitem [{\citenamefont {Golinelli}\ \emph {et~al.}(1998)\citenamefont
  {Golinelli}, \citenamefont {Jolicoeur},\ and\ \citenamefont
  {Sorensen}}]{golinelli_incommensurate}%
  \BibitemOpen
  \bibfield  {author} {\bibinfo {author} {\bibfnamefont {O.}~\bibnamefont
  {Golinelli}}, \bibinfo {author} {\bibfnamefont {T.}~\bibnamefont
  {Jolicoeur}},\ and\ \bibinfo {author} {\bibfnamefont {E.}~\bibnamefont
  {Sorensen}},\ }\bibfield  {title} {\bibinfo {title} {Incommensurability in
  the magnetic excitations of the bilinear-biquadratic spin-1 chain},\
  }\href@noop {} {\bibfield  {journal} {\bibinfo  {journal} {Eur. Phys. J. B}\
  }\textbf {\bibinfo {volume} {11}},\ \bibinfo {pages} {199} (\bibinfo {year}
  {1998})}\BibitemShut {NoStop}%
\bibitem [{\citenamefont {Uimin}(1970)}]{uimin}%
  \BibitemOpen
  \bibfield  {author} {\bibinfo {author} {\bibfnamefont {G.}~\bibnamefont
  {Uimin}},\ }\href@noop {} {\bibfield  {journal} {\bibinfo  {journal} {JETP
  Lett.}\ }\textbf {\bibinfo {volume} {12}},\ \bibinfo {pages} {225} (\bibinfo
  {year} {1970})}\BibitemShut {NoStop}%
\bibitem [{\citenamefont {Lai}(1974)}]{lai}%
  \BibitemOpen
  \bibfield  {author} {\bibinfo {author} {\bibfnamefont {C.}~\bibnamefont
  {Lai}},\ }\href@noop {} {\bibfield  {journal} {\bibinfo  {journal} {J. Math.
  Phys.}\ }\textbf {\bibinfo {volume} {15}},\ \bibinfo {pages} {1675} (\bibinfo
  {year} {1974})}\BibitemShut {NoStop}%
\bibitem [{\citenamefont {Sutherland}(1975)}]{sutherland}%
  \BibitemOpen
  \bibfield  {author} {\bibinfo {author} {\bibfnamefont {B.}~\bibnamefont
  {Sutherland}},\ }\href@noop {} {\bibfield  {journal} {\bibinfo  {journal}
  {Phys. Rev. B}\ }\textbf {\bibinfo {volume} {12}},\ \bibinfo {pages} {3795}
  (\bibinfo {year} {1975})}\BibitemShut {NoStop}%
\bibitem [{\citenamefont {Zamolodchikov}\ and\ \citenamefont
  {Fateev}(1985)}]{zamolodchikov_1985}%
  \BibitemOpen
  \bibfield  {author} {\bibinfo {author} {\bibfnamefont {A.~B.}\ \bibnamefont
  {Zamolodchikov}}\ and\ \bibinfo {author} {\bibfnamefont {V.~A.}\ \bibnamefont
  {Fateev}},\ }\bibfield  {title} {\bibinfo {title} {Nonlocal (parafermion)
  currents in two-dimensional conformal quantum field theory and self-dual
  critical points in $z_n$-symmetric statistical systems},\ }\href@noop {}
  {\bibfield  {journal} {\bibinfo  {journal} {Sov. Phys. JETP}\ }\textbf
  {\bibinfo {volume} {62}},\ \bibinfo {pages} {215} (\bibinfo {year}
  {1985})}\BibitemShut {NoStop}%
\bibitem [{\citenamefont {Jos{\'e}}\ \emph {et~al.}(1977)\citenamefont
  {Jos{\'e}}, \citenamefont {Kadanoff}, \citenamefont {Kirkpatrick},\ and\
  \citenamefont {Nelson}}]{jose_planar_2d}%
  \BibitemOpen
  \bibfield  {author} {\bibinfo {author} {\bibfnamefont {J.~V.}\ \bibnamefont
  {Jos{\'e}}}, \bibinfo {author} {\bibfnamefont {L.~P.}\ \bibnamefont
  {Kadanoff}}, \bibinfo {author} {\bibfnamefont {S.}~\bibnamefont
  {Kirkpatrick}},\ and\ \bibinfo {author} {\bibfnamefont {D.~R.}\ \bibnamefont
  {Nelson}},\ }\href@noop {} {\bibfield  {journal} {\bibinfo  {journal} {Phys.
  Rev. B}\ }\textbf {\bibinfo {volume} {16}},\ \bibinfo {pages} {1217}
  (\bibinfo {year} {1977})}\BibitemShut {NoStop}%
\bibitem [{\citenamefont {Kaeding}(1995)}]{kaeding_su3_1995}%
  \BibitemOpen
  \bibfield  {author} {\bibinfo {author} {\bibfnamefont {T.~A.}\ \bibnamefont
  {Kaeding}},\ }\bibfield  {title} {\bibinfo {title} {Tables of {SU}(3)
  {Isoscalar} {Factors}},\ }\href {https://doi.org/10.1006/adnd.1995.1011}
  {\bibfield  {journal} {\bibinfo  {journal} {Atomic Data and Nuclear Data
  Tables}\ }\textbf {\bibinfo {volume} {61}},\ \bibinfo {pages} {233} (\bibinfo
  {year} {1995})},\ \bibinfo {note} {nucl-th/9502037}\BibitemShut {NoStop}%
\bibitem [{\citenamefont {Rajaraman}(1982)}]{rajaraman_instanton}%
  \BibitemOpen
  \bibfield  {author} {\bibinfo {author} {\bibfnamefont {R.}~\bibnamefont
  {Rajaraman}},\ }\href@noop {} {\emph {\bibinfo {title} {Solitons and
  Instantons: An Introduction to solitons and Instantons in Quantum Field
  Theory}}}\ (\bibinfo  {publisher} {North Holland},\ \bibinfo {address}
  {Amsterdam},\ \bibinfo {year} {1982})\BibitemShut {NoStop}%
\bibitem [{\citenamefont {Dashen}\ \emph {et~al.}(1975)\citenamefont {Dashen},
  \citenamefont {Hasslacher},\ and\ \citenamefont {Neveu}}]{dashen_sinegordon}%
  \BibitemOpen
  \bibfield  {author} {\bibinfo {author} {\bibfnamefont {R.~F.}\ \bibnamefont
  {Dashen}}, \bibinfo {author} {\bibfnamefont {B.}~\bibnamefont {Hasslacher}},\
  and\ \bibinfo {author} {\bibfnamefont {A.}~\bibnamefont {Neveu}},\ }\bibfield
   {title} {\bibinfo {title} {Particle spectrum in model field theories from
  semiclassical functional integral techniques},\ }\href
  {https://doi.org/10.1103/PhysRevD.11.3424} {\bibfield  {journal} {\bibinfo
  {journal} {Phys. Rev. D}\ }\textbf {\bibinfo {volume} {11}},\ \bibinfo
  {pages} {3424} (\bibinfo {year} {1975})}\BibitemShut {NoStop}%
\bibitem [{\citenamefont {Babujian}\ \emph {et~al.}(1999)\citenamefont
  {Babujian}, \citenamefont {Fring}, \citenamefont {Karowski},\ and\
  \citenamefont {Zapletal}}]{babujian99_ff_sg_1}%
  \BibitemOpen
  \bibfield  {author} {\bibinfo {author} {\bibfnamefont {H.}~\bibnamefont
  {Babujian}}, \bibinfo {author} {\bibfnamefont {A.}~\bibnamefont {Fring}},
  \bibinfo {author} {\bibfnamefont {M.}~\bibnamefont {Karowski}},\ and\
  \bibinfo {author} {\bibfnamefont {A.}~\bibnamefont {Zapletal}},\ }\bibfield
  {title} {\bibinfo {title} {Exact form factors in integrable quantum field
  theories: the sine-gordon model},\ }\href@noop {} {\bibfield  {journal}
  {\bibinfo  {journal} {Nucl. Phys. B}\ }\textbf {\bibinfo {volume} {538}},\
  \bibinfo {pages} {535} (\bibinfo {year} {1999})},\ \bibinfo {note}
  {hep-th/9805185}\BibitemShut {NoStop}%
\bibitem [{\citenamefont {Barber}\ and\ \citenamefont
  {Batchelor}(1989)}]{barber_1989}%
  \BibitemOpen
  \bibfield  {author} {\bibinfo {author} {\bibfnamefont {M.~N.}\ \bibnamefont
  {Barber}}\ and\ \bibinfo {author} {\bibfnamefont {M.~T.}\ \bibnamefont
  {Batchelor}},\ }\bibfield  {title} {\bibinfo {title} {Spectrum of the
  biquadratic spin-1 antiferromagnetic chain},\ }\href
  {https://doi.org/10.1103/PhysRevB.40.4621} {\bibfield  {journal} {\bibinfo
  {journal} {Physical Review B}\ }\textbf {\bibinfo {volume} {40}},\ \bibinfo
  {pages} {4621} (\bibinfo {year} {1989})}\BibitemShut {NoStop}%
\bibitem [{\citenamefont {Klumper}(1990)}]{klumper_1990}%
  \BibitemOpen
  \bibfield  {author} {\bibinfo {author} {\bibfnamefont {A.}~\bibnamefont
  {Klumper}},\ }\bibfield  {title} {\bibinfo {title} {The spectra of q-state
  vertex models and related antiferromagnetic quantum spin chains},\ }\href
  {https://doi.org/10.1088/0305-4470/23/5/023} {\bibfield  {journal} {\bibinfo
  {journal} {Journal of Physics A: Mathematical and General}\ }\textbf
  {\bibinfo {volume} {23}},\ \bibinfo {pages} {809} (\bibinfo {year}
  {1990})}\BibitemShut {NoStop}%
\bibitem [{\citenamefont {Sandvik}(2010)}]{sandvik_computational_2010}%
  \BibitemOpen
  \bibfield  {author} {\bibinfo {author} {\bibfnamefont {A.~W.}\ \bibnamefont
  {Sandvik}},\ }\bibfield  {title} {\bibinfo {title} {Computational {Studies}
  of {Quantum} {Spin} {Systems}},\ }in\ \href
  {https://doi.org/10.1063/1.3518900} {\emph {\bibinfo {booktitle} {{Lectures}
  {on} {the} {Physics} {of} {Strongly} {Correlated} {Systems}}}},\ \bibinfo
  {series} {{AIP} {Conference} {Proceedings}}, Vol.\ \bibinfo {volume} {1297},\
  \bibinfo {editor} {edited by\ \bibinfo {editor} {\bibnamefont {{Adolfo
  Avella}}}\ and\ \bibinfo {editor} {\bibnamefont {{Ferdinando Mancini}}}}\
  (\bibinfo  {publisher} {AIP Publishing},\ \bibinfo {address} {Vietri sul
  Mare, (Italy)},\ \bibinfo {year} {2010})\ pp.\ \bibinfo {pages} {135--338},\
  \bibinfo {note} {arXiv:1101.3281 [cond-mat, physics:hep-lat]}\BibitemShut
  {NoStop}%
\bibitem [{\citenamefont {White}(1992)}]{white_dmrg_letter}%
  \BibitemOpen
  \bibfield  {author} {\bibinfo {author} {\bibfnamefont {S.~R.}\ \bibnamefont
  {White}},\ }\href@noop {} {\bibfield  {journal} {\bibinfo  {journal} {Phys.
  Rev. Lett.}\ }\textbf {\bibinfo {volume} {69}},\ \bibinfo {pages} {2863}
  (\bibinfo {year} {1992})}\BibitemShut {NoStop}%
\bibitem [{\citenamefont {White}(1993)}]{white_dmrg}%
  \BibitemOpen
  \bibfield  {author} {\bibinfo {author} {\bibfnamefont {S.~R.}\ \bibnamefont
  {White}},\ }\href@noop {} {\bibfield  {journal} {\bibinfo  {journal} {Phys.
  Rev. B}\ }\textbf {\bibinfo {volume} {48}},\ \bibinfo {pages} {10345}
  (\bibinfo {year} {1993})}\BibitemShut {NoStop}%
\bibitem [{\citenamefont {Schollwock}(2005)}]{schollwock_2005}%
  \BibitemOpen
  \bibfield  {author} {\bibinfo {author} {\bibfnamefont {U.}~\bibnamefont
  {Schollwock}},\ }\bibfield  {title} {\bibinfo {title} {The density-matrix
  renormalization group},\ }\href@noop {} {\bibfield  {journal} {\bibinfo
  {journal} {Rev. Mod. Phys.}\ }\textbf {\bibinfo {volume} {77}},\ \bibinfo
  {pages} {259} (\bibinfo {year} {2005})}\BibitemShut {NoStop}%
\bibitem [{\citenamefont {Hallberg}(2006)}]{hallberg_2006}%
  \BibitemOpen
  \bibfield  {author} {\bibinfo {author} {\bibfnamefont {K.~A.}\ \bibnamefont
  {Hallberg}},\ }\bibfield  {title} {\bibinfo {title} {New trends in density
  matrix renormalization},\ }\href@noop {} {\bibfield  {journal} {\bibinfo
  {journal} {Adv. Phys.}\ }\textbf {\bibinfo {volume} {55}},\ \bibinfo {pages}
  {477} (\bibinfo {year} {2006})}\BibitemShut {NoStop}%
\bibitem [{\citenamefont {Assaad}\ and\ \citenamefont
  {Evertz}(2008)}]{fehske_2008}%
  \BibitemOpen
  \bibfield  {author} {\bibinfo {author} {\bibfnamefont {F.}~\bibnamefont
  {Assaad}}\ and\ \bibinfo {author} {\bibfnamefont {H.}~\bibnamefont
  {Evertz}},\ }\bibfield  {title} {\bibinfo {title} {World-line and
  {Determinantal} {Quantum} {Monte} {Carlo} {Methods} for {Spins}, {Phonons}
  and {Electrons}},\ }in\ \href {https://doi.org/10.1007/978-3-540-74686-7_10}
  {\emph {\bibinfo {booktitle} {Computational {Many}-{Particle} {Physics}}}},\
  \bibinfo {series} {Lecture {Notes} in {Physics}}, Vol.\ \bibinfo {volume}
  {739},\ \bibinfo {editor} {edited by\ \bibinfo {editor} {\bibfnamefont
  {H.}~\bibnamefont {Fehske}}, \bibinfo {editor} {\bibfnamefont
  {R.}~\bibnamefont {Schneider}},\ and\ \bibinfo {editor} {\bibfnamefont
  {A.}~\bibnamefont {Weiße}}}\ (\bibinfo  {publisher} {Springer Berlin
  Heidelberg},\ \bibinfo {address} {Berlin, Heidelberg},\ \bibinfo {year}
  {2008})\ p.\ \bibinfo {pages} {277}\BibitemShut {NoStop}%
\bibitem [{\citenamefont {James}\ \emph {et~al.}(2018)\citenamefont {James},
  \citenamefont {Konik}, \citenamefont {Lecheminant}, \citenamefont
  {Robinson},\ and\ \citenamefont {Tsvelik}}]{james_2018}%
  \BibitemOpen
  \bibfield  {author} {\bibinfo {author} {\bibfnamefont {A.~J.~A.}\
  \bibnamefont {James}}, \bibinfo {author} {\bibfnamefont {R.~M.}\ \bibnamefont
  {Konik}}, \bibinfo {author} {\bibfnamefont {P.}~\bibnamefont {Lecheminant}},
  \bibinfo {author} {\bibfnamefont {N.~J.}\ \bibnamefont {Robinson}},\ and\
  \bibinfo {author} {\bibfnamefont {A.~M.}\ \bibnamefont {Tsvelik}},\
  }\bibfield  {title} {\bibinfo {title} {Non-perturbative methodologies for
  low-dimensional strongly-correlated systems: {From} non-abelian bosonization
  to truncated spectrum methods},\ }\href
  {https://doi.org/10.1088/1361-6633/aa91ea} {\bibfield  {journal} {\bibinfo
  {journal} {Reports on Progress in Physics}\ }\textbf {\bibinfo {volume}
  {81}},\ \bibinfo {pages} {046002} (\bibinfo {year} {2018})},\ \bibinfo {note}
  {arXiv:1703.08421 [cond-mat]}\BibitemShut {NoStop}%
\bibitem [{\citenamefont {Lecheminant}\ and\ \citenamefont
  {Orignac}(2002)}]{lecheminant02_magnet}%
  \BibitemOpen
  \bibfield  {author} {\bibinfo {author} {\bibfnamefont {P.}~\bibnamefont
  {Lecheminant}}\ and\ \bibinfo {author} {\bibfnamefont {E.}~\bibnamefont
  {Orignac}},\ }\bibfield  {title} {\bibinfo {title} {Magnetization and
  dimerization profiles of the cut two-leg spin ladder and spin-1 chain},\
  }\href@noop {} {\bibfield  {journal} {\bibinfo  {journal} {Phys. Rev. B}\
  }\textbf {\bibinfo {volume} {65}},\ \bibinfo {pages} {174406} (\bibinfo
  {year} {2002})},\ \Eprint {https://arxiv.org/abs/cond-mat/0111177}
  {cond-mat/0111177} \BibitemShut {NoStop}%
\bibitem [{\citenamefont {Ng}(1994)}]{ng_edges}%
  \BibitemOpen
  \bibfield  {author} {\bibinfo {author} {\bibfnamefont {T.~K.}\ \bibnamefont
  {Ng}},\ }\bibfield  {title} {\bibinfo {title} {Edge states in
  antiferromagnetic quantum spin chains},\ }\href@noop {} {\bibfield  {journal}
  {\bibinfo  {journal} {Phys. Rev. B}\ }\textbf {\bibinfo {volume} {50}},\
  \bibinfo {pages} {555} (\bibinfo {year} {1994})}\BibitemShut {NoStop}%
\bibitem [{\citenamefont {Jiang}\ \emph {et~al.}(2020)\citenamefont {Jiang},
  \citenamefont {Zaanen}, \citenamefont {Devereaux},\ and\ \citenamefont
  {Jiang}}]{jiang_2020}%
  \BibitemOpen
  \bibfield  {author} {\bibinfo {author} {\bibfnamefont {Y.-F.}\ \bibnamefont
  {Jiang}}, \bibinfo {author} {\bibfnamefont {J.}~\bibnamefont {Zaanen}},
  \bibinfo {author} {\bibfnamefont {T.~P.}\ \bibnamefont {Devereaux}},\ and\
  \bibinfo {author} {\bibfnamefont {H.-C.}\ \bibnamefont {Jiang}},\ }\bibfield
  {title} {\bibinfo {title} {Ground state phase diagram of the doped {Hubbard}
  model on the four-leg cylinder},\ }\href
  {https://doi.org/10.1103/PhysRevResearch.2.033073} {\bibfield  {journal}
  {\bibinfo  {journal} {Physical Review Research}\ }\textbf {\bibinfo {volume}
  {2}},\ \bibinfo {pages} {033073} (\bibinfo {year} {2020})},\ \bibinfo {note}
  {arXiv: 1907.11728}\BibitemShut {NoStop}%
\bibitem [{\citenamefont {Chung}\ \emph {et~al.}(2020)\citenamefont {Chung},
  \citenamefont {Qin}, \citenamefont {Zhang}, \citenamefont {Schollwöck},\
  and\ \citenamefont {White}}]{chung_2020}%
  \BibitemOpen
  \bibfield  {author} {\bibinfo {author} {\bibfnamefont {C.-M.}\ \bibnamefont
  {Chung}}, \bibinfo {author} {\bibfnamefont {M.}~\bibnamefont {Qin}}, \bibinfo
  {author} {\bibfnamefont {S.}~\bibnamefont {Zhang}}, \bibinfo {author}
  {\bibfnamefont {U.}~\bibnamefont {Schollwöck}},\ and\ \bibinfo {author}
  {\bibfnamefont {S.~R.}\ \bibnamefont {White}},\ }\bibfield  {title} {\bibinfo
  {title} {Plaquette versus ordinary \$d\$-wave pairing in the \$t'\$-{Hubbard}
  model on a width 4 cylinder},\ }\href {http://arxiv.org/abs/2004.03001}
  {\bibfield  {journal} {\bibinfo  {journal} {Phys. Rev. B}\ }\textbf {\bibinfo
  {volume} {102}},\ \bibinfo {pages} {041106} (\bibinfo {year} {2020})},\
  \bibinfo {note} {arXiv: 2004.03001}\BibitemShut {NoStop}%
\bibitem [{\citenamefont {Ehlers}\ \emph {et~al.}(2018)\citenamefont {Ehlers},
  \citenamefont {Lenz}, \citenamefont {Manmana},\ and\ \citenamefont
  {Noack}}]{ehlers_2018}%
  \BibitemOpen
  \bibfield  {author} {\bibinfo {author} {\bibfnamefont {G.}~\bibnamefont
  {Ehlers}}, \bibinfo {author} {\bibfnamefont {B.}~\bibnamefont {Lenz}},
  \bibinfo {author} {\bibfnamefont {S.~R.}\ \bibnamefont {Manmana}},\ and\
  \bibinfo {author} {\bibfnamefont {R.~M.}\ \bibnamefont {Noack}},\ }\bibfield
  {title} {\bibinfo {title} {Anisotropy crossover in the frustrated {Hubbard}
  model on four-chain cylinders},\ }\href
  {https://doi.org/10.1103/PhysRevB.97.035118} {\bibfield  {journal} {\bibinfo
  {journal} {Physical Review B}\ }\textbf {\bibinfo {volume} {97}},\ \bibinfo
  {pages} {035118} (\bibinfo {year} {2018})},\ \bibinfo {note}
  {arXiv:1705.04450 [cond-mat]}\BibitemShut {NoStop}%
\bibitem [{\citenamefont {Jiang}\ \emph {et~al.}(2018)\citenamefont {Jiang},
  \citenamefont {Weng},\ and\ \citenamefont {Kivelson}}]{jiang_2018}%
  \BibitemOpen
  \bibfield  {author} {\bibinfo {author} {\bibfnamefont {H.-C.}\ \bibnamefont
  {Jiang}}, \bibinfo {author} {\bibfnamefont {Z.-Y.}\ \bibnamefont {Weng}},\
  and\ \bibinfo {author} {\bibfnamefont {S.~A.}\ \bibnamefont {Kivelson}},\
  }\bibfield  {title} {\bibinfo {title} {Superconductivity in the doped
  \$t\$-\${J}\$ model: results for four-leg cylinders},\ }\href
  {https://doi.org/10.1103/PhysRevB.98.140505} {\bibfield  {journal} {\bibinfo
  {journal} {Physical Review B}\ }\textbf {\bibinfo {volume} {98}},\ \bibinfo
  {pages} {140505} (\bibinfo {year} {2018})},\ \bibinfo {note}
  {arXiv:1805.11163 [cond-mat]}\BibitemShut {NoStop}%
\bibitem [{\citenamefont {{von Delft}}\ and\ \citenamefont
  {Schoeller}(1998)}]{delft_bosonization}%
  \BibitemOpen
  \bibfield  {author} {\bibinfo {author} {\bibfnamefont {J.}~\bibnamefont {{von
  Delft}}}\ and\ \bibinfo {author} {\bibfnamefont {H.}~\bibnamefont
  {Schoeller}},\ }\bibfield  {title} {\bibinfo {title} {Bosonization for
  beginners - refermionization for experts},\ }\href@noop {} {\bibfield
  {journal} {\bibinfo  {journal} {Ann. Phys. (Leipzig)}\ }\textbf {\bibinfo
  {volume} {7}},\ \bibinfo {pages} {225} (\bibinfo {year} {1998})}\BibitemShut
  {NoStop}%
\end{thebibliography}
\end{document}